\documentclass[aps,pre,amsmath,amssym,amsthm,superscriptaddress,reprint]{revtex4-1}
\pdfoutput=1
\usepackage[colorlinks=true,
linkcolor=blue,
urlcolor=blue,
citecolor=blue]{hyperref}
\usepackage{cancel}
\usepackage{graphicx}
\usepackage{subcaption}
\usepackage{dcolumn}
\usepackage{bm}
\usepackage{url}
\usepackage{bm}   
\usepackage{bbm}   
\usepackage{verbatim}
\usepackage{stmaryrd}
\usepackage{amsthm}
\usepackage{amssymb}
\usepackage{array,multirow}
\usepackage{empheq}
\usepackage[usenames,dvipsnames]{color}
\usepackage[normalem]{ulem}
\usepackage{tikz}
\usepackage{algorithm}
\usepackage{algcompatible}
\usepackage{algpseudocode}

\newcommand{\cob}{\color{blue}}

\newcommand{\LS}{Szil\'{a}rd}

\newcommand{\kleinV}{v}
\DeclareMathOperator{\sign}{sign}

\begin{document}

\title{The physical observer in a \LS\ engine with uncertainty}

\author{Dorian Daimer}
\email{ddaimer@hawaii.edu}
\affiliation{Department of Physics and Astronomy\\ University 
of Hawaii at M\=anoa, Honolulu, HI 96822, USA}

\author{Susanne Still}
\email{sstill@hawaii.edu}
\affiliation{Department of Physics and Astronomy\\ University 
of Hawaii at M\=anoa, Honolulu, HI 96822, USA}

\begin{abstract}
Information engines model ``Maxwell's demon" mechanistically. However, the demon's strategy is pre-described by an external experimenter, and information engines are conveniently designed such that observables contain complete information about variables pertinent to work extraction. In real world scenarios, it is more realistic to encounter partial observability, which forces the {\em physical} observer, an integral part of the information engine, to make inferences from incomplete knowledge.  
Here, we use the fact that an algorithm for computing optimal strategies can be directly derived from maximizing overall engine work output \cite{CB}. For a simple binary decision problem, we discover interesting optimal strategies that differ notably from naive coarse graining. They inspire a model class of simple, yet compelling, parameterized soft partitionings of the observable.

We analyze and compare three different observer classes: (1) optimal observers, (2) observers limited to the parameterized soft partitionings introduced here and (3) observers limited to coarse graining. While coarse graining based observers are outperformed by the other two types of observers, there is no difference in performance between unconstrained, optimal observers and those limited to soft partitionings. The parameterized soft partitioning strategies allow us to compute key quantities of the decision problem analytically.

\end{abstract}

\maketitle

\section{Introduction}\label{Sec:Intro}
Thermodynamics provides a physical foundation for quantifying information. This was first made explicit by \LS's 1929 thought experiment \cite{szilard-german}, motivated by thoughts about Maxwell's ``demon" \cite{maxwell-demon, maxwell1871theory}. The Gedankenexperiment can be understood as an ``information engine" that converts between information and work.
\LS\ suggested that the accessible volume of a one-particle gas in a container can be reduced by insertion of a divider without doing work, instead of compression with a piston \cite{szilard-german}. Compression costs at least $kT \ln(V_i/V_f)$ joules if the gas is compressed from volume $V_i$ to volume $V_f < V_i$ in an isothermal quasi-static process at temperature $T$ ($k$ denotes the Boltzmann constant), yet if the insertion of the divider is sufficiently slow, it does not require work \cite{zurek1986maxwell}.
Both insertion and compression create a work potential, because work can be coupled out of the system via isothermal, quasi-static expansion of the gas at temperature $T$, using the partition as a piston. If the work potential was created by inserting the partition, knowledge of the particle's location is needed in order to reliably extract work from the system. Thereby, the demon can leverage information to turn thermal fluctuations into work. 

The demon's function as a real world observer, embedded and itself a part of the information engine, is thus reduced to correlating an observation to a binary decision about the protocol to be applied, contingent on which side of the container is empty. The demon requires a physical memory for this task, because the location of the particle, before the piston starts to move, has to remain correlated to the protocol by which the piston moves throughout the work extraction process. If the divider is inserted in the middle of the container, then one bit of information can be converted to $kT \ln(2)$ joules of work. 

The past two decades have seen enormous progress in measuring and manipulating microscopic objects, which has enabled experimental verification, and spurred further analysis of information engines \cite{toyabe2010experimental, berut2012experimental, mandal2012work, barato2013autonomous, horowitz2013imitating, diana2013finite, koski2014experimental, koski2014SEszilard, jun2014high, koski2015chip, chapman2015autonomous, martinez2016brownian, hong2016experimental, camati2016experimental, gavrilov2016erasure, boyd2016identifying, boyd2017correlation, boyd2017leveraging, strasberg2017quantum, mcgrath2017biochemical, gavrilov2017direct, chida2017power, cottet2017observing, ciampini2017experimental,  kumar2018nanoscale, paneru2018lossless, admon2018experimental, masuyama2018information, stopnitzky2019physical, brittain2019biochemical, stevens2019quantum, ribezzi2019large, peterson2020implementation, paneru2020efficiency, paneru2020colloidal, saha2021maximizing, dago2021information, saha2023information}.  

But thermodynamics also provides a physical foundation for intelligent {\em processing} of information. Strategies for representing available data without unnecessarily racking up thermodynamic costs 
can directly be derived from physical bounds on dissipation \cite{CB}. 
This requires careful thought about the demon's choices.

In \LS’s Gedankenexperiment, the demon's choice of how to represent observations in memory reflects how ``intelligent" the demon is, together with its actions on the system, the latter being based on a decision about the direction of piston movement required for volume expansion. These choices are made {\em a priori} by an external experimenter, and the information engine is carefully designed such that the observable contains complete information about the quantity that needs to be know for successful work extraction, and therefore the best data representation and action strategy is rendered trivial by design. 

A common characteristic of most situations involving {\it real world} observers, biological and artificial alike, is partial observability. It arises from intrinsic limitations on what  
can be observed and what can be controlled by a physical agent. Usually knowledge of the variables pertinent to control
has to be {\em inferred} from available data. Observations are typically correlated with, but not identical to, neither in a one-to-one relationship with, the quantities that need to be known for successful control. 
Partial observability is so common, because most real world observers are subject to physical constraints dictating what can be measured and what can be controlled, and these constraints typically exclude the convenient, one-to-one relationship between observables and variables pertinent to control, that is built into most models of information engines \cite{mandal2012work, barato2013autonomous, diana2013finite, boyd2016identifying, boyd2017correlation, boyd2017leveraging, strasberg2017quantum, stopnitzky2019physical}.
Examples abound: Animals with limited sensors and limited actuators, robots in complex environments, networks of neurons that have access to visual data and need to infer what is in the image, so that down the information processing stream, a useful action can be taken based on this inference. 

Together with partial observability, another generalization is required to study the thermodynamics of decision making under uncertainty. An information engine based on \LS's ideas uses information to extract work in an isothermal process, 
it does not use a temperature gradient to convert heat to work, as a regular heat engine does. Therefore, in a cyclic process that includes the physical memory, such an information engine cannot produce net work output, on average, because running the memory converts at least as much work to information as information is re-converted to work when it is being used \cite{szilard-german, zurek1986maxwell, koski2014experimental, parrondo2015thermodynamics, CB, landauer1961irreversibility}.  
Generalizing to engine processes that allow information to be created at a lower temperature than the temperature at which it is converted to work, lets information engines be treated in one framework together with heat engines \cite{CB}. In the case of \LS's engine, the result is a Carnot process, if the demon (or observer) uses a data encoding and decoding that {\it maximizes} average net engine work output at each value of the temperature ratio \footnote{\label{FN:1}This is further discussed in \cite{stilldaimer2022} and in a forthcoming article.}.
However, if the observer uses a suboptimal data representation, then Carnot efficiency cannot be achieved.  

Extension of the information engine paradigm to generalized, partially observable engines provides a foundation to the study of the physics of information processing and decision making under uncertainty. 
Optimal observer strategies for data representation and inference can be derived from maximization of the engine's average net work output \cite{CB}. 
The resulting thermodynamically optimal 
memory making 
strategies  
have complex, nontrivial physical characteristics, even for simple examples \cite{stilldaimer2022}.
While in \LS's original setup coarse graining is optimal (all observations of the particle to the left/right of the divider can be lumped together, which is equivalent to coarse graining the configuration space accordingly), 
in the general case, thermodynamically optimal data representations do not always result in coarse graining, because they can be probabilistic maps. The resulting ``soft" partitioning of observables is much less common in physics, and perhaps less intuitive. 

To improve intuition, we investigate here a stylizedly simple example: a partially observable \LS\ engine with two distinct types of observations, those without uncertainty, and those with maximum uncertainty. 
The analysis of this rudimentary binary decision problem with uncertainty sheds light on emerging optimal memories in the simplest case, and thus improves the most basic understanding of the thermodynamics involved in decision making under uncertainty. 
It is important to fully understand the physics of binary decision making under uncertainty in the presence of thermodynamic constraints, as this  constitutes one of the crucial building blocks of energy efficient computation. Previously thermodynamic frameworks have been used to model bounded rational decision making via maximization of a free energy functional \cite{ortega2013thermodynamics, ortega2015information}, resulting in a different optimization problem than the one considered here and they fail to include a physical model of the observer.

Our analysis pays attention to a fact \LS\ pointed out: the full model of an information engine (and therefore also the accounting of entropy production) has to entail not only the engine's work medium, but also an explicit, physical model of the engine's memory.
In Sec. \ref{Sec:Model} we describe our model, which is a 
parameterized extension of the example introduced in \cite{CB}. 

We find that the physical process governing optimal observer memories is indeed easy to understand (Sec. \ref{Sec:Results}). The optimal strategies we find algorithmically inspire the use of a parameterized model class, consisting of soft partitions, outperforming naive coarse graining of the observable (Sec. \ref{Sec:Approx}). These parameterized soft partitions are as good as optimal strategies, while computationally less cumbersome. Importantly, they capture the intuition behind optimal strategies in a clearly interpretable form and allow analytical calculation of key quantities characterizing optimal observers.

\section{Model}
\label{Sec:Model}

We consider here an extremely simple model problem, in which an observation either leads to certainty or carries with it maximal uncertainty. 

As the information engine's work medium, we use a single-particle gas in a container of unit length in all three spatial directions, endowed with a divider that can function as a piston. This contraption is used as the work extraction device, together with some machinery capable of implementing isothermal, quasi-static expansion of the one-particle gas, in accordance with \LS's original idea \cite{szilard-german} (see also \cite{zurek1986maxwell, parrondo2015thermodynamics, CB, stilldaimer2022}). The difference is that the divider shape is chosen such that it models partial observability when only the particle's $x$ position is available to the observer. We elaborate on this below. 
As the memory, we use another one-particle gas in a different container, endowed with dividers which can also function as pistons.
Our physical memory making protocol can implement not only deterministic, but also probabilistic memories \cite{stilldaimer2022}.

Thermodynamic costs and gains are in reference to volume changes of a single-particle gas \footnote{In order to apply thermodynamic concepts to the analysis of the single-particle engine, we may imagine arbitrarily many such engines running in parallel or consider many runs of a single engine.}: isothermal, quasi-static compression from $V_i$ to $V_f$ at temperature $T$ requires, on average, work input of \mbox{$W = kT \ln(V_i/V_f)$} joules ($V_i > V_f$), and, conversely, volume expansion allows, on average, for the extraction of \mbox{$-W = kT \ln(V_f/V_i)$} joules ($V_i < V_f$). We adopt the convention that energy flowing into the gas is positive. Thus, work done on either, the work medium, or the memory, is reflected by a positive value of $W$, while work done by the engine is reflected by a positive value of $-W$.  

The data available to the observer is the $x$ position of the particle in the work medium, at time $t_M$, $x(t_M)$, which we abbreviate by $x$. The probability density is constant inside the container, and since the container has unit length, we have $\rho(x) = 1$.

To convert thermal motion of the particle in the work medium to work, in an isothermal process, the observer needs to know which side of the container is empty when the work extraction protocol is run. Define a random variable, $u$, which has outcome $1$ whenever the left side of the container is empty, and $-1$ when the right side is empty. 

The work medium's geometry then sets the correlation between $x$ and $u$ \cite{CB, stilldaimer2022}. 
Let the divider split the volume in half, so that $p(u) = 1/2$, which is, without knowledge of $x$, the observer's best guess regarding the {\em a priori} probability of either side being empty. 

Let the work medium container have two regions of equal width along the $x$-axis, one at either side of the divider, within which measurements provide certainty about $u$, and a region of width $w$ in the middle with maximal uncertainty (henceforth referred to as ``uncertain region"):
\begin{subequations}
\label{Eq:Regions}
\begin{align}
&\text{left certain region:} \qquad\qquad\quad  \mathcal{X}_L \equiv [-1/2, -w/2] \\
&\text{uncertain region (middle):} \qquad\!\! \mathcal{X}_M \equiv (-w/2, w/2) \\
&\text{right certain region:} \qquad\qquad\quad\!\!\! \mathcal{X}_R \equiv [w/2, 1/2].
\end{align}
\end{subequations} 
The work medium is characterized by $p(u|x)$, the probability of the left (or the right) side being empty, given measurement outcome $x$. Since $u$ is binary, one of these two functions suffices, as $p(u=-1|x) = 1-p(u=1|x)$. Our model is given by 
\begin{equation} 
\label{Eq:pugx}
    p(u=1\vert x) = 
    \begin{cases}
    \quad 0~ &  x \in \mathcal{X}_L\\
    \quad {1 \over 2} & x \in \mathcal{X}_M\,. \\ 
    \quad 1~ & x \in \mathcal{X}_R \\
    \end{cases}
\end{equation}

\begin{figure}[h]
\includegraphics[width=0.95\linewidth]{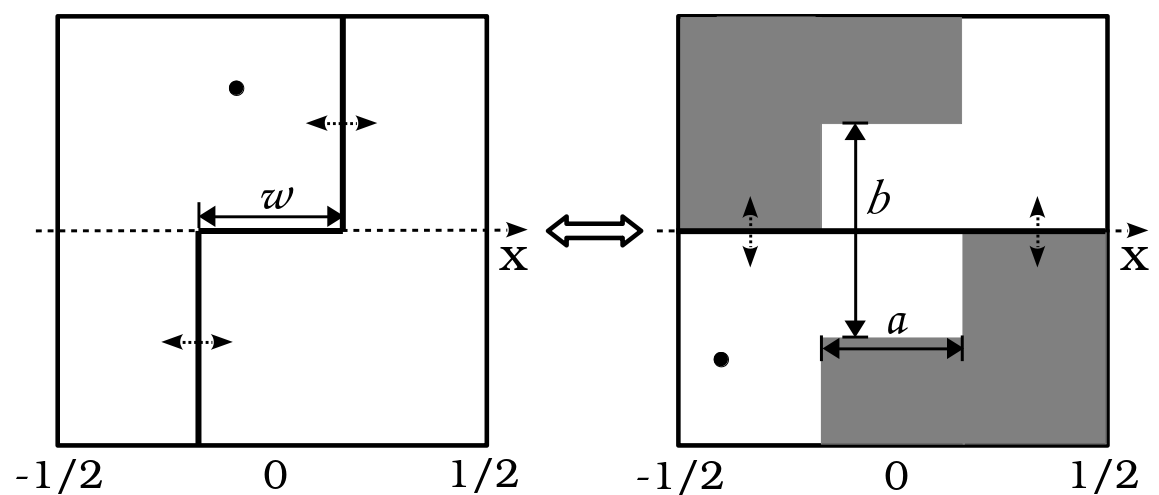}
\caption{
Two realizations of the information engine's work medium described by Eq. (\ref{Eq:pugx}). They are equivalent to each other for $2ab/(1-a) = w/(1-w)$. Sketched: $w=1/3$ (left), $a=1/3$ and $b=1/2$ (right).}
\label{Fig:CB_box}
\end{figure}

This model describes a parameterized version of the modified \LS\ engine introduced in \cite{CB}, shown on the right in Fig. \ref{Fig:CB_box}. It also describes an equivalent \LS\ engine, depicted on the left of Fig. \ref{Fig:CB_box}, that uses a divider, the shape of which reflects Eq. (\ref{Eq:pugx}). Conveniently, any binary decision problem, described by $p(u|x)$, can be modelled by a partially observable \LS\ engine with a divider shaped to follow $p(u=1|x)$, see App. \ref{App:BDP} for a detailed discussion of the mapping. We find this intuitive to think about, and therefore use it for the analysis performed in this paper. But let us take a brief look at the two equivalent models and their relation to each other.

The example of \cite{CB} constrains a single gas particle to an accessible volume (white) of the container (inaccessible regions are shaded dark gray in Fig. \ref{Fig:CB_box}). The divider (black line) can move up and down, e.g. parallel to the $y$-axis, but the observer only has access to the particle's $x$ position. The empty side of the container (above or below the divider) has to be inferred from partial information, because the particle's $y$ position cannot be directly observed. Manipulable and observable degrees of freedom are correlated by the chosen shape of the accessible region. 
This model can be parameterized by the volume of the uncertain region in the middle, that is, by its width, $a$, and height, $b$, ($a,b \in [0,1]$).

To map between the two equivalent models depicted in Fig. \ref{Fig:CB_box}, 
note that the two resulting information engines will behave identically whenever the ratio of the volume of the uncertain region, $V_u$, to the total volume of both certain regions, $V_c$, is the same for both engines. 
For the one parameter engines on the left of Fig. \ref{Fig:CB_box}, this ratio is given by $V_u/V_c = w/(1-w)$. For the two parameter engines on the right, the ratio is 
$V_u/V_c = 2ab/(1-a)$. Therefore, any choice of $a$ and $b$ that satisfies \mbox{$2ab/(1-a) = w/(1-w)$} for any $w \in (0,1)$ produces two physically equivalent engines \footnote{We only consider $w \in (0,1)$ since $w=0$ corresponds to a regular \LS\ engine while $w=1$ corresponds to an engine with no usable information at all. Both of these cases have trivial optimal strategies (coarse grain into two symmetric states and do nothing respectively).}. 
One easy way to think about the mapping is to fix $a=1/3$, then $b=w/(1-w)$ maps any one parameter engine with $w \leq 1/2$ to a corresponding two parameter engine. For $w > 1/2$, we fix $b=1$ and choose $a= w/(2-w)$ to map from one type of engine to the other. 

\subsection{Available usable information}
We use theory from \cite{CB}, and closely follow the analysis performed in \cite{stilldaimer2022}. 
In the interest of readability, we will briefly review pertinent quantities below. 
For further details we refer the reader to \cite{CB, stilldaimer2022}.

The observer uses the knowledge it has stored in memory to extract work via isothermal expansion of the one-particle gas in the work medium. Before we address how much information the observer actually keeps in memory, we want to know the baseline: how much usable information is available from the observable? That is, how much information does the observable data, $x$, contain about the relevant quantity, $u$? 
The mutual information between $u$ and $x$ \footnote{Brackets, $\langle \cdot \rangle_p$ denote averages over the probability distribution $p$.},
\begin{equation} \label{Eq:Iux}
    I[u,x] := \left\langle \ln\left[{p(x,u) \over p(x) p(u)} \right] \right\rangle_{p(x,u)}~,
\end{equation}
quantifies the total reduction in uncertainty the observer has about which side of the container is empty, upon receiving the particle's $x$ position:
\begin{equation} \label{Eq:Iux_ent}
    I[u,x] := H[u] - H[u \vert x],
\end{equation}
with Shannon entropy $H[u] := - \langle \ln\left[{p(u)} \right] \rangle_{p(u)}$, and conditional entropy $H[u\vert x] := - \langle \ln\left[{p(u\vert x)} \right] \rangle_{p(u,x)}$. 
Since $u$ is a binary random variable with $p(u)=1/2$, its entropy is given by $H[u]=\ln(2)$ nats, or equivalently one bit. 
Using Eq. (\ref{Eq:pugx}), the conditional entropy is  
\begin{eqnarray}
   H[u \vert x]
    &=& 2 \int_{-w/2}^{w/2} \!\!\!dx\,{1 \over 2} \ln(2) = w \ln(2).
\end{eqnarray}
The maximum available usable information is thus linear in the total volume of both certain regions together (the volume reduces to their combined width along the $x$-axis, which is $1-w$, because the container has unit transverse area):
\begin{equation}\label{Eq:TotalIrel}
I[u, x] \equiv I_{\rm u}^{\rm max}(w) = (1-w)\ln(2).
\end{equation}

\subsection{Usable and total information in memory} \label{Sec:CB}
To capture any of this information, a memory needs to be made that is stable on the timescale over which it is needed (namely the duration of the work extraction protocol), because the measurement outcome is available only transiently. Thus the available information $I_{\rm u}^{\rm max}(w)$ is lost, unless it, or part of it, is saved in memory. Let the physical memory's state be given by $m$. 
The information that the memory captures about the observable,
\begin{equation}
I_{\rm m} \!\equiv\! I[m,x] \!=\! H[m] \!-\! H[m|x] \!=\! \left\langle \!\ln\left[ {p(m\vert x) \over p(m)} \right]\right\rangle_{\!\!p(m,x)}\!, \label{Eq:Imem}
\end{equation}
depends on the conditional probability distributions $p(m\vert x)$ that characterize the stochastic map from observables to memory states. 

If this map contains no randomness, we can write 
\begin{equation}
\label{Eq:pmgx_det}
    p(m\vert x) = \delta_{m f(x)} = 
    \begin{cases}
    \quad 1 & {\rm if}\;\; m=f(x)\\
    \quad 0 & {\rm else}
    \end{cases}
\end{equation}
where $\delta$ is a function $\mathbb{Z} \times \mathbb{Z} \to \{0,1\}$ inspired by the Kronecker-Delta, and $f(x)$ is a function $\mathbb{R} \to \mathbb{Z}$ that maps observations $x$ to memory states $m$.  We call memories with this property {\em deterministic}. Their conditional entropy is zero, $H[m\vert x]=0$, and thus $I[m,x] = H[m]$. Deterministic memories are equivalent to coarse graining of the observable space.

How much of the memorized information is predictive of the relevant quantity and can thus be used to extract work? It is the mutual information retained in memory about which side of the container is empty:
\begin{equation}
\label{Eq:Irel-def}
I_{\rm u} \!\equiv\! I[m,u] \!=\! H[u] \!-\! H[u|m] \!=\! \left\langle \!\ln\left[ {p(u\vert m) \over p(u)} \right] \right\rangle_{\!\!p(m,u)}\!.
\end{equation}
Here, $u$ is inferred from $m$ by calculating
\begin{equation}
p(u\vert m) = \langle p(m|x)p(u|x) \rangle_{\rho(x)}/p(m), \label{Eq:pum_1} 
\end{equation}
but since $p(u|x)$ is symmetric around $x=0$, we have $p(u) = 1/2$.
The probability of left/right side being empty, given memory state $m$, is thus determined from the ratio of two different averages: the memory-making rule $p(m|x)$ is averaged over the likelihood of observing $x$, {\em given} $u$, and then compared to the probability of memory state $m$ occurring given data $x$, averaged over $\rho(x)$
\begin{equation}\label{Eq:pm}
    p(m) =\langle p(m|x)\rangle_{\rho(x)},
\end{equation}
also called the ``weight" of state $m$.

It is important to remember that usable information \footnote{In previous papers \cite{CB, stilldaimer2022} we called usable information about the relevant quantity ``relevant information'', in accordance with the wording of \cite{IB}.} retained in memory, $I_u$, is a functional of $p(m\vert x)$ (see Appendix \ref{App:Info}). It is also a functional of $p(u|x)$, which depends on the physical constraints imposed by the work medium and by what the observer has access to. In our model, it depends on the geometry of the divider in the work medium, specifically on the width $w$ of the uncertain region.

\subsection{Physical encoding and decoding---cost and gain of a physical memory} \label{Sec:PhysEnc}
While the memory making strategy, abstractly, is a probabilistic rule for assigning data to memory, which is mathematically specified by $p(m|x)$, physically, this encoding is implemented by a sequence of actions, ${\bf s}(x, t)$, performed on a system that is used as memory. This sequence of operations results in the creation of distinct memory states. The initialization of the sequence of actions depends on the observed value of $x$, but the final memory states can be read out by a pre-specified physical decoding mechanism independent of $x$.
Therefore, after the data-triggered sequence of actions on the physical memory is complete, a measurement has been committed to memory, and the observer then has continuing access to the memory's state until it is deleted after work extraction, to close the engine's cycle. 

\begin{figure*}[t!]
\centering 
\includegraphics[width=0.95\textwidth]{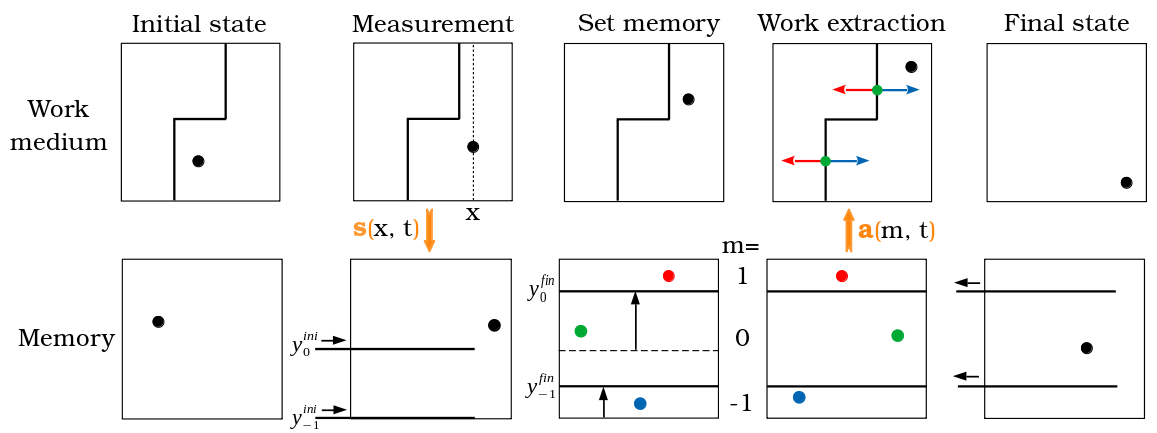}
\caption{Physical encoding (memory making) procedure, ${\bf s}(x, t)$, is shown in the bottom panel, in response to the availability of a measurement (second panel). Initial (ini) and final (fin) divider positions delineating memory states $m$ are denoted by $y_m$ (see text). Physical decoding (work extraction) procedure consists of a memory-dependent protocol ${\bf a}(m,t)$, applied to the work medium (fourth panel).  Parameters: $w=0.3$; $\tau=2\!+\!\epsilon$; $x \in \mathcal{X}_R$.}
\label{Fig:MemoryMaking}
\end{figure*}

There are many possible choices for physical memories, and obviously an electronic memory would be a practical choice. However, for simplicity of the exposition we choose a one-particle gas in a container with unit length along the $y$-axis and volume $V$, into which dividers can be inserted perpendicular to the $y$-axis. These dividers can act as pistons, moving parallel to the $y$-axis. We choose the $y$-axis simply for semantic convenience, to avoid confusion with the work medium, in which the particle's $x$ position is measured. The same assumptions are made as in the typical Gedankenexperiment involving any \LS\ engine: the dividers are assumed to be very thin, and are inserted slow enough, so that the probability of doing work on the single particle during incision is zero.

The process of creating a $K$ state memory involves the following steps: at time $t_M$, the $x$ position of the particle is made available to the observer, and depending on the value of $x$, the observer selects $K-1$ positions $y^{\rm ini}_m (x)$ at which dividers are inserted into the memory container. While the positions are chosen at time $t_M$, the process of insertion can occur sufficiently slowly. After the dividers are fully inserted, they are moved quasi-statically, parallel to the $y$-axis, to positions $y^{\rm fin}_m$, which do not depend on $x$. The memory is then stable for the remaining cycle and can be read out at any time by accessing the $y$ position of the particle in the memory container. This physical encoding and decoding mechanism, illustrated in Fig \ref{Fig:MemoryMaking}, is the same mechanism as proposed in \cite{stilldaimer2022}. 
Depicted are the work medium and the memory. In the second panel, the work medium's state is sketched at time $t_M$. The resulting $K-1$ insertion positions in the memory, $y^{\rm ini}_m (x)$, are shown in the bottom panel, second drawing, and the final positions in the memory, $y^{\rm fin}_m$, are shown in the third and fourth drawing in the bottom panel. 

The initial positions are chosen to be \mbox{$y^{\rm ini}_m (x) = \sum_{m'\leq m} p(m'|x)$}. This choice ensures that the relative volumes between dividers are initially equal to the probabilities $p(m|x)$ that characterize the observer's strategy.
The final positions are $x$-independent, set to $y^{\rm fin}_m = \sum_{m'\leq m} p(m')$, ensuring that the relative volumes between dividers are equal to the average probabilities $p(m) = \langle p(m|x) \rangle_{\rho(x)}$, which are the same for any outcome $x$. Only the initial step of this protocol requires $x$, and this moment thus sets the time of measurement, $t_M$.

In this process, there are different possibilities regarding which dividers the particle initially gets trapped in between (indicated by different colors in Fig. \ref{Fig:MemoryMaking}). If the particle happens to be trapped between two dividers, whose distance shrinks during memory making, then the resulting compression costs energy. The average work associated with isothermal volume changes of the one-particle gas at temperature $T$ is $W = kT \ln(V_i/V_f)$, where $V_i = \kleinV_i V$ denotes the initial volume between the dividers, expressed in terms of the memory container's total volume $V$. Similarly, $V_f = \kleinV_fV$ denotes the final volume. 
If $v_i > v_f$, then compression occurs, which costs work in the amount of $W = kT \ln(\kleinV_i/\kleinV_f)$. 
If, on the other hand, the distance between the dividers increases, then work in the amount of $-W = kT \ln(\kleinV_f/\kleinV_i)$ can be extracted in this process. 

With probability $p(m\vert x)$ the particle initially gets trapped between dividers that, at the end of the memory making process, enclose the volume corresponding to memory state $m$.
Thus, the volume fraction changes from $\kleinV_i=p(m \vert x)$ to \mbox{$v_f=p(m)$} with probability $p(m\vert x)$, and therefore, the average work cost, given this particular realization of $x$ is \mbox{$W(x) = kT \sum_m p(m \vert x) \ln[p(m \vert x)/ p(m)]$}. The overall average cost of running this memory is thus \mbox{$W = \langle W(x) \rangle_{\rho(x)} = kTI_m$} \cite{CB, stilldaimer2022}. 

Labels associated with memory states, $m$, i.e. the values associated to the outcomes of random variable $m$, can be chosen freely in the mathematical description. In Fig. \ref{Fig:MemoryMaking}, and throughout the paper, we use the labels \mbox{$m\in \{-1,1\}$} for memories with two states, and  the labels \mbox{$m\in \{-1,0,1\}$} for three-state memories. 

Since $K-1$ dividers are needed to create $K$ partitions of the volume, two parameters determine the physical memory making process for any two state memory, namely $y^{\rm \alpha}_{-1}$, where $\alpha$ can be either the initial (ini) or the final (fin) position of the divider. The distances to the walls are $\Delta y^{\rm \alpha}_{-1} =  y^{\rm \alpha}_{-1}$ and $\Delta y^{\rm \alpha}_{1} = 1-y^{\rm \alpha}_{-1}$. In certain contexts it is convenient to display these initial and final distances, because that type of display compactly conveys information about the volume changes of the partitions created in the memory container, which correspond to the memory states. This provides intuition about how the associated costs appear. We use this type of display in Sec. \ref{Sec:PhysicalCoding}, in Fig. \ref{Fig:PhysicalCB_c03}.
Memories with three states, are determined by the four parameters $y^{\rm \alpha}_{-1}$ and $y^{\rm \alpha}_{0}$, and the distances to the walls are given by \mbox{$\Delta y^{\rm \alpha}_{-1} = y^{\rm \alpha}_{-1}$}, and \mbox{$\Delta y^{\rm \alpha}_{1} = 1 - y^{\rm \alpha}_{0}$}, while the distance between the two dividers is \mbox{$\Delta y^{\rm \alpha}_{0} = y^{\rm \alpha}_{0} - y^{\rm \alpha}_{-1}$}. 

After the dividers have reached their final positions, the observation is committed to a stable memory, to which the observer has access henceforth. The memory state, $m$, can then be read off from the particle's $y$ position.
For three-state memories, as depicted in Fig. \ref{Fig:MemoryMaking}: if \mbox{$0 < y \leq y^{\rm fin}_{-1}$}, then $m=-1$; if \mbox{$y^{\rm fin}_{-1} < y \leq y^{\rm fin}_{0}$}, then $m=0$; and if \mbox{$y^{\rm fin}_{0} < y \leq 1$}, then $m=1$. Two-state memories work analogously: if \mbox{$0 < y \leq y^{\rm fin}_{-1}$}, then $m=-1$; if \mbox{$y^{\rm fin}_{-1} < y \leq 1$}, then $m=1$.

This proposed physical method is a concrete, rudimentary way of ``writing down" $x$ to some precision.
Note that while information about the particle position in the work medium, $x$, is required for the choice of the locations $y^{\rm ini}_{m} (x)$, it is not required when the memory is read out. 
Thus, two parameters $y^{\rm fin}_{-1}$ and $y^{\rm fin}_{0}$ fully determine the read-out for three-state memories, as depicted in Fig. \ref{Fig:MemoryMaking}. Similarly, the read-out of two state memories is fully specified by one parameter, $y^{\rm fin}_{-1}$. 

Depending on the given memory state, $m$, the observer then chooses the protocol ${\bf a}(m, t)$ to apply to the work medium (fourth panel in Fig. \ref{Fig:MemoryMaking}).
The state of the memory lets the observer infer the probability of either side of the work medium being empty using Eq. (\ref{Eq:pum_1}).  
The inference is turned into an action on the work medium, in such a way that it utilizes information to extract work from thermal fluctuations.
Physically, this is implemented by a mechanism that applies the work extraction protocol, ${\bf a}(m, t)$, to the work medium. This procedure can be understood as the decoding part of the physical code book.

In our context, this enables isothermal expansion towards the side of the work medium container that has the larger probability of being empty, $u^*(m) =$ \mbox{arg$\max_u p(u|m)$}, up to a residual volume, $\gamma(m) V'$, where $V'$ is the volume of the work medium container, attached to a heat bath at temperature $T'$. 
The purpose of leaving a residual volume is to avoid compression to zero volume due to an inference error.
The optimal size of the residual volume is given by the probability that the inference is wrong, i.e. $\gamma(m) = p(u \neq u^*|m)$ (recall that $u$ is binary) \cite{stilldaimer2022, sagawa2012thermodynamics}.

One can conceptualize the movement of the divider during work extraction as proceeding in two parts. The two vertical parts of the divider move from positions $x=\pm w/2$ to $x=0$. Thereafter all parts of the divider move together until the vertical part reaches its final position, leaving a residual volume of $\gamma(m)V'$ on the side that is most likely to be empty. 

This mechanistic implementation of the work extraction protocol requires only the specification of (i) direction, and (ii) residual volume on the smaller side. Those can be condensed into one variable, for example the volume on the left side at the end of protocol ${\bf a}(m, t)$, $V_{\ell}(m)$, in units of the work medium container's total volume, $V'$: $\kleinV_{\ell}(m) = V_{\ell}(m) / V'$ \footnote{Due to symmetry there is an obvious choice to specify either the left or right final volume fraction, however, one of those suffices.}. 
If the inference suggests that the left side of the container is most likely empty, 
$u^*(m) = 1$, then $\kleinV_{\ell}(m) = \gamma(m)$. If the inference suggests that the right side of the container is most likely empty, $u^* = -1$, then $\kleinV_{\ell}(m) = 1-\gamma(m)$. 
The $y^{\rm fin}_{m}$, together with $\kleinV_{\ell}(m)$, fully specify the $x$-independent physical decoding scheme. 

For the example depicted in Fig. \ref{Fig:MemoryMaking}, the protocol ${\bf a} (m=1, t)$, applied to the work medium, results in volume reduction on the left side of the divider, with final volume $V_{\ell}(m=1) = 0$, while the protocol ${\bf a} (m=-1, t)$ results in volume reduction on the right side (with $V_{\ell}(m=-1) = V'$), and ${\bf a} (m=0, t)$ corresponds to no change in the work medium.

When the work medium is manipulated in this way, information saved in memory is leveraged to extract work, on average. We allow the engine to perform work extraction at a higher temperature $T'>T$ than the temperature at which the memory is formed and destroyed.  
At the beginning of the work extraction protocol, the particle in the work medium container occupies a volume ${V}_{\rm i}'$. The volume relative to the total volume, $V'$, of the container, $\kleinV_{\rm i}' = {V}_{\rm i}'/V'$, is exactly equal to the {\it a priori} probability that either left or right side of the work medium is empty, $\kleinV_{\rm i}' = p(u) = 1/2$. The final occupied volume depends on whether the inference was correct or not. 
If the observer correctly inferred the empty side, then the occupied volume expands to ${V}_{\rm f}' = (1-\gamma(m))V'$, that is, the relative final volume is \mbox{$\kleinV_{\rm f}' = p(u^*|m) \geq 1/2$}, implying an average work extraction of \mbox{$-W'(m,u^*) = kT' \ln\left(\kleinV_{\rm f}'/\kleinV_{\rm i}'\right) = kT' \ln\left[ p(u^*|m)/p(u^*) \right]$}. This case occurs with probability $p(u^*|m)$. The other case occurs with probability $1-p(u^*|m)$, and results in a volume reduction to $\kleinV_{\rm f}' = 1- p(u^*|m)$, and hence work done on the work medium, on average, in the amount of $-W'(m,u \neq u^*) = kT' \ln\left[ p(u\neq u^*|m)/p(u \neq u^*) \right]$. Thus, for each outcome of $m$, we have, $-W'(m) = \sum_u p(u|m) \ln \left[ p(u|m) / p(u)\right]$. Averaging over $m$, we see that the average work derived in this way is proportional to the usable information captured in memory: $-W' = kT I_{u}$ \cite{CB,stilldaimer2022}.

The chosen physical memory encoding and decoding procedure saturates limits on dissipation. 
Note that, due to the fact that the operations are isothermal, quasi-static processes, dissipated (absorbed) heat equals work done on (by) each system (memory/work medium), i.e., $Q = -W$ and $Q' = -W'$ (all quantities are averages). Thus, heat is dissipated in the amount of $-Q-Q' = k \left(T I_m - T' I_u \right)$. This is the general lower bound for any partially observable information engine \cite{CB}, as average heat generated during the memory making process is lower bound by $-Q \geq kT I_m$, and average heat absorbed during work extraction is upper bound by $Q' \leq kT' I_u$. Therefore, the overall dissipation is no less than $-Q-Q' \geq k \left(T I_m - T' I_u \right)$. 

\subsection{Optimal memory-making strategies}
The maximum net average work output deliverable by this type of partially observable information engine is thus achievable by our memory making and work extraction protocol design. It is a positive quantity when $-W'$ is positive, after subtracting memory costs, $W$: we define $W^{\rm net}_{\rm out}:= -W'-W$, which is given by \mbox{$W^{\rm net}_{\rm out} = kT'I_u - kTI_m$}. In units of $kT'$ it is
\begin{eqnarray}
{1 \over kT'} W^{\rm net}_{\rm out} = I_u - {T\over T'} I_m.
\label{Eq:Woutengine}
\end{eqnarray}
The average work output depends not only on the temperature ratio, but also, on the statistical structure of the partial observability, characterized by $p(u|x)$, and on the data representation, characterized by $p(m|x)$. 
While $p(u|x)$ is given and fixed by the physical constraints of the setup, $p(m|x)$ has to be chosen by the observer, and it is this choice that we address here \footnote{It helps to take another careful look at these dependencies, which we do in Appendix \ref{App:Info}}.

The optimal strategy can simply be calculated by maximizing the engine's net average work output, Eq. (\ref{Eq:Woutengine}), over all possible data representation schemes, characterized by the probabilistic mapping from data to memory, $p(m|x)$ (alternatively, dissipated heat can be minimized, resulting in an equivalent optimization problem with the same solutions).
This argument directly justifies \cite{CB} the ``Information Bottleneck Method" (IB) \cite{IB}. The optimization is: 
\begin{eqnarray}
\label{Eq:IB}
&&\max_{p(m\vert x)} \left(I[m,u] - {T \over T'} I[m,x] \right)\\
&&{\rm subject \; to:\;} \sum_m p(m\vert x) = 1; \; \forall x \notag,
\end{eqnarray}
where normalization of the probability measure is ensured by the constraint $\sum_m p(m\vert x) = 1; \; \forall x$. 

The temperature ratio sets the trade-off between energetic cost of the memory and potential energetic gains \cite{CB}. As a shorthand, define the ratio of higher to lower temperature as $\tau = T'/T >1$. For any fixed $\tau$, there is an optimal solution. All solutions have to fulfill a set of self-consistent equations. The optimal assignments at each $\tau$, $p_{\rm opt}^{\tau}(m\vert x)$, are easily calculated from the saddle-point condition, and are consistent with average memory state probabilities, $p_{\rm opt}^{\tau}(m)$, and optimal inferences, $p_{\rm opt}^{\tau}(u\vert m)$, computed with Eqs. (\ref{Eq:pm}) and (\ref{Eq:pum_1}), respectively \cite{IB}:
\begin{subequations}
\label{Eq:IBalg}
\begin{align}
p_{\rm opt}^{\tau}(m\vert x) &= {p_{\rm opt}^{\tau}(m) e^{-\tau {\cal D}[p(u\vert x) || p_{\rm opt}^{\tau}(u\vert m)]} \over \sum_m  p_{\rm opt}^{\tau}(m) e^{-\tau {\cal D}[p(u\vert x) || p_{\rm opt}^{\tau}(u\vert m)]}}, \label{Eq:pmx_opt}\\
p_{\rm opt}^{\tau}(m) &= \langle p_{\rm opt}^{\tau}(m\vert x) \rangle_{\rho(x)}, \label{Eq:pm_opt}\\
p_{\rm opt}^{\tau}(u\vert m) &= {p(u)\over p_{\rm opt}^{\tau}(m)} \langle p_{\rm opt}^{\tau}(m\vert x) \rangle_{\rho(x\vert u)}. \label{Eq:pum_opt}
\end{align}
\end{subequations}
These solutions can be computed numerically, by iteration, with the Information Bottleneck algorithm \cite{IB}. 

The $\tau$-dependent optimal code book, consisting of the encoding, $p_{\rm opt}^{\tau}(m|x)$, $p_{\rm opt}^{\tau}(m)$, and the decoding, $p_{\rm opt}^{\tau}(u|m)$, is computed once, at the beginning of the information engine's run time. It is used to set the physical code book parameters, $y^{\rm ini}_{m}(x)$, $y^{\rm fin}_{m}$, and $\kleinV_{\ell}(m)$, which then remain fixed for all $N$ cycles of operation. The energetic costs for executing the algorithm are not considered in this analysis, since the focus is on $N \rightarrow \infty$, whereby the upfront computational costs per cycle would become negligible. 

\subsection{Full engine cycle}
An important aspect of the engine model we consider is that work extraction is allowed to occur at a higher temperature $T'>T$ than memory making \cite{CB}. To enable this, isentropic compression along the $z$-axis is employed, after the memory is formed, leaving correlations between memory and work medium intact, and leaving the mutual information unchanged.  

Work is then extracted from the work medium by an isothermal transformation at temperature $T'$, using the inference-based, memory dependent protocol ${\bf a}(m,t)$ discussed in Sec. \ref{Sec:PhysEnc}. To close the cycle, isentropic expansion along the $z$-axis is used, which recovers precisely the work needed for the isentropic compression, whereby the two isentropic steps of the cycle together contribute nothing to the overall engine work output \cite{CB,stilldaimer2022}. 

In summary, one engine cycle consists of the following steps: 
\begin{enumerate}
\item Memory preparation at temperature $T$, in response to an observation of the $x$ position of the particle in the work medium. The container of the one-particle gas implementing the work medium is assumed to have unit length along all three spatial dimensions.
The container used as the memory has unit length along the $y$-axis, and volume $V$.
\item Change of temperature, $T\rightarrow T'$, by isentropic compression along the $z$-axis. 
\item The work medium container has volume $V'$ and is attached to a heat bath at temperature $T'$. Work is extracted by a memory-dependent protocol ${\bf a}(m,t)$ applied to the work medium.
\item Change of temperature back to starting temperature, $T'\rightarrow T$, by isentropic expansion along the $z$-axis; restoration of the memory to its initial state, by pulling out dividers; restoration of the work medium to its initial state, by inserting divider in the centered starting position.

\end{enumerate}

\section{Results}
\label{Sec:Results}
The example we study here is constructed such that one expects memories to have at most three states, because coarse graining the observations into three groups captures all available usable information, i.e., mapping $x \in \mathcal{X}_L$ to one memory state (without loss of generality, we can label it by $m=-1$), mapping $x \in \mathcal{X}_M$ to $m=0$, and mapping $x \in \mathcal{X}_R$ to $m=1$. Which side of the container is empty is then known with probability one, whenever $m=\pm 1$. This case occurs with probability $1-w$, and allows the observer to extract up to $kT' \ln(2)$ joules whenever it occurs. With probability $w$, nothing can be said about $u$, and therefore no work can be extracted on average. In total, at most $kT' (1-w) \ln(2)$ joules of work can be extracted, on average, when the observer uses this data representation.  

This specific memory is deterministic and has three memory states; we label information quantities that depend on this memory with superscript $(d3)$. All available usable information is retained, 
\begin{equation} \label{Eq:Irel_d3}
 I_{\rm u}^{(d3)} =  (1-w) \ln(2) = I_{\rm u}^{\rm max},
\end{equation}
and the maximum work that can be extracted, on average, is thus $-W' = kT' I_u^{\rm max}$ joules.

The minimum amount of work necessary to run this memory is proportional to the total information stored, $W = kT \,I_{\rm m}^{(d3)}$, with
\begin{equation} \label{Eq:3cg}
    I_{\rm m}^{(d3)} = (1-w) \ln(2) + h(w).
\end{equation}
We abbreviate by $h$ a non-negative binary entropy function \mbox{$\mathbb{R} \to \mathbb{R}$}: 
\begin{eqnarray}
h(x) &\equiv& - \left(1- x \right) \ln\left(1-x\right) - x \ln(x) > 0, 
\label{Eq:Entq}
\end{eqnarray}
defined for $x \in (0,1)$, with \mbox{$h(x=0)=h(x=1)=0$}.

When the engine is run using this deterministic, three-state memory, it produces an average net work output (in units of $kT'$) of
\begin{equation}
    \!\!{ W_{\rm out}^{(d3)} \over k\,T'} = \eta_C (1-w) \ln(2) + \left(\eta_C - 1 \right) h(w),
\end{equation}
where $\eta_C = 1-1/\tau = (T'-T)/T'$ is the Carnot efficiency, which depends only on the ratio of high to low temperature, $\tau = T'/T$. 

From the fact that there is a zero crossing 
\footnote{The zero crossing starts at $\tau_{zc} = 1$, when there is no uncertainty ($w=0$), and diverges as full uncertainty is approached ($w\rightarrow 1$). When the uncertain region spans half of the box, $\tau_{zc}(w = 1/2) = 3$, and when it spans a third, $\tau_{zc}(w = 1/3) = 3\ln(3)/(2\ln(2)) \approx 2.77$.} 
at 
\begin{eqnarray} \label{Eq:tau_zc}
\tau_{zc}(w) = 1+ { h(w) \over (1-w) \ln(2) }  \geq 1,
\end{eqnarray}
we see immediately that for all smaller values, \mbox{$\tau < \tau_{zc}(w)$}, using this particular memory is worse than doing nothing. The reason for this is that for small $\tau$, the costs of having this deterministic three-state memory are not outweighed by the resulting amount of extractable work.

The thermodynamic value of information, which is, in this simple example, proportional to the ratio of temperatures at which the isothermal transformations are performed, dictates the appropriate detail with which information ought to be stored. Therefore, even if data are abundant, and statistical overfitting \cite{dietterich1995overfitting} is not an issue, the model can still be too complicated in the sense that it retains information about the observed system, which the observer cannot use to derive any net benefit, because of given physical constraints.

An observer that memorizes non-zero useful information, $I_{\rm u} > 0$, can run the information engine successfully with positive net work output only if the ratio of memorized information, $I_{\rm m}$,
to usable information, \mbox{$\xi \equiv I_{\rm m}/I_{\rm u}$}, is less than the temperature ratio, $\xi < \tau$. For a fixed strategy, $\bar{p}(m|x)$, the ratio is given by $\bar{\xi}$, and we know that the temperature ratio must exceed $\bar{\xi}$ to make it worthwhile for the observer to employ this strategy.
Note that information ratios, such as $\xi$, or measures found in e.g. \cite{proesmans2015efficiency, masuyama2018information} do not help us to decide if observer $A$ is doing better in terms of net work output than observer $B$, because $W^{\rm net}_{\rm out}[A] > W^{\rm net}_{\rm out}[B]$ is equivalent to 
$(I_{\rm m}[A]- I_{\rm m}[B])/(I_{\rm u}[A]-I_{\rm u}[B]) < \tau$.

While we expect the deterministic three-state solution to maximize the average net work output for large $\tau$, we expect there to be other solutions that maximize the net work output at smaller $\tau$. 
In Sec. \ref{sec:optmem} we explore and analyze these optimal memories, found by maximizing the net engine work output over all probabilistic maps, $p(m \vert x)$. Careful analysis of the optimal strategies inspires a simple yet interesting class of parameterized soft partitionings, which we introduce in Sec. \ref{Sec:PSP}. The new model class constrains the observer to a much smaller set of possible encoding strategies, yet it results in no tangible engine performance loss. In Sec. \ref{Sec:CG} we constrain possible observer models even further, limiting them to basic coarse graining (hard partitioning) of the observable. 

\subsection{Optimal observers - numerical results} \label{sec:optmem}
The data representations which allow for maximization of the engine's average net work output at any value of the trade-off parameter $\tau$ are found algorithmically as the solutions to Eqs. (\ref{Eq:IBalg}). For $\tau >> 1$ these optimal memories are identical to the deterministic three-state memory discussed above, but
for smaller $\tau$, the optimal memories are not deterministic. 

\begin{figure}[ht!]
\centering 
\includegraphics[width=0.95\linewidth]{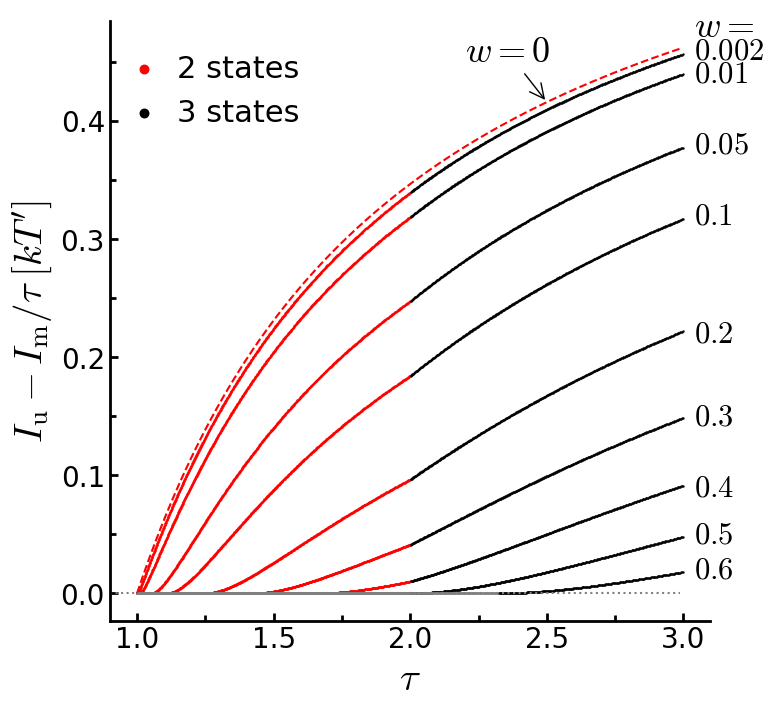}
\caption{Maximum net engine work output in units of $kT'$, Eq. (\ref{Eq:Woutengine}), as a function of $\tau$ for different engine geometries (different $w$). Red marks two-state, black three-state, and gray one-state memories. Dashed red: Unmodified \LS\ engine, for comparison.}
\label{Fig:Obj}
\end{figure}

For each fixed geometry (fixed value of $w$), it is the case that with increasing temperature ratio, $\tau$, each bit of usable information captured in memory becomes increasingly valuable, in the sense that it can be turned back into increasingly more work output. It thus becomes worthwhile to keep more bits of information in memory. 
Figure \ref{Fig:Obj} shows, for different values of $w$, the average net work output of the engine, Eq. (\ref{Eq:Woutengine}), evaluated for the memories that maximize it at each $\tau$.

At $\tau = 1$, an optimal solution is always to do nothing, because the engine cannot possibly produce positive net work output, on average, when $T'=T$. 
Now, to do nothing, no decision at all is required, and this corresponds to a zero bit memory, mapping all values of the observed quantity onto one memory state. 

\begin{figure*}[t!]
\centering 
\begin{subfigure}{.45\linewidth}
    \centering
    \includegraphics[width=\linewidth]{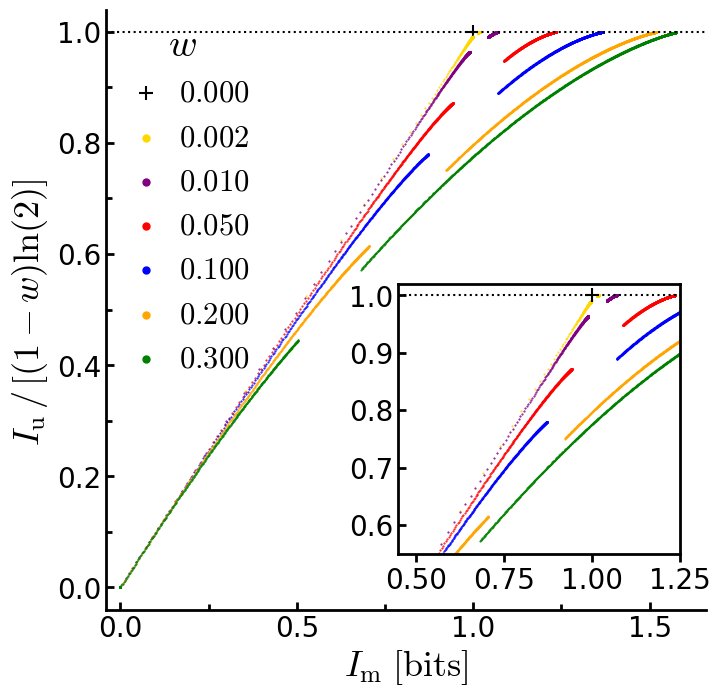}
    \phantomsubcaption
    \label{Fig:InfoplaneSmall}
\end{subfigure}
\hspace{1em}
\begin{subfigure}{.45\linewidth}
    \centering
    \includegraphics[width=\linewidth]{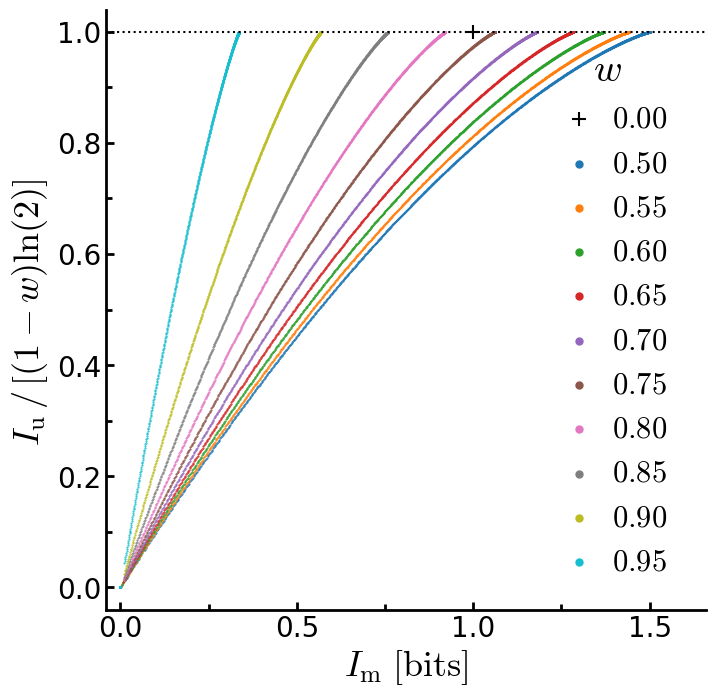}
    \phantomsubcaption
    \label{Fig:InfoplaneLarge}
\end{subfigure}
\vspace{-2em}
\caption{Information plane representation of optimal memories for select engines with $w \leq 0.3$ (left) and $w \geq 0.5$ (right). Fraction of the maximum usable information as a function of total information retained. The gaps in the curves occur at the phase transition from two- to three-state memories. Black cross: \LS\ engine ($w=0$).}
\label{Fig:Infoplane}
\end{figure*}

When $\tau$ increases, it becomes worthwhile to do something. Then, a decision is required, whereby the memory has to have at least two states. 
Those solutions for which the optimal memories have two states are displayed in red in Fig. \ref{Fig:Obj}, those with three states in black. Points at which the one-state solution is optimal are plotted in grey, with zero average net work output.
When we talk about the number of states, we mean the smallest number of realizations for random variable $m$, compatible with a mapping for which the engine's net average work output has the value $I_u[p_{\rm opt}^{\tau}(m\vert x)] - I_m[p_{\rm opt}^{\tau}(m\vert x)]/\tau$ \footnote{It is usually possible to artificially construct degeneracies, as has been pointed out e.g. in \cite{marzen2016predictive}, however, those are uninformative about the structure of the optimal solutions.}.

We observe two different $w$ regimes. 
For $w < w_c$ smaller than a critical value $w_c$ two changes occur: from one to two memory states, at a critical value of $\tau^*_{1\to 2}(w)$, and from two to three memory states at a critical value of $\tau^*_{2\to 3}(w)$. While the one-state solution is optimal up to larger values of $\tau$, as $w$ increases, the phase transition \footnote{Following the definition introduced in \cite{wu2020phase}, we refer to a ``phase transition'' as a qualitative change in the objective function landscape of the Information Bottleneck algorithm. The behaviour at the critical $\tau$ value is discussed in more detail in Appendix \ref{App:CritTau}.} from two to three states happens at the critical value $\tau^*_{2\to 3}(w) = 2$, regardless of the value of $w$. Analytical calculation of the critical $\tau$ values is possible, if optimal strategies are expressed in terms of the parametric soft partitionings introduced in Sec. \ref{Sec:PSP}. 
At $w=0$, we have the original \LS\ engine, for which we know that a deterministic two-state memory is optimal for all values of $\tau$ (yielding, effectively, a Carnot engine). It is plotted as a dashed red line in Fig. \ref{Fig:Obj} for comparison. 

For $w \geq w_c$, there is so much uncertainty in the observations, that a two-state memory is never optimal, and we have only one phase transition, from one to three memory states, at $\tau^*_{1\to 3}(w)$. The value of $w_c$ corresponds exactly to an uncertain region spanning half of the work medium container. This can also be shown analytically using the parametric soft partitions (Sec. \ref{Sec:PSP}) and detailed calculations of $w_c$, $\tau^*_{1\to 2}(w)$ and $\tau^*_{1\to 3}(w)$ can be found in Appendix \ref{App:FirstPT}.

Henceforth, we focus most of our analysis on the more interesting region $w < w_c$, with two phase transitions in the number of memory states. 

\subsubsection{Cost-benefit trade-off}
How does the expected output work scale with the minimum input work necessary to run the memory? To compare between different geometries with uncertain regions of different sizes (different values of $w$), we have to normalize the optimal 
work output, 
$kT' I_{\rm u}^*(w, \tau)$,  
by the maximum possible work output, $kT'(1-w)\ln(2)$. The resulting measure is $I_u^*(w, \tau) / (1-w) \ln(2)$. How does this scale with the corresponding thermodynamic cost encountered for the optimal memory? In units of $kT$, the cost is simply the total information kept in memory, $I_{\rm m}^*(w, \tau)$.
The minimum input to maximum output work relationship, i.e. the thermodynamic cost-benefit relation, of the least dissipative data representation possible at each $\tau$, thus corresponds to a (properly scaled) plot in what is often called the information plane (e.g. in \cite{slonim1999agglomerative}).

This is displayed for select $w<w_{c}$ in Fig. \ref{Fig:InfoplaneSmall}, and for $w \geq w_{c}$ in Fig. \ref{Fig:InfoplaneLarge}.
Cost-benefit relationships to the left of these curves are unattainable (this is the ``infeasible" region in rate-distortion theory), while regions to the right of the curves are suboptimal.

For $w<w_{c}$ (Fig. \ref{Fig:InfoplaneSmall}), there is a gap at the phase transition from two- to three-state memories, because both the necessary work input and the derivable work output change discontinuously.
As the uncertain region vanishes ($w\rightarrow 0$), the solution for the regular \LS\ engine is approached: coarse graining into two symmetric regions, which costs one bit and yields one bit, indicated by the black cross in Fig. \ref{Fig:Infoplane}. 
The curve for $w=0.4$ largely overlaps with the curve for $w=0.3$, and is thus omitted in the plot.

For $w \geq w_c$ (Fig. \ref{Fig:InfoplaneLarge}), optimal memories immediately transition from one state to three states.  
Since there are no optimal two-state memories, there are no visible gaps in the curves.

The endpoints of the optimal curves in the information plane are reached once all usable information is captured in memory, 
which corresponds to 1 on the $y$-axis of the plots in Fig. \ref{Fig:Infoplane}. 
These endpoints overlap with the information plane coordinates of the deterministic three-state coarse graining, characterized by 
Eqs. (\ref{Eq:Irel_d3}) and (\ref{Eq:3cg}). It is worthwhile to emphasize the fact that those deterministic memories, which retain all of the available usable information, are only one point (for every value of $w$) in the information plane. 
In contrast, solving the full optimization problem, Eq. (\ref{Eq:IB}), yields a solution (and thus a point in the information plane) for every $\tau$, resulting in the much richer representation depicted in Fig. \ref{Fig:Infoplane}, which reflects the underlying statistical nature of the problem, and which reveals how useful data compression is \cite{still2007structure}.

The total amount of available usable information decreases linearly with increasing uncertainty, Eq. (\ref{Eq:TotalIrel}). Engines with very large uncertain regions (last four curves in Fig. \ref{Fig:InfoplaneLarge}, $w \geq 0.8$) require less than 1 bit of memory to encode all usable information. For these engines the endpoints of their information curves thus lie to the left of the \LS\ engine marker (black cross) in Fig. \ref{Fig:InfoplaneLarge}. While this might seem surprising at first, note that these memories capture much less than 1 bit of usable information. Their memory cost is so low, because the whole uncertain region, which spans at least $80\%$ of the full engine volume, gets mapped onto a single memory state, $p(m=0)=w$. Consequently the two other memory states, which are associated with the two certain regions of the engine are realized much more infrequently, with $p(m= \pm 1) = (1-w)/2$, and the entropy of the memory, $H[m]$, {which upper bounds $I_m = H[m]-H[m|x] \leq H[m]$}, is thus less than 1 bit.

\vspace{-0.2cm}
\subsubsection{Strategies to maximize net engine work output}
\label{Sec:OptimalStrategies}
Optimal memories are deterministic only for small and for very large temperature ratios, $\tau$. For $\tau < \tau^*_{1\rightarrow 2}(w)$, the thermodynamic value of usable information is so low, that there is no incentive to memorize or to do anything. For very large $\tau$, the relative value of usable information, compared to the cost of memorizing, is so large that coarse graining into three regions, and thereby capturing all usable information, is optimal despite the thermodynamic cost it entails. For other $\tau$ values, memories that maximize the net average engine work output are not coarse grainings, but rather characterized by those probabilistic assignments, $p_{\rm opt}^\tau(m|x)$, which solve Eqs. (\ref{Eq:IBalg}). What do they look like?   

\begin{figure}[h]
\centering 
\begin{subfigure}{.95\linewidth}
    \centering
    \includegraphics[width=\linewidth]{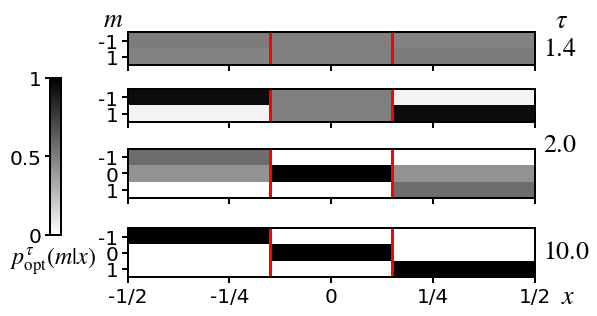}
    \label{Fig:pmgxc03}
\end{subfigure}
\vspace{1em}
\begin{subfigure}{.95\linewidth}
    \centering
    \includegraphics[width=\linewidth]{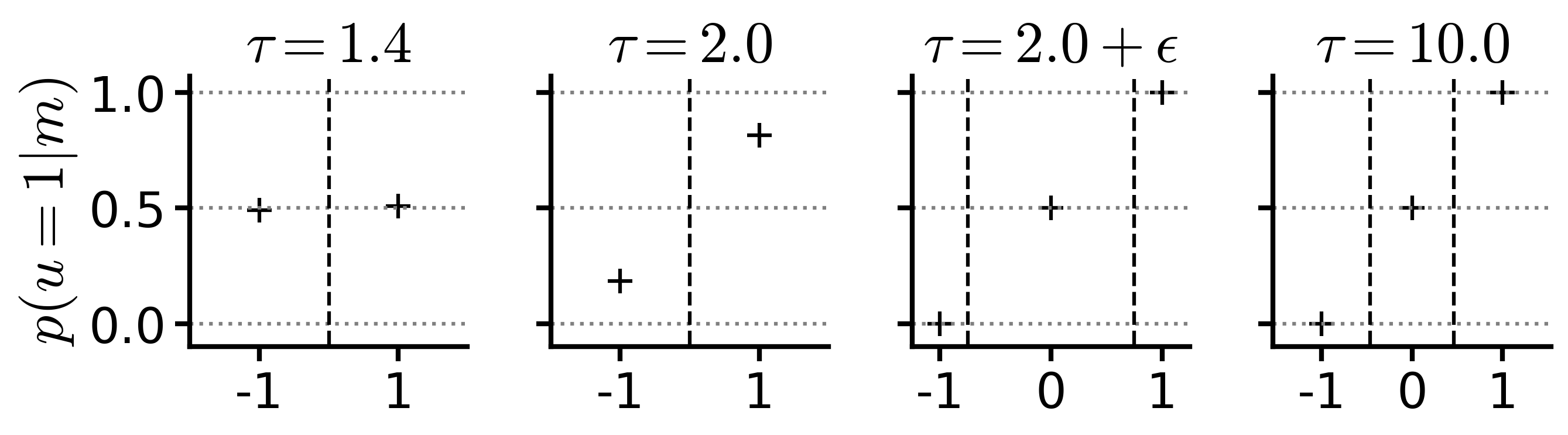}
    \label{Fig:pugmc03}
\end{subfigure}
\vspace{-2em}
\caption{Memory assignments $p_{\rm opt}^\tau(m|x)$ (upper panel) and inference derived from the memory states $p_{\rm opt}^\tau(u \vert m)$ (lower panel) for $w=0.3$ (red lines at $x=\pm w/2$). Dashed lines indicate state boundaries (see text).}
\label{Fig:Memoryc03}
\end{figure}

For the geometry with $w=0.3$, we visualize them in Fig. \ref{Fig:Memoryc03}, plotting $p_{\rm opt}^\tau(m|x)$ in grey scale for the four memories found around the phase transitions. The top plot in the upper panel is the first memory after the transition from one state (doing nothing) to two states. This memory captures only a tiny amount of information, thus costing only ever so slightly more than zero joules, but also yields only a negligible amount of usable information, and thus negligible work potential. This is reflected by the inference being close to chance, as can be read off from the left plot in the lower panel, which displays $p_{\rm opt}^\tau(u|m)$ for this memory. 

Recall that each memory state corresponds to a work-extraction protocol performed on the work medium, which allows for isothermal expansion in the direction of the most likely empty side, up to a residual volume, determined by the remaining uncertainty in the inference. The work extraction protocol uses $m$ to infer the most likely empty side. Mathematically, this inference is characterized by $p_{\rm opt}^\tau(u|m)$. This is plotted in the lower panel of Fig. \ref{Fig:Memoryc03} for all memories depicted in the upper panel, labeled by the value of $\tau$ at which the memories are optimal. The black dashed lines separate memory states, and the distance between them corresponds to the probability, $p_{\rm opt}^\tau(m)$, with which the individual memory states occur. Their labels, $m$, are plotted under the $x$-axis. 

As $\tau$ increases, the optimal memory eventually approaches the two-state memory which occurs right before the phase transition to three states, displayed in the second plot of the top panel in Fig. \ref{Fig:Memoryc03}. The employed strategy is to assign measurements in the certain region to the respective memory state, almost with probability one, and to assign measurements in the uncertain region to either state at random. This memory making strategy, which was found algorithmically, is an elegant solution that optimizes the cost-benefit trade-off for $\tau=2$ \footnote{At $\tau=2$, optimal two- and three-state memories produce exactly the same amount of net engine work output (see Appendix \ref{App:SecondPT}). The optimal solution is degenerate at $\tau=2$, yet we choose to discuss the two-state memory at $\tau=2$, because it achieves the maximum net work output with less memory states.}, i.e. when usable bits result in twice the thermodynamic gain, compared to what it costs to memorize them. Despite its simplicity, it is not immediately obvious that it would have been guessed ``by hand" without knowledge of the analysis presented in this paper. This strategy is quite distinct from a naive coarse graining of $x$ into two connected regions that correspond to two memory states. 

To gain intuition for this strategy, consider the limit in which measurements in the certain region are assigned to the respective memory state with probability one. In this limit, the total information stored by this memory is $(1-w)\ln(2)$ nats \footnote{This is calculated in Appendix \ref{App:PSP}, and depicted in Sec. \ref{Sec:PSP}, where one has to set $q=0$ in Eq. \ref{Eq:Imem_s2}.}, so it costs at least $kT (1-w)\ln(2)$ joules to encode on average. Compare that to coarse graining into positive vs. negative $x$-values, $m=\sign(x)$, which captures one bit, or $\ln(2)$ nats, of information and costs at least $kT \ln(2)$ joules on average. The inference error is $w/2$ in each case \footnote{This is calculated in detail in Appendix \ref{App:DCG} and \ref{App:PSP}.}, whereby the memory-dependent work extraction protocol is the same (leaving a residual volume of $w/2$, extracting up to $kT' \ln(2-w)$ joules). Therefore, in this limit, the reason why the optimal two-state memory is better than naive coarse graining into positive and negative $x$-values is that it saves encoding costs without reducing the amount of extractable work. The savings grow with the relative size of the uncertain region: on average $kT w \ln(2)$ joules can be saved in each cycle. Note that in every cycle, the observer performs an action on the work medium (as $p(u|m) \neq 1/2$). 

For the optimal two-state memory found at $\tau=2$, the corresponding $p_{\rm opt}^\tau(u|m)$ is plotted in the second plot from the left in the lower panel of Fig. \ref{Fig:Memoryc03}. There is significant remaining uncertainty in the inference corresponding to either state, due to the fact that observations in the uncertain region contain no usable information, yet they are assigned to either state at random.

We get an approximation of how certain the inference is by considering again the limit in which measurements in the certain region are assigned to the respective memory state with probability one. The probability of the left side being empty is then
\begin{equation} 
\label{Eq:pugm_approx}
    p(u=1\vert m) = 
    \begin{cases}
    \quad w/2 & m=-1\\
    \quad 1 - w/2 & m=1
    \end{cases}
\end{equation}
For $w=0.3$, there is 15\% remaining inference uncertainty for this approximation. This is close to the numerical value for the optimal memory depicted in Fig. \ref{Fig:Memoryc03}, which is approximately $18 \%$.

\begin{figure*}[t!]
\centering 
\begin{subfigure}{.47\linewidth}
    \centering
    \begin{subfigure}{\linewidth}
    \centering
    \includegraphics[width=\linewidth]{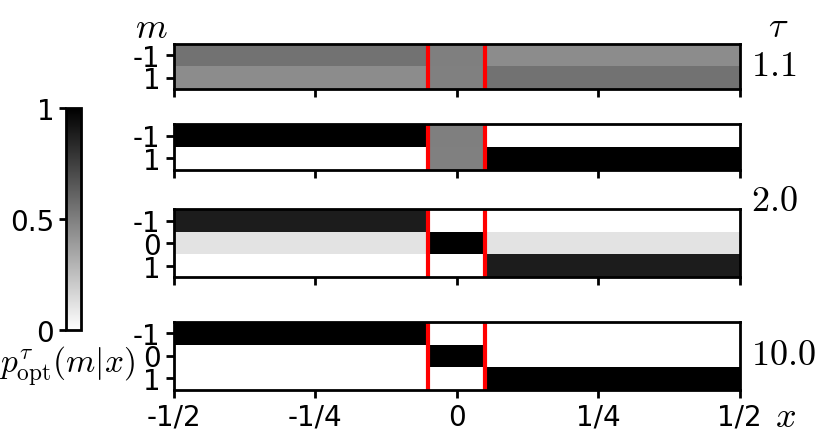}
    \label{Fig:pmgxc01}
    \end{subfigure}
    \vspace{1em}
    \begin{subfigure}{\linewidth}
    \centering
    \includegraphics[width=\linewidth]{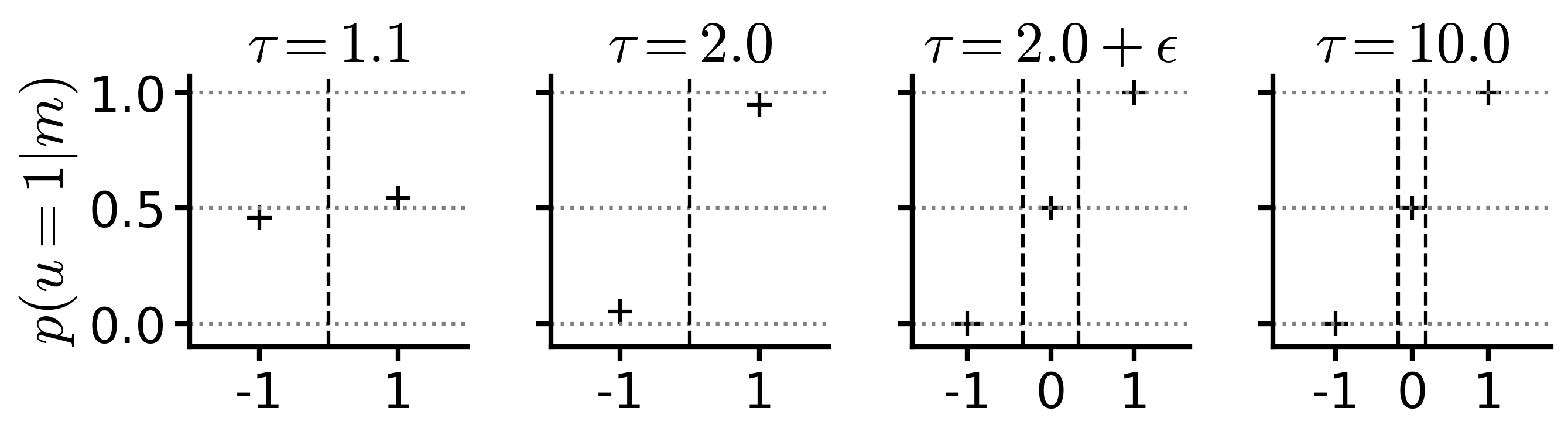}
    \label{Fig:pugmc01}
    \end{subfigure}
    \phantomsubcaption
    \label{Fig:Memoryc01}
\end{subfigure}
\hspace{1em}
\begin{subfigure}{.47\linewidth}
    \centering
    \begin{subfigure}{\linewidth}
    \centering
    \includegraphics[width=\linewidth]{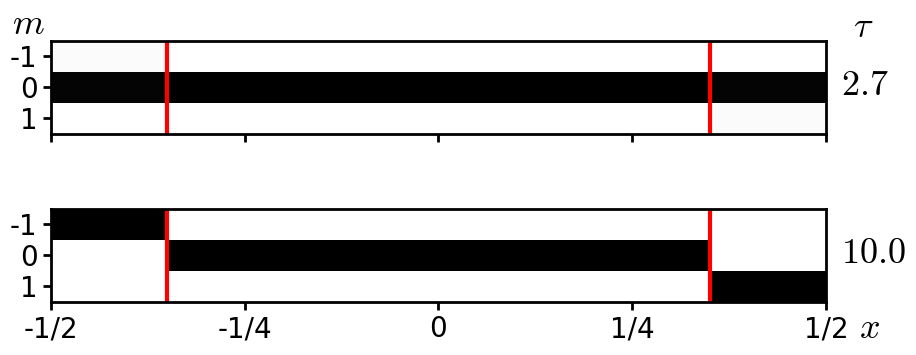}
    \label{Fig:pmgxc07}
    \end{subfigure}
    \par\bigskip
    \begin{subfigure}{\linewidth}
    \centering
    \includegraphics[scale=0.35]{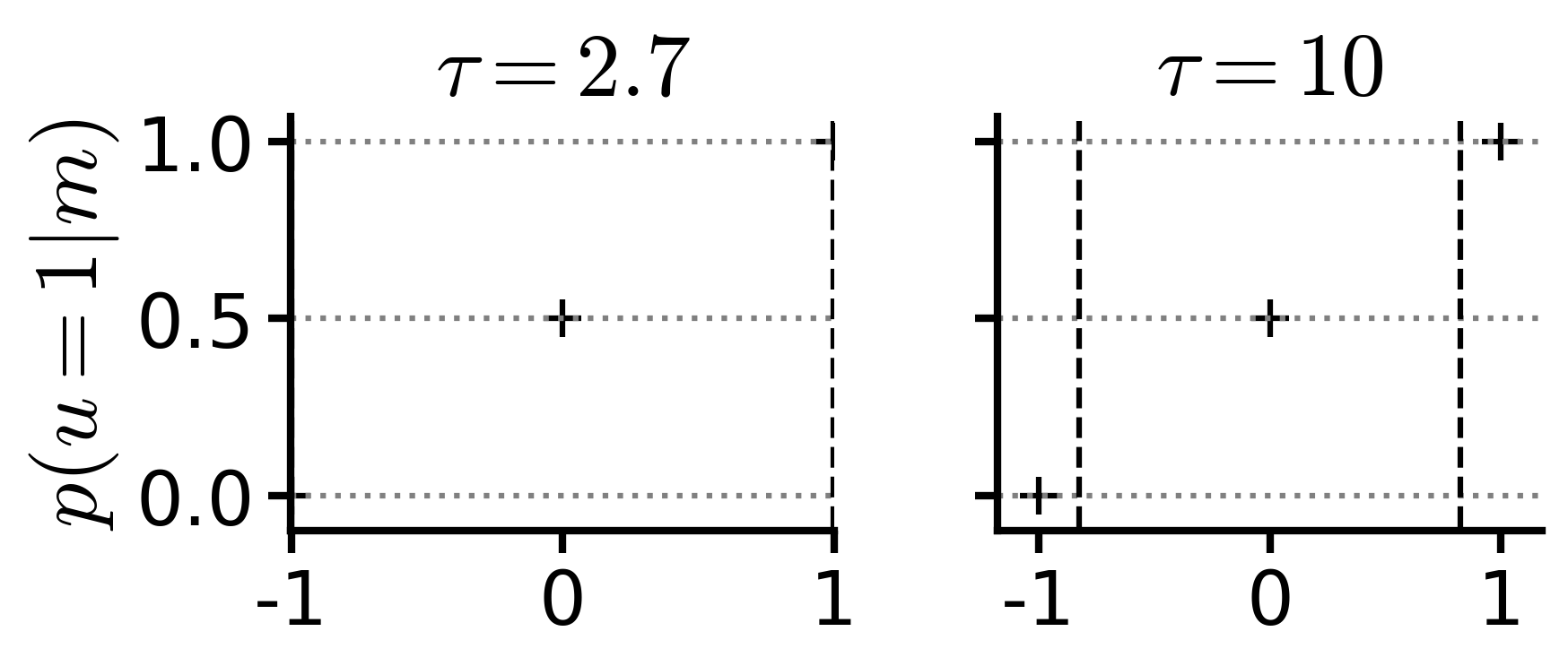}
    \label{Fig:pugmc07}
    \end{subfigure}
    \phantomsubcaption
    \label{Fig:Memoryc07}
\end{subfigure}
\vspace{-2em}
\caption{Memory assignments $p_{\rm opt}^\tau(m|x)$ (upper panel) and inference derived from the memory states $p_{\rm opt}^\tau(u \vert m)$ (lower panel) for $w=0.1$ (left) and $w=0.7$ (right); red lines at $x=\pm w/2$. Dashed lines indicate state boundaries (see text).}
\label{Fig:Memories0107}
\end{figure*}

When the stakes are high enough, in the sense that it is proportionally more important to have usable information, compared to saving on costs, because $\tau$ is large enough, then it makes sense to afford the use of a third state, one in which the observer admits to knowing nothing, and hence also can do nothing with the work medium that would, on average, result in engine work output. This happens when $\tau > 2$.

The first memory after the phase transition (third plot from top in upper panel of Fig. \ref{Fig:Memoryc03}) assigns measurements in the uncertain region to the state that corresponds to the decision to do nothing with probability one, while assigning the certain regions with higher probability ($\approx 57 \%$) to the state that corresponds to the protocol on the work medium which lets the gas expand into the empty region, and with remaining probability ($\approx 43\%$) to the state that corresponds to doing nothing, but never to the state that corresponds to the decision to compress the gas. Importantly, for those two memory states that result in actions on the work medium, the inference $p_{\rm opt}^\tau(u|m)$ is now one, if $u=u^*(m)$, and zero otherwise.  
This is shown in the third plot from the left in the lower panel of Fig. \ref{Fig:Memoryc03}. The main gain of this type of memory is that the observer can act on the work medium with certainty.

As $\tau$ increases, the benefit of having more usable information increasingly outweighs the cost of running the memory, and for large $\tau$, the deterministic three-state solution is found (fourth panel from top in Fig. \ref{Fig:Memoryc03}). 
Note that the inference associated with each memory state does not change as optimal three-state memories become more deterministic. What changes instead is that the observer exploits the two certain memory states more often, as $\tau$ increases, and more rarely does nothing. This can be seen in the third and fourth plot in the lower panel of Fig. \ref{Fig:Memoryc03}. The distance between the two dashed lines decreases between the third and fourth plot. This distance is $p_{\rm opt}^\tau(m=0)$. The observer less frequently does nothing as $\tau$ increases in the range from $\tau > 2$ to $\tau = 10$. At $\tau = 10$ the memory coarse grains, and $p_{\rm opt}^\tau(m=0) = w$.

Fig. \ref{Fig:Memoryc01} shows memory assignments, $p_{\rm opt}^\tau(m|x)$, and derived inferences, 
$p_{\rm opt}^\tau(u \vert m)$, for an engine with less uncertainty ($w=0.1$). 
The optimal memories do not qualitatively differ from the ones depicted in Fig. \ref{Fig:Memoryc03}, but since the uncertain region in Fig. \ref{Fig:Memoryc01} is smaller, the deterministic coarse graining into three states is approached at smaller values of $\tau$. This can be seen by comparing the first three-state memory after the phase transition for the two different engines (third panel from top in Figs. \ref{Fig:Memoryc03} and \ref{Fig:Memoryc01}). While the change to three memory states happens at the same value of $\tau$ for both engines, the corresponding optimal memory assignments right after this phase transition are closer to deterministic assignments for the engine with less uncertainty  (Fig. \ref{Fig:Memoryc01}).

For large uncertainties, $w \geq 0.5$, there are no optimal two-state memories. The assignments and inferences for $w=0.7$ are shown in Fig. \ref{Fig:Memoryc07}. The first optimal memory that produces positive average work output looks like a one-state memory, but it has two additional, infrequently realized, states, $p(m=\pm 1) = \mathcal{O}(10^{-5})$, for which there is no uncertainty in the inference. As $\tau$ increases, those two states are realized more frequently until the deterministic three-state memory is reached, with $p(m=\pm 1) = (1-w)/2$, as can be seen in the bottom panel of Fig. \ref{Fig:Memoryc07}.
This is less interesting than what we observed for smaller uncertain regions ($w<1/2$).

In summary, maximization of net average engine work output over all possible data representations finds the following strategies for $w < 1/2$:
\begin{enumerate}
\item One-state memory at low $\tau$ values: all data points are mapped onto the same state, no information is captured and nothing can be done.
\item Two-state memory at intermediate $\tau$ values: data in the left certain region is mapped with high probability to one state and data in the right certain region with high probability to the other state. Data in the uncertain region is mapped to either state at random (with probability $1/2$). The resulting inference retains a residual uncertainty, which is dealt with by leaving a residual volume in the work medium at the end of the work extraction protocol.
\item Three-state memory at larger $\tau$ values: The strategy is to be certain about when there is complete uncertainty. To that end, all data in the uncertain region are mapped deterministically (with probability one) to the same memory state, which results in no action on the work medium. This enables the cost efficient creation of two other states that carry no uncertainty in the resulting inference, and thus enable complete work extraction. The encoding saves costs by mapping all data in the left certain region with some probability (larger than $1/2$ for $w \leq 0.3$, see Figs. \ref{Fig:Memoryc03} and \ref{Fig:Memoryc01}) to one of these memory states that result in full work extraction, and, by symmetry, mapping all data in the right certain region to the other state, with the same probability. With remaining probability, those data are mapped to the inactive state. As the temperature ratio increases, the assignments to memory states that enable complete work extraction become more and more likely, until data in the certain regions is assigned with probability one, which results in the deterministic three-state solution that captures all usable information, which is optimal for large $\tau$.
\end{enumerate}
If the uncertain region spans less than half of the work medium container, one-state memories prevail for \mbox{$1 < \tau \leq \tau^*_{1\rightarrow 2}(w)$}. Two-state memories are found for \mbox{$\tau^*_{1\rightarrow 2}(w) < \tau \leq 2$}, and three-state memories are found for $\tau > 2$. If the uncertain region spans more than half of the work medium container, then two-state memories are never optimal, and there is a transition from a one-state to a three-state memory, which occurs at $\tau^*_{1\rightarrow 3}(w)$.

In the following, we focus on the engine with \mbox{$w=0.3$}, as an example that displays the full spectrum of the effects of partial observability in our model class. 

\subsubsection{Physical encoding and decoding of the memory} \label{Sec:PhysicalCoding}
Let us recall the physical manipulations used to create a memory with $K$ states: $K-1$ dividers are inserted at time $t_M$ perpendicular to the $y$-axis in the container that is used as the memory. Each divider is inserted at an $x$-dependent distance \mbox{$\Delta y^{\rm ini}_m(x) = y^{\rm ini}_{m}(x) - y^{\rm ini}_{m-1}(x) = p_{\rm opt}^\tau(m|x)$} from the previous divider (for the first divider, it is the distance from the container's wall at $y=0$). 

Each divider is then moved quasi-statically, parallel to the memory container's $y$-axis, to an $x$-independent distance \mbox{$\Delta y^{\rm fin}_m = p_{\rm opt}^\tau(m)$} from the last divider (or the wall). If both distances are the same, $\Delta y^{\rm fin}_m = \Delta y^{\rm ini}_m$, then the volume does not change, even if the dividers move. The volume shrinks when $\Delta y^{\rm ini}_m > \Delta y^{\rm fin}_m$, corresponding to $p_{\rm opt}^\tau(m|x) > p_{\rm opt}^\tau(m)$, and it expands when $\Delta y^{\rm ini}_m < \Delta y^{\rm fin}_m$. 

\begin{figure}[h!]
\centering 
\includegraphics[width=0.95\linewidth]{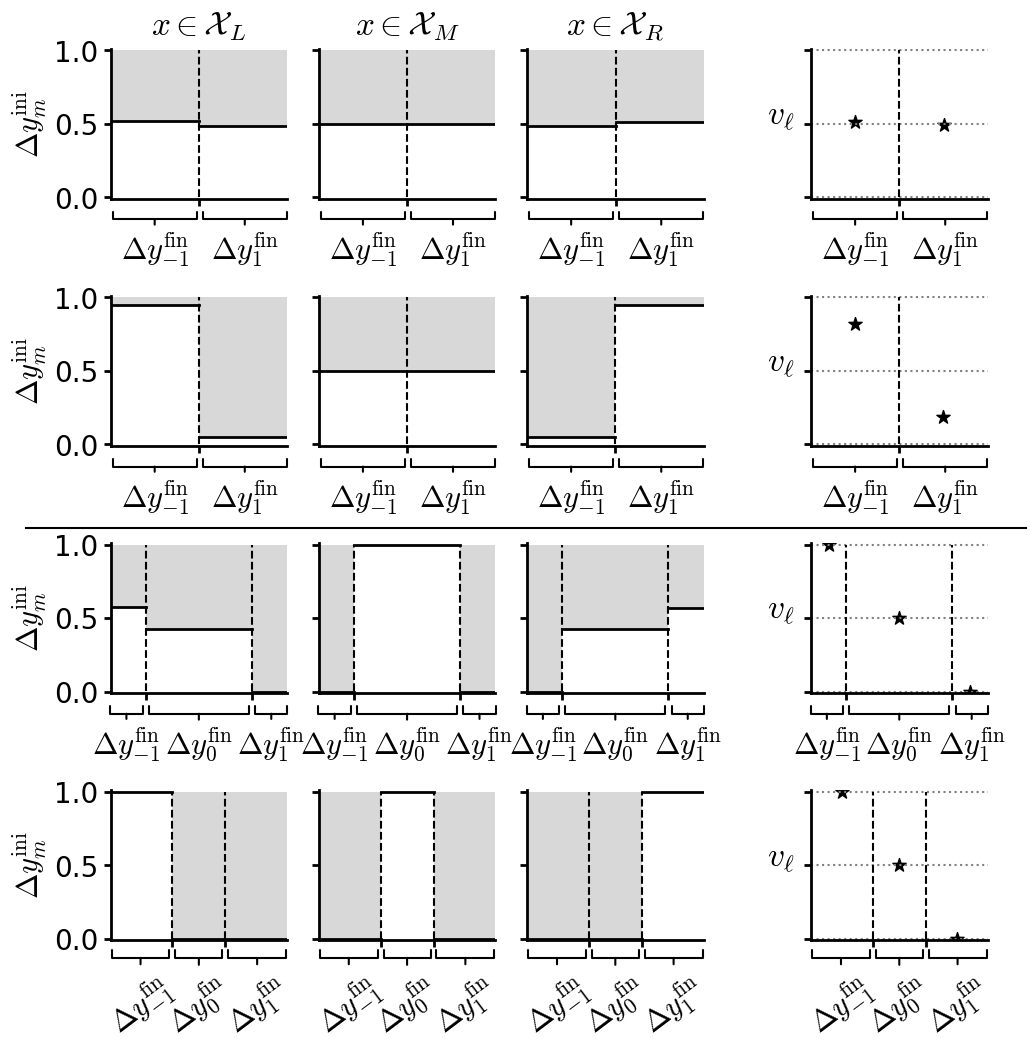}
\caption{Physical memory code book at $\tau = 1.4$ (top row), $\tau$ just below/above 2 (second/third row), $\tau = 10$ (bottom row); work medium geometry: $w=0.3$. 
First three plots in each row: initial distances $\Delta y^{\rm ini}_m$ vs. final distances, $\Delta y^{\rm fin}_m$ ($x$-axis), given observation in left, ($\mathcal{X}_L$), center, ($\mathcal{X}_M$), or right region, ($\mathcal{X}_R$). Last column: volume fraction remaining after work extraction in the work medium, left of the divider, $v_{\ell}(m)$, when memory is in state $m$. Dashed lines at positions $y^{\rm fin}_m$.}
\label{Fig:PhysicalCB_c03}
\end{figure}

In Fig. \ref{Fig:PhysicalCB_c03}, we visualize the physical memory making process by displaying the initial and final distances between divider(s) (and walls). This is visualized by a horizontal line at position $\Delta y^{\rm ini}_m$ that has a width of $\Delta y^{\rm fin}_m$.
The resulting white space under the horizontal line tells us if the volume that the particle occupies, should it happen to be trapped in this region, remains unchanged in the memory making process (if the white space is a square), gets compressed (if the white space is a rectangular standing on its shorter side), or expands (white space is a rectangular lying on its longer side), resulting in thermodynamic costs (or gains) during memory making.

Whenever the particle in the memory container is trapped in volume $\Delta y^{\rm fin}_m$,
the work extraction protocol ends with 
leaving the volume fraction $\kleinV_{\ell}(m) = V_{\ell}(m)/V'$ on the left side of the divider's final position in the work medium. This residual volume fraction is displayed in the last column in Fig. \ref{Fig:PhysicalCB_c03}. 

Together, $\Delta y^{\rm ini}_m$, $\Delta y^{\rm fin}_m$, and $\kleinV_{\ell}(m)$ constitute the physical code book, visualized in Fig. \ref{Fig:PhysicalCB_c03} at $\tau$ values just before and after the phase transitions (corresponding to the memories shown in Fig. \ref{Fig:Memoryc03})
\footnote{The physical code book at the $\tau$ values before and after the phase transitions is fully specified by Fig. \ref{Fig:PhysicalCB_c03}, as it suffices to visualize the $x$-dependent values $\Delta y^{\rm ini}_m(x)$ for any one $x$ in one of the three distinct regions of the work medium.}. 
Instructions for the physical process depicted in Fig. \ref{Fig:PhysicalCB_c03} can be entirely hard wired into machinery, without any further ``external" intelligence---all decision making is encapsulated. The physical code book is thus a complete physical model and implementation of the real world observer necessary for the operation of the information engine.  

To gain intuition, let us first consider the strategy shown in the second panel from the top in Fig. \ref{Fig:PhysicalCB_c03}. 
The memory is set into one of two memory states by inserting one divider relatively close to the wall (that is, either at position $\Delta y_{-1}^{\rm ini} \approx 0.95$ when the particle in the work medium is found in the certain region to the left ($x \in \mathcal{X}_L$, see Eq. (\ref{Eq:Regions})), or at $\Delta y_{-1}^{\rm ini} \approx 0.05$ for the right certain region, $x \in \mathcal{X}_R$). The divider is then moved to the middle, which almost always compresses the single-particle gas in the memory container and thus results in work costs. However, when the $x$ position of the particle in the work medium falls into the uncertain region ($x \in \mathcal{X}_M$), then the divider is inserted into the middle and not moved, requiring no work.

The $y$ position of the particle in the memory container after the memory making process indicates the memory state, which can be directly coupled to the work extraction protocol using the following rule: if $1 \geq y > 1-\Delta y^{\rm fin}_{1}$ (corresponding to $m=1$, with $p(u=1|m=1) > p(u=-1|m=1)$), then the divider in the work medium can move towards the left until 
the remaining volume  fraction on the left side is $\kleinV_{\ell}(m=1) \approx 0.18$. If $0 < y \leq \Delta y^{\rm fin}_{-1}$ (corresponding to $m=-1$), then, by symmetry, the same as above happens, but with movement towards the right instead of the left, with $\kleinV_{\ell}(m = -1) \approx 0.82$.

Recall that when the temperature ratio $\tau$ is increased slightly, a new strategy appears, one which uses three states. This strategy completely changes the behaviour of the physical observer. Now, whenever the observation falls into the uncertain region, pistons are moved from both walls towards the center of the memory container, until the distance between them has shrunk to $\Delta y_{0}^{\rm fin} \approx 0.6$.
This action ensures that the state $m=0$ is recorded, at an average work cost of $kT \ln(5/3) \approx 0.51\,kT$, by compressing the one-particle gas in the memory container.
The final distance between pistons is approximately $2w$, not just in this case, but also for all other values of $w$ that we considered \footnote{All values of $w$ shown in Fig. \ref{Fig:InfoplaneSmall}, including $w=0.4$ and $w=0.45$. The difference between the numerical value of $\Delta y_{0}^{\rm fin}$ and $2w$ is never greater than $\mathcal{O}(10^{-3})$ and as small as $\mathcal{O}(10^{-6})$ for the smallest values of $w$.}. 
It is no coincidence, that $\Delta y^{\rm fin}_{0}$ for the first optimal three state memory is roughly $2w$. Using the parametric soft partitionings (Sec. \ref{Sec:PSP}) we can show that for any $w < 1/2$ the first optimal three state memory has $\Delta y^{\rm fin}_{0}=2w$ \footnote{For three-state soft partitionings $p^{(s3)}(m=0) = w+(1-w)q_3$ (see Appendix \ref{App:PSP}) and inserting Eq. (\ref{Eq:q3*_main}) evaluated at the critical value $\tau=2$, we have $p^{(s3)}(m=0) = 2w = \Delta y^{\rm fin}_{0}$ for the first optimal three-state observer in engines with $w < 1/2$.}.
 
The memory state $m=0$ corresponds to inaction during work extraction. That means, whenever the data offers no information about the relevant quantity, the observer makes sure to take no action on the work medium. 

The addition of the third memory state unlocks the option to do nothing and while it may seem counter-intuitive to reserve a costly state for inaction, this additional state ensures that the observer can reduce its inference error in the other two states to zero, allowing it to act with certainty whenever it does act. This is reflected by the fact that for the two memory states associated with actions, $\kleinV_{\ell}(m=\pm 1)$ is either 0 or 1, meaning that the maximum amount of work (that is, on average, $kT' \ln(2)$), gets extracted from the wok medium (see Fig. \ref{Fig:PhysicalCB_c03}, last column, rows three and four).

When $x \in \mathcal{X}_L$, a divider is inserted at distance $\Delta y^{\rm ini}_{-1} \approx 0.57$ from the wall at $y=0$, and a piston is placed at the wall at $y=1$, because $\Delta y^{\rm ini}_{1} = 0$. Therefore, the particle never gets trapped in the volume that will eventually correspond to $m=1$. 
Both, divider and piston are moved towards $y=0$, until they reach their final positions, where $\Delta y^{\rm fin}_{-1} \approx 0.2$, and $\Delta y^{\rm fin}_{0} \approx 0.6$. 

When the particle is initially trapped in the volume between $y=0$ and the divider, then the gas is compressed. This happens with probability equal to $\Delta y^{\rm ini}_{-1}$. Whenever this happens, average work costs of $kT \ln(\Delta y^{\rm ini}_{-1} / \Delta y^{\rm fin}_{-1}) \approx 1.05 \,kT$ are recorded. However, $kT' \ln(2)$, can be extracted from the work medium by coupling $0 < y \leq \Delta y^{\rm fin}_{-1}$ (corresponding to $m=-1$) to complete volume expansion towards the right.

When the particle is initially trapped between divider and piston, then the gas expands, because the distance between them increases. This occurs with probability equal to $\Delta y^{\rm ini}_{0} \approx 0.43$. 
Whenever this happens, gains are recorded during memory making, in the amount of $kT \ln(\Delta y^{\rm fin}_{0} / \Delta y^{\rm ini}_{0}) \approx 0.33\,kT$. This reduces the average cost of the memory making process to $\approx 0.46\,kT$. However, the final $y$ position of the particle then corresponds to $m=0$, which results in no action on the work medium, whereby no gain can be derived from the work medium at the higher temperature. 

This strategy of choosing inaction in the majority of cases ($\Delta y^{\rm fin}_{0} \approx 0.6$), may seem suboptimal at first, but actually it is optimal because it significantly reduces the costs of the encoding, while simultaneously ensuring that the work medium is never compressed, and thereby reducing the loss in potential work gain, due to inference errors, to zero. 
The strategy for $x \in \mathcal{X}_R$ is symmetrically analogous to that for $x \in \mathcal{X}_L$. 

As the temperature ratio is further increased, the relative work costs of running the memory become less compared to the work derivable from the inferred information. Therefore, the dividers are inserted closer and closer to the walls (see Fig. \ref{Fig:PhysicalCB_c03}, fourth row, for $\tau = 10$, where the resulting memory is indistinguishable from a deterministic coarse graining into three regions). This increases the average cost of memory making, but it also increases the frequency with which a memory state is realized that enables maximum work extraction. This is visualized by the different positions of the dashed lines in the third and fourth row in Fig. \ref{Fig:PhysicalCB_c03}, as $\Delta y^{\rm fin}_{\pm 1}$ have increased to $(1-w)/2 \!=\! 0.35$ in the fourth row, while the inactive state with $\Delta y^{\rm fin}_{0} \!=\! w \!=\! 0.3$ is realized half as frequently as in the third row.

\subsection{Constrained observers}\label{Sec:Approx}
To find optimal observer strategies using the Information Bottleneck Algorithm the net engine work output is maximized over all possible data representation strategies. The optimal memories of Sec. \ref{sec:optmem} inspire a parameterized function class of observer strategies, which we introduce in Sec. \ref{Sec:PSP}. The new model class constrains the observer to a subset of all possible encoding strategies. This simplifies the algorithmic procedure for finding optimal memories (parametric optimization instead of using the Information Bottleneck Algorithm), yet it does not result in any engine performance loss. Additionally it allows analytical results to be derived for key quantities characterizing optimal observers, such as the critical value of the trade-off parameter, $\tau$, at transitions to more memory states. 

What happens when an observer can only coarse grain the observable into connected regions? This corresponds to an even more limited function class for the observer's possible representations of the observed system. How large of a loss, compared to optimal observers, will the constraint result in? We quantify this in Sec. \ref{Sec:CG}. 
 
\subsubsection{Parametric model for optimal observers}\label{Sec:PSP}

Having a memory that uses optimal solutions endows the observer with a qualitatively different strategy, compared to having a memory based on naive coarse graining. The analysis of optimal memories (Sec. \ref{Sec:OptimalStrategies}), computed with the Information Bottleneck Algorithm, points us towards a simple parameterized function class, which we will explore here. Detailed calculations of the quantities presented here can be found in Appendix \ref{App:PSP} and \ref{App:CritTau}.

Recall that within the $\tau$ regime that has optimal two state memories, the probability of assigning observations in the certain regions to only one memory state increases gradually. In the $\tau$ regime, where three-state solutions are optimal, the probability of assigning observations in the certain regions to the state that corresponds to doing nothing, gradually declines. 

\begin{figure}[h]
\centering 
\includegraphics[width=0.95\linewidth]{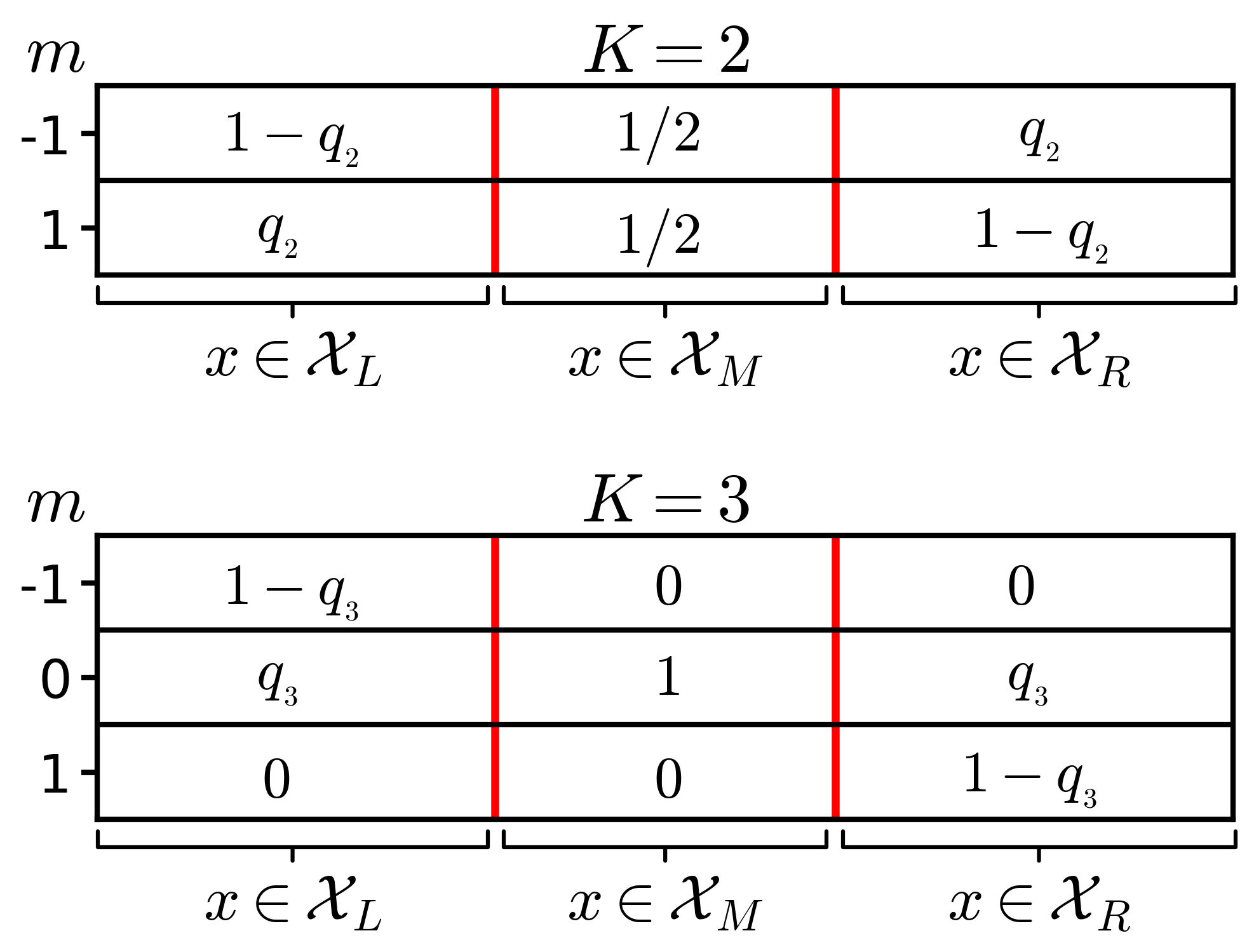}
\caption{Parameterization of $p(m \vert x)$ for soft partitioning models of optimal memories for $w=0.3$ (red lines at $\pm w/2$). Free parameters are $K$ and $q_K$.}
\label{Fig:SP_param}
\end{figure}

This suggests a parameterized model for the solutions of Eqs. (\ref{Eq:IBalg}), depicted in Fig. \ref{Fig:SP_param}. The free parameters are the number of states, $K$, ranging from one to three, and, for $K=\{2,3\}$, the residual probability $q_K$.

For $K=2$, we have a two-state {\em soft partitioning}, fully characterized by the assignments \footnote{Quantities relating to the two-state soft partitioning strategy are labeled with the superscript $(s2)$ and those relating to the three-state strategy with $(s3)$ for better readability.},
\begin{equation} \label{Eq:pmgx_2sp}
    p^{\rm (s2)}(m=-1 \vert x) =
    \begin{cases}
    \quad 1-q_2 & x \in \mathcal{X}_L \\ 
    \quad \;1/2 & x \in \mathcal{X}_M  \\ 
    \quad \;\;\,q_2 & x \in \mathcal{X}_R \\
    \end{cases}
\end{equation}
(with intervals $\mathcal{X}_L$, $\mathcal{X}_M$ and $\mathcal{X}_R$ defined in Eq. (\ref{Eq:Regions})).
This parameterization reflects the fact that optimal memories, at those intermediate $\tau$ values where two memory states emerge as optimal, use a memory making strategy that is markedly different from coarse graining: Each memory state gets all data from one of the certain regions with probability larger than one half (increasing as $\tau$ increases), while data in the uncertain region are assigned to either memory state at random (see Sec. \ref{Sec:OptimalStrategies}). 

It is sufficient to consider $0 \leq q_2 \leq 1/2$, because any memory with $q_2 > 1/2$ is identical to a memory with the labels of the memory states switched and $q_2 \leq 1/2$.

The resulting inference error associated with each memory state determines the fractional volume \mbox{$\gamma^{\rm (s2)} (m) =  p^{\rm (s2)}(u\neq m\vert m)$}, limiting the observer's work extraction. 
Due to symmetry, the error is the same for both memory states, $\gamma^{(s2)} := \gamma^{(s2)}(m=-1) = \gamma^{(s2)}(m=1)$. As a function of the parameters, it is given by
\begin{equation} \label{Eq:pugm_s2}
    \gamma^{(s2)} = {w \over 2} + (1-w)q_2. 
\end{equation} 
The inference error is thus lower bound by $w/2$, even if $q_2=0$. Naive coarse graining along the mid-line results in the same inference error (see Appendix \ref{App:DCG}, Eq. (\ref{Eq:pugm_d2})), but costs more. For $q_2 \neq 0$, there is an additional $q_2$ dependent term (second term in Eq. (\ref{Eq:pugm_s2})) accounting for the uncertainty in the encoding of the two certain regions.

The total memorized information and the usable part of it are
\begin{eqnarray}
    I_{\rm m}^{\rm (s2)} &=& (1-w) \left[\ln(2)-h(q_2)\right], \label{Eq:Imem_s2} \\
    I_{\rm u}^{\rm (s2)} &=& \ln(2)-h\!\left(\gamma^{(s2)}\right). \label{Eq:Irel_s2}
\end{eqnarray}

For $\tau >2$, optimal observers use the strategy to assign data in the uncertain region with probability one to one state that does not result in any action on the work medium (Sec. \ref{Sec:OptimalStrategies}). This memory state then implies that the observer's inference contains no knowledge of which side of the work medium container is empty. Thus, the strategy with three memory states is to be sure about knowing nothing. 

Data from each certain region is assigned to a corresponding state with probability $1-q_3$, with $0 \leq q_3 \leq 1$.
These two memory states result in complete volume expansion in the work medium (Sec. \ref{Sec:PhysicalCoding}). With remaining probability, $q_3$, data from each certain region is assigned to the state that results in no action on the work medium. As $\tau$ increases, three-state memories with increasingly smaller values of $q_3$ become optimal. For $K=3$, we thus have a three-state soft partitioning characterized by
\begin{eqnarray}
\label{Eq:pmgx_3sp0}
    p^{\rm (s3)}(m=0 \vert x) &=&
    \begin{cases}
    \quad q_3 & \quad\;\; x \in \mathcal{X}_L \\
    \quad 1 & \quad\;\; x \in \mathcal{X}_M \\ 
    \quad q_3 & \quad\;\; x \in \mathcal{X}_R \\ 
    \end{cases}\\
    p^{\rm (s3)}(m=1 \vert x) &=&
    \begin{cases}
    \quad 1-q_3 & x \in \mathcal{X}_R \\ 
    \quad 0 & {\rm otherwise},
    \end{cases}\\
    p^{\rm (s3)}(m=-1 \vert x) &=&
    \begin{cases} \label{Eq:pmgx_3sp}
    \quad 1-q_3 & x \in \mathcal{X}_L \\ 
    \quad 0 & {\rm otherwise}.
    \end{cases}
\end{eqnarray}

Like the three-state coarse graining, discussed in the beginning of Sec. \ref{Sec:Results}, this soft partitioning has no uncertainty in the inference whenever $m=\pm1$, but total uncertainty for $m=0$. Total and usable information retained by this memory are
\begin{eqnarray}
    I_{\rm m}^{\rm (s3)}  &=& (1-w)(1-q_3) \ln(2) - (1-w)h(q_3) \notag \\
    &&+ h(w+(1-w)q_3), \label{Eq:Imem_s3}\\
     I_{\rm u}^{\rm (s3)} &=& (1-w)(1-q_3)\ln(2), \label{Eq:Irel_s3}
\end{eqnarray}
which shows that the usable information carried by this encoding is reduced from the maximally available usable information by $q_3$ percent.

The optimal strategy for each value of $\tau$ can then be found by maximizing $I_u - I_m/\tau$ over the parameters ($K$ and $q_{K}$), an algorithmically more straightforward 
procedure than the Information Bottleneck algorithm. The resulting soft partitionings are optimal \footnote{The comparison between optimal solutions and the soft partitionings is discussed in more detail in Appendix \ref{App:PSP}.}. 

\paragraph{Minimal number of memory states in the $w$--$\tau$ parameter plane.}

The expressions for total and usable information retained by soft partitioning solutions can be used to determine the critical value of the temperature ratio at which it becomes worthwhile for optimal observers to use more than a single memory state (more detailed calculations in Appendix \ref{App:FirstPT}). With one memory state an observer cannot make a decision and is thus limited to doing nothing. For $q_2=1/2$, the two-state soft partitioning corresponds to a one-state memory with $I_{\rm m} = 0 = I_{\rm u}$, thus for $q_2 \to 1/2$ the transition from one to two memory states is approached. Using a two-state soft partitioning becomes worthwhile at 
\begin{equation} \label{Eq:taustar1-2}
    \tau^*_{1 \to 2}(w) = \lim_{q_2 \to {1 \over 2}} {I_{\rm m}^{\rm (s2)} \over I_{\rm u}^{\rm (s2)}} = {1 \over 1-w}.
\end{equation}
Similarly, the temperature ratio for which a three-state soft partitioning outperforms doing nothing can be computed:
\begin{equation} \label{Eq:taustar1-3}
    \tau^*_{1 \to 3}(w) = \lim_{q_3 \to 1}  {I_{\rm m}^{\rm (s3)} \over I_{\rm u}^{\rm (s3)}} =1 - {\ln(1-w) \over \ln(2)}.
\end{equation}

Recall that with optimal memories we found two distinct $w$ regimes with qualitatively different behavior. For small $w$, memories changed from using one to two to three states, while for large $w$, there were no optimal two-state memories (see Sec. \ref{Sec:OptimalStrategies}). Comparing Eqs. (\ref{Eq:taustar1-2}) and (\ref{Eq:taustar1-3}), we see that $\tau^*_{1 \to 2}(w < 1/2) < \tau^*_{1 \to 3}(w < 1/2)$, thus if less than half of the possible observation outcomes are uninformative, optimal observers transition from one memory state to two at $\tau^*_{1 \to 2}(w)$. However if more than half of the outcomes are uninformative ($w>1/2$), optimal observers switch from one memory state immediately to three states, since $\tau^*_{1 \to 2}(w > 1/2) > \tau^*_{1 \to 3}(w > 1/2)$. Both Eqs. (\ref{Eq:taustar1-2}) and (\ref{Eq:taustar1-3}) are increasing functions of $w$. The less certainty the observer has, the larger the temperature ratio needs to be, before it becomes worthwhile for the observer to do anything at all. 

\begin{figure}[h!]
\centering 
\includegraphics[width=\linewidth]{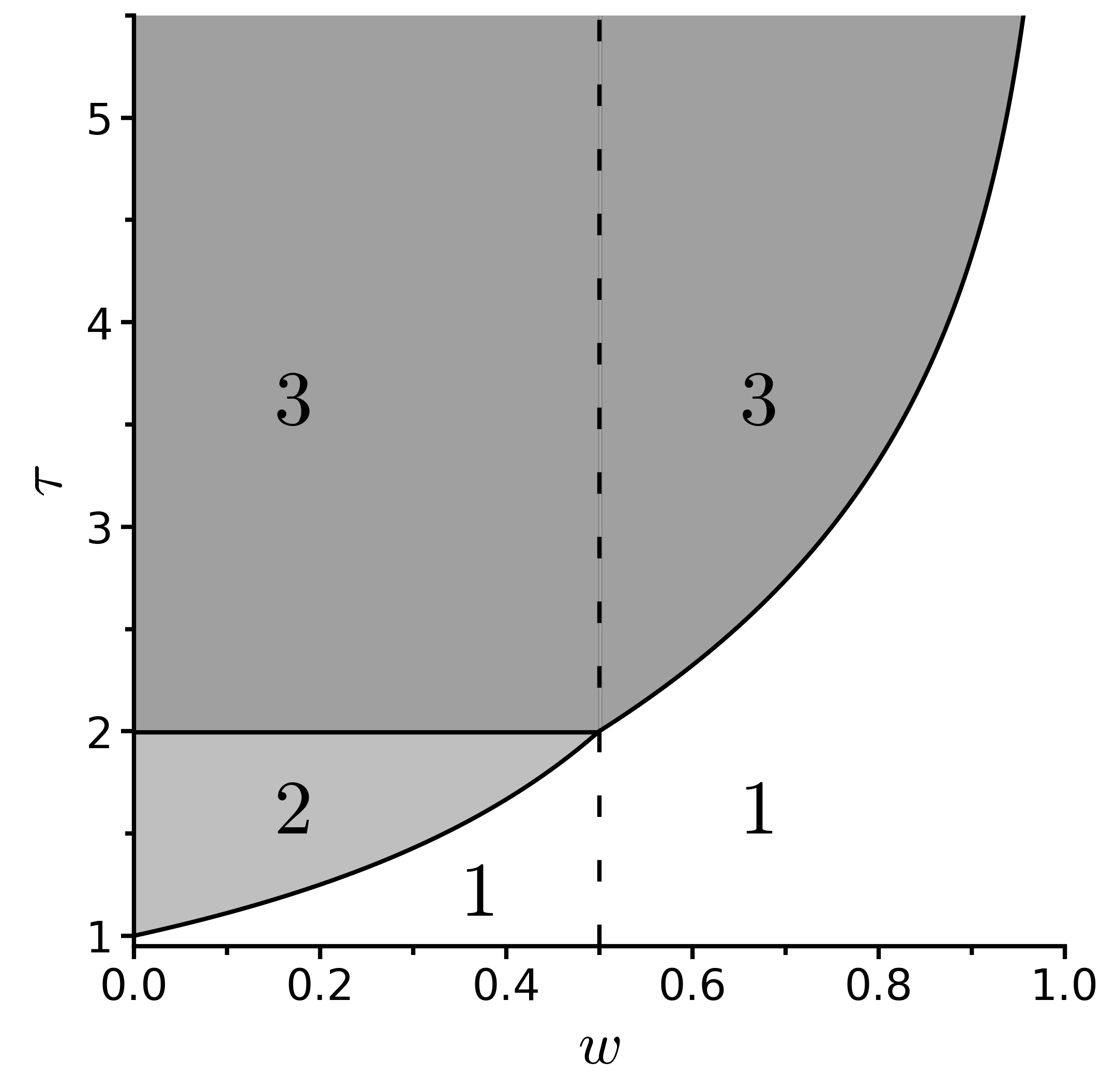}
\caption{Number of memory states as a function of the size of the uncertain region $w$ and the temperature ratio $\tau$. Shaded regions correspond to the number of states: one (white), two (light grey), three (dark grey). Black lines mark critical $\tau$ values at transitions in the number of states (Eqs. (\ref{Eq:taustar1-2})-(\ref{Eq:taustar2-3})). Dashed line at $w_c\!=\!1/2$.}
\label{Fig:PT}
\end{figure}

For $w < 1/2$ there is a second transition from two to three states, which happens at the critical value of 
\begin{equation} \label{Eq:taustar2-3}
    \tau^*_{2\rightarrow 3} = 2.
\end{equation} 
For all engines in this regime, the second transition happens precisely when the temperature at which energy is harvested is twice the temperature at which the memory is encoded. To see this we evaluate the net engine work output of the minimally dissipative two-state observer and show that it is exactly equal to the net engine work output of the minimally dissipative three-state observer at $\tau = 2$. To determine the residual probability $q_K^*$ that maximizes the net engine work output, we take the derivative of $W_{\rm out}^{(sK)}(q_K, w, \tau)$ with respect to $q_K$ and find the roots of the resulting equations. For optimal two-state memories we find,
\begin{equation} \label{Eq:q2*_main}
    q_2^*(w, \tau=2) = \frac{1}{2} - \frac{\sqrt{1+5w^2-4w-2w^3}}{2(1-w)^2}.
\end{equation}
Note that this expression is only valid at $\tau=2$. For general values of $\tau$ the optimal residual probability for two-state memories, $q_2^*(w,\tau)$, cannot be expressed analytically (see Appendix \ref{App:FullCalc}). For three memory states the optimal residual probability is given by
\begin{equation} \label{Eq:q3*_main}
    q_3^*(w, \tau) = \frac{w}{(1-w)(2^{\tau-1}-1)}.
\end{equation}
In contrast to the two-state solution, $q_3^*$ maximizes the net engine work output at any value of $\tau$. We show this in Appendix \ref{App:FullCalc}.

Using Eqs. (\ref{Eq:q2*_main}) and (\ref{Eq:q3*_main}), the maximum net engine work output of optimal two- and three-state observers at $\tau=2$ can be expressed as a function of $w$. In units of $kT'$ it is:
\begin{equation}
    {W_{\rm out}^{(s2)}(q_2^*, w) \over kT'} = \frac{1}{2}\biggl(\ln(2) - h(w)\biggr) = {W_{\rm out}^{(s3)}(q_3^*, w) \over kT'} ~.
\end{equation}

At the critical point $(w=1/2, \tau=2)$, optimal strategies with either two or three memory states are doing nothing. We have $q_3^*(w=1/2, \tau=2) = 1$, i.e. all observations are assigned to $m=0$, while $q_2^*(w=1/2, \tau=2) = 1/2$, i.e. all observations are assigned at random to one of the two states (which is equivalent to using one state). In both cases this results in no movement of the divider in the work medium and thus no work extraction.

Optimal three-state observers are outperformed by optimal two-state observers for $\tau < 2$ and in turn outperform optimal two-state observers for $\tau > 2$ (see Appendix \ref{App:SecondPT} and \ref{App:FullCalc} for details).

These results can be visualized with a phase diagram. In Fig. \ref{Fig:PT} the optimal number of memory states used by minimally dissipative observers is shown as a function of the size of the uncertain region, $w$, and the temperature ratio, $\tau$. In white regions observers are limited to using one memory state, in light gray regions observers with two memory states maximize the net engine work output and in dark gray areas, three-state memories allow for the maximum net work output. The critical values of the temperature ratio $\tau$ at which transitions in the optimal number of memory states occur, Eqs. (\ref{Eq:taustar1-2})--(\ref{Eq:taustar2-3}), are plotted as black lines in Fig. \ref{Fig:PT}.

\subsubsection{Naive coarse graining (hard partitions)}\label{Sec:CG}
An observer can choose to coarse grain the observable into either two, or three connected regions, which are then mapped to the respective memory states, $m$, deterministically, i.e. with probability one. 

\begin{figure}[h]
\centering 
\includegraphics[width=0.95\linewidth]{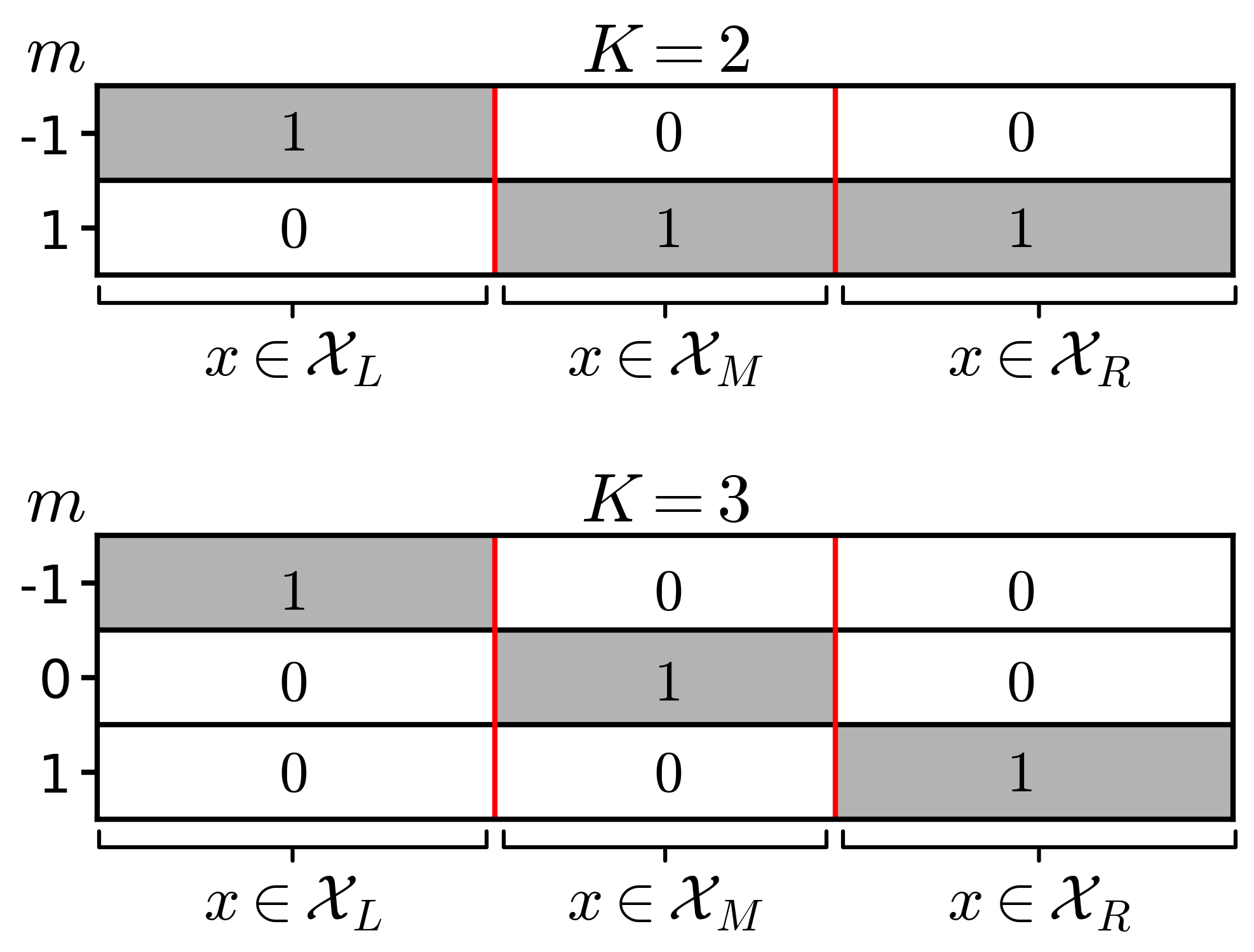}
\caption{Minimally dissipative encoding strategies for observers limited to coarse graining for $w=0.3$ (red lines at $\pm w/2$). Connected grey regions mark coarse grained states. Values correspond to $p(m \vert x)$.} 
\label{Fig:CG}
\end{figure}

For two memory states, a completely naive observer might choose to map the left/right portions of the $x$-axis to the respective memory state. 
But the observer could draw the line at a different position, and obtain an asymmetric memory. A numerical search over dividing positions reveals that, interestingly, the map 
\begin{equation} \label{Eq:pmgx_da2}
    p^{{\rm (da2)}}(m=1 \vert x) =
    \begin{cases}
    \quad 0~ &  x \in \mathcal{X}_L \\
    \quad 1~ & x \in (\mathcal{X}_M \cup \mathcal{X}_R),
    \end{cases}
\end{equation} 
results in the largest average net engine work output, using two coarse grained states.
This map is sketched in the top panel of Fig. \ref{Fig:CG} \footnote{Mirror imaging this mapping around $x=0$ results in an equivalent memory with identical performance.}. 

The $\tau$ values at which the changes occur from one to two, $\tau_{1\to 2}^{\rm det}(w)$, and two to three, $\tau_{2\to 3}^{\rm det}(w)$, memory states can be computed analytically from the total information, and the usable information, kept in the coarse grained memories. A one state memory corresponds to doing nothing at all and thus produces zero average engine work output. So to calculate $\tau_{1\to 2}^{\rm det}(w)$, we have to calculate the lowest value of the temperature ratio at which the asymmetrical two-state coarse graining yields positive net work output. For this memory, the minimum average input work is 
\begin{equation} \label{Eq:Imem_da2}
   kT \; I_{\rm m}^{\rm (da2)} = kT \; h\!\left({1+w \over 2} \right), 
\end{equation}
and the maximum average output work is  
\begin{equation} \label{Eq:Irel_det2acg}
    kT' \; I_{\rm u}^{\rm (da2)} = kT' \left[ \ln(2) - {1 + w \over 2}\, h\!\left( {w \over 1+w} \right) \right].
\end{equation}
The critical temperature 
is thus 
\begin{equation}\label{Eq:taudagger}
    \tau_{1\to 2}^{\rm det}(w) = {I_{\rm m}^{\rm (da2)} \over I_{\rm u}^{\rm (da2)}} = {h\!\left({1+w \over 2} \right) \over \ln(2) - {1 + w \over 2}\, h\!\left( {w \over 1+w}\right)}.
\end{equation}

For $\tau > \tau_{2\to 3}^{\rm det}(w)$ minimally dissipative observers limited to coarse graining use the three-state coarse graining that captures all available usable information (discussed in the beginning of Sec. \ref{Sec:Results} and shown in the bottom panel of Fig. \ref{Fig:CG}). This strategy would also be used by observers that maximize the average work output per cycle, regardless of the costs. In App. \ref{App:MaxWout} we compare this strategy to optimal observers and provide an analytical expression for the advantage derived from using thermodynamically optimal encodings.

It is never beneficial to use any other three-state memory, if the observer is limited to coarse graining \footnote{We verified this numerically.}. The critical temperature ratio, $\tau_{2\to 3}^{\rm det}(w)$, at which it becomes worthwhile for a deterministic observer to switch to three states, can be found by calculating the $\tau$ value for which the two-and three-state memories result in the same average net engine work output, i.e. 
\begin{equation} \label{Eq:taustar}
    \tau_{2\to 3}^{\rm det}(w) = {I_{\rm m}^{\rm (d3)} - I_{\rm m}^{\rm (da2)} \over I_{\rm u}^{\rm (d3)} - I_{\rm u}^{\rm (da2)}}.
\end{equation}
With Eqs. (\ref{Eq:Irel_d3}), (\ref{Eq:3cg}), (\ref{Eq:Imem_da2}), and (\ref{Eq:Irel_det2acg}), we obtain 
\begin{eqnarray} \label{Eq:tau_det_2to3}
   \!\!\!\!\tau_{2\to 3}^{\rm det}(w) \!=\!  1\!+\! { w \ln(w) \!+\! (1\!-\!w) \left[ \ln(1\!-\!w) \!+\! 2\ln(2) \right] \over
    w \!\left[\ln(w) \!+\! 2 \ln(2)\right] \!-\!(1\!+\!w) \ln(1\!+\!w)
    }.
\end{eqnarray}
\begin{figure}[ht!]
\centering
\includegraphics[width=0.95\linewidth]{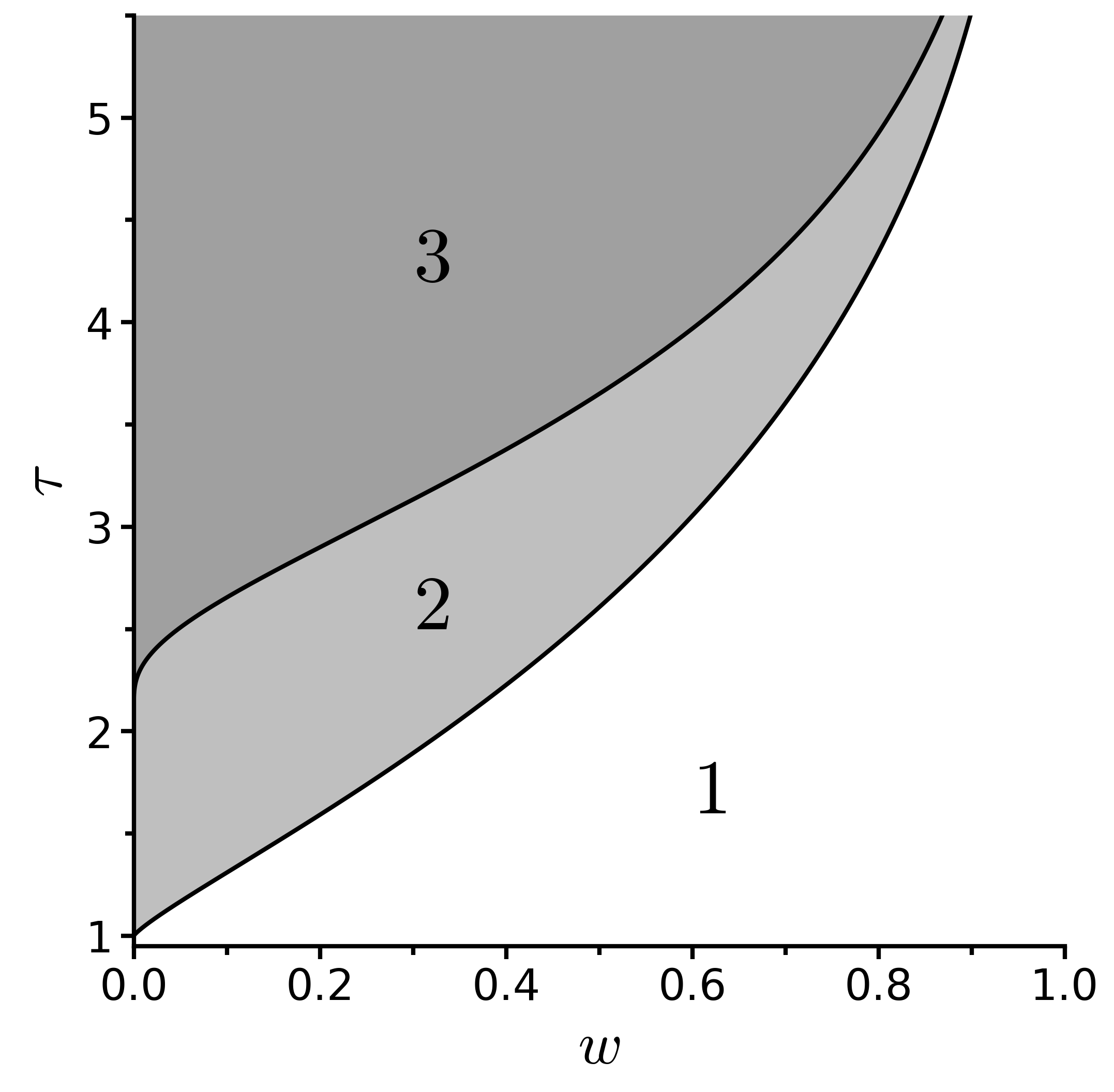}
\caption{Phase diagram for observers limited to coarse graining. Shaded regions correspond to the number of states: one (white), two (light grey), three (dark grey). Black lines mark critical $\tau$ values at transitions in the number of states (Eqs. (\ref{Eq:taudagger}) and (\ref{Eq:tau_det_2to3})).}
\label{Fig:Det_PT}
\end{figure}

The resulting behaviour can again be summarized in a phase diagram. The best deterministic observers use asymmetric two-state memories for temperature ratios $\tau^{\rm det}_{1 \to 2} \leq \tau < \tau^{\rm det}_{2 \to 3}$. For greater temperature ratios they use three memory states (dark grey region in Fig. \ref{Fig:Det_PT}). In contrast to optimal observers, observers limited to coarse graining derive no benefit from skipping two-state memories, regardless of the size of the uncertain region, since $\tau_{1\to 2}^{\rm det}(w) < \tau_{2\to 3}^{\rm det}(w)$ for all $w < 1$. For engines with vanishing certain regions, $w \to 1$, the temperature ratio, at which the first phase transition occurs, diverges for both optimal and deterministic observers, because there is no longer any usable information whereby doing nothing is the optimal strategy, even for $\tau \to \infty$.

Detailed calculations leading to Eqs. (\ref{Eq:Irel_d3}), (\ref{Eq:3cg}), (\ref{Eq:Imem_da2}), and (\ref{Eq:Irel_det2acg}), as well as a comparison between the performance of different coarse grainings can be found in Appendix \ref{App:DCG}.

\begin{figure}[h]
\centering 
\includegraphics[width=0.95\linewidth]{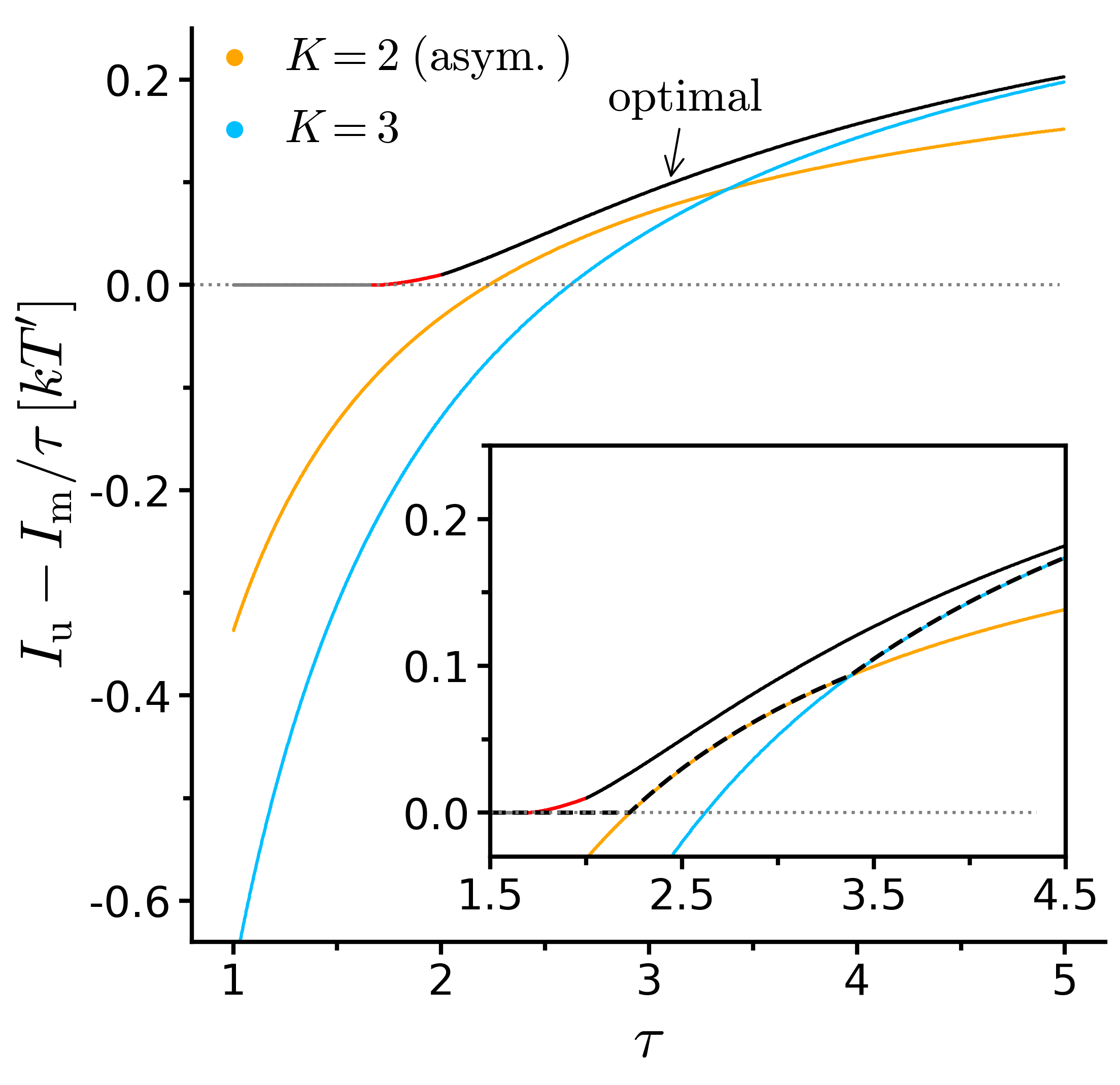}
\caption{Average net work output as a function of $\tau$ for optimal and deterministic memories of an engine with $w=0.4$. Optimal memories with two states are plotted in red, with three states in black and with one state in grey. The dashed, black line in the inset shows the best deterministic solution at each value of $\tau$.}
\label{Fig:opt04}
\end{figure}

To gauge the extent to which net work output is lost when observers are restricted to coarse graining, we plot, in Fig. \ref{Fig:opt04}, for $w=0.4$, the engine's expected net work output as a function of $\tau$, Eq. (\ref{Eq:Woutengine}), comparing between engines run with memories that coarse grain, and engines run with optimal memories (those that solve Eqs. (\ref{Eq:IBalg})).

For temperature ratios below $\tau_{1\to 2}^{\rm det}(0.4) \approx 2.2$, observers limited to coarse graining can derive no positive net work output from the information engine. Instead, they would have to invest resources to run the engine. This can be seen by the total average engine work output going negative in Fig. \ref{Fig:opt04} (orange and blue curves), which happens when the thermodynamic costs of running the memory outweigh the gain derivable from the memorized information. In this situation the best strategy for the observer is to do nothing, i.e. to memorize no information at all. The inset of Fig. \ref{Fig:opt04} shows, as a dashed line, the maximum achievable net work output with coarse graining at each $\tau$.

For intermediate temperature ratios, $\tau_{1\to 2}^{\rm det}(0.4) \leq \tau < \tau_{2\to 3}^{\rm det}(0.4) \approx 3.4$, minimally dissipative deterministic observers use the asymmetric two-state coarse graining, Eq. (\ref{Eq:pmgx_da2}). For higher temperature ratios, $\tau \geq \tau_{2\to 3}^{\rm det}(0.4)$, they use a three-state coarse graining that captures all available usable information. Optimal observers have more flexibility in their encoding, consequently they can derive positive work output from the engine at temperature ratios for which the best observers limited to coarse graining are still doing nothing. For large values of $\tau$, the advantage of optimal observers diminishes, as encoding costs become less and less important, and already at $\tau=5$, the difference in net engine work output shown in Fig. \ref{Fig:opt04} is minuscule.

\begin{figure*}[ht!]
\centering 
\includegraphics[width=0.95\linewidth]{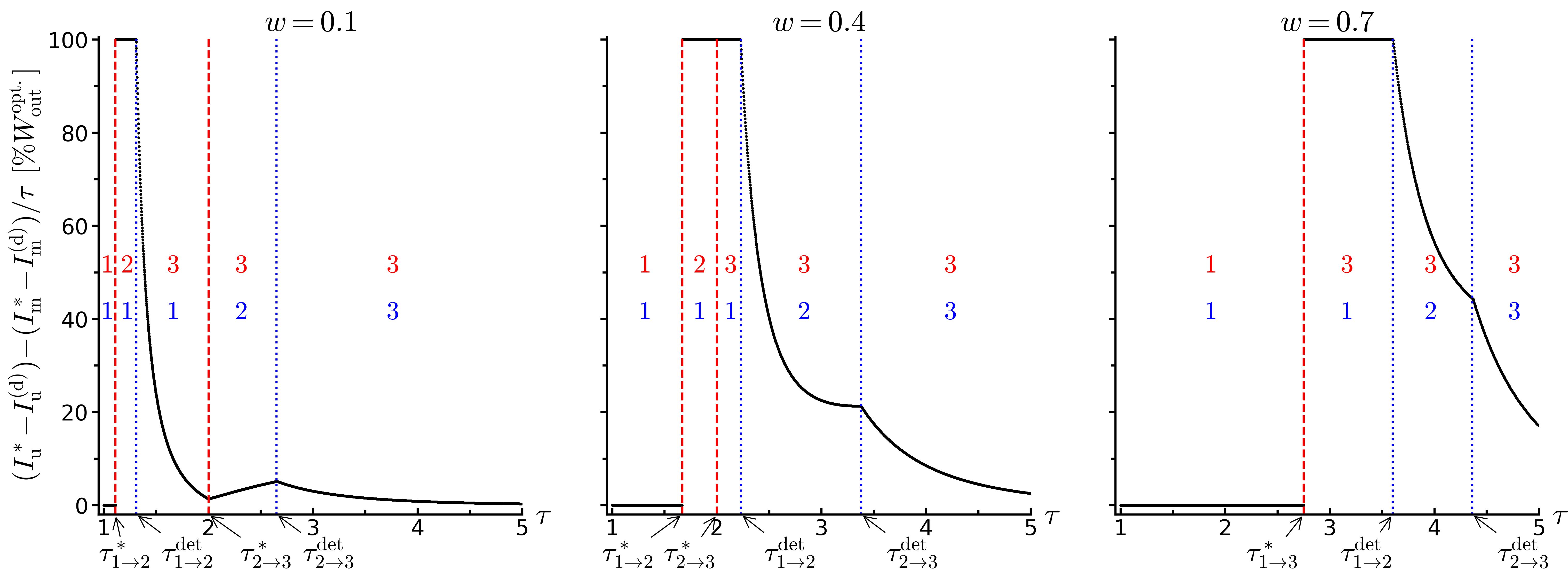}
\caption{Difference in average net engine work output between optimal and best deterministic observers in percent of the optimal work output, $W_{\rm out}^{\rm opt.}\!=\!kT' I_{\rm u}^{*} - kT I_{\rm m}^*$, for three selected engine geometries (left: $w=0.1$, center: $w=0.4$, right: $w=0.7$) as a function of $\tau$. Critical $\tau$ values and number of memory states for optimal (dashed red lines, red numbers) and best deterministic observers (dotted blue lines, blue numbers).}
\label{Fig:Det_diff}
\end{figure*}

To quantify the difference in engine performance we plot in Fig. \ref{Fig:Det_diff} the minimum amount of average net work output lost due to using coarse graining, instead of using optimal memories, measured in percent of the net work output achieved with optimal memories, \mbox{$W_{\rm out}^{\rm opt.} = I_{\rm u}^{*} - I_{\rm m}^*/\tau$},
\begin{equation}
\frac{(I_{\rm u}^* - I_{\rm u}^{\rm (d)}) -(I_{\rm m}^* - I_{\rm m}^{\rm (d)})/\tau}{I_{\rm u}^{*} - I_{\rm m}^*/\tau}, \nonumber
\end{equation}
as a function of $\tau$ for three select engine geometries. (All work quantities are measured in units of $kT'$.)
Discontinuities in the curves are due to phase transitions in the number of memory states used by deterministic observers. Critical $\tau$ values at which transitions in the number of memory states occur are marked by dashed red lines for optimal observers and by dotted blue lines for observers limited to coarse graining and the optimal number of memory states used by both types of observers is denoted by colored numbers (red for optimal observers and blue for coarse graining) in the regions delineated by the critical $\tau$ values.

The difference between optimal observers and those limited to coarse graining is clearest for engines with substantial, but not overwhelming uncertainty ($1/3 < w < 1/2$). In this regime, for $\tau < \tau_{1\rightarrow 2}^*$ neither observer can operate the engine successfully, which is equivalent to using one memory state; in the range $\tau_{1\rightarrow 2}^* \leq \tau < \tau_{1\rightarrow 2}^{\rm det}$ optimal observers successfully harvest energy, relying on two memory states when $\tau \leq 2$, and on three when $2 < \tau$, while observers whose model class is restricted to coarse graining cannot harvest energy, see center panel of Fig. \ref{Fig:Det_diff}. For $\tau_{1\rightarrow 2}^{\rm det} \leq \tau < \tau_{2\rightarrow 3}^{\rm det}$ optimal observers rely on three memory states, but restricted observers coarse grain into two regions, having lower engine performance. For $\tau \geq \tau_{2\rightarrow 3}^{\rm det}$, coarse graining observers employ the three-state memory that captures all available usable information, and the difference in engine performance becomes small, disappearing entirely in the limit of large $\tau$. 

\section{Conclusion} \label{Sec:Conclusion} 
We use the framework of generalized, partially observable information engines to study optimal encoding strategies for three different classes of observers in a \LS\ engine with uncertainty from a physical standpoint. 

Each bit of information captured in a physical memory incurs a thermodynamic cost of $kT \ln(2)$. In general, not all information retained in memory, $I_m$, can be traded back into work gain. At best, $I_u \leq I_m$ usable bits of information about the work medium can be leveraged to extract $kT' \,I_u$ joules of work from a heat bath in an information engine process which allows the observer to make the memory at a lower temperature, $T<T'$, than the temperature at which work is extracted. The temperature ratio $\tau = T'/T$ then determines the thermodynamic value of the usable information, and thereby sets the trade-off between thermodynamic costs and benefits of the memory. 

Optimal memories waste as little energy as possible by retaining as little useless information as possible. They can be derived from maximization of the information engine's average net work output, optimized over all possible data encodings. An iterative algorithm (called ``Information Bottleneck method") can be used to find the optimal memory making strategies. Once those are known, a physical code book can be constructed, that enables machinery to implement not only a pre-described function of Maxwell's demon, as is done in \LS's engine, but also the demon's strategy of how to execute its function (which is predetermined by an ``external" experimenter in \LS's engine, as well as most current information engines \cite{toyabe2010experimental, berut2012experimental, mandal2012work, barato2013autonomous, horowitz2013imitating, diana2013finite, koski2014experimental, koski2014SEszilard, jun2014high, koski2015chip, chapman2015autonomous, hong2016experimental, camati2016experimental, gavrilov2016erasure, boyd2016identifying, boyd2017correlation, boyd2017leveraging, strasberg2017quantum, mcgrath2017biochemical, gavrilov2017direct, chida2017power, cottet2017observing, ciampini2017experimental, paneru2018lossless, admon2018experimental, masuyama2018information, stopnitzky2019physical, brittain2019biochemical, stevens2019quantum, ribezzi2019large, peterson2020implementation, paneru2020efficiency, paneru2020colloidal, saha2021maximizing, dago2021information, saha2023information}).

For unconstrained observers optimal encoding strategies can be found numerically, but what happens if the physical observer is limited to a subset of all possible encoding strategies? Analysis of the algorithmically discovered optimal solutions lead us to introduce parameterized soft partitions. We found that the observer's model class need only include these parameterized soft partitioning strategies in order to contain thermodynamically optimal strategies, it is not necessary to consider observers with access to the much larger model class containing {\em all} probabilistic maps. On the other hand, we showed that it is not sufficient to consider observers limited to coarse graining of the observable (hard partitioning).
For both classes of constrained observers analytic expressions for key quantities characterizing thermodynamically optimal strategies, such as memorized and usable information and critical $\tau$ values at transitions to more memory states, are derived.

The soft partitioning strategies allow for an easy interpretation of emerging optimal strategies. We find that optimal memory making strategies, in this simple decision problem, do not coarse grain for most values of $\tau$. Only for small $\tau$ and for very large $\tau$, is it optimal to coarse grain the observable. For small $\tau$, the observable space is coarse grained into one region, capturing no information and thus no action is taken on the work medium, resulting in zero work output. For very large $\tau$, the thermodynamic value of usable information relative to the cost of memory making is vast. Then it is optimal to coarse grain the observable into three regions: the two regions in which it is certain that the left/right side of the work medium is empty, and the region (of size $w$) in which we have complete uncertainty. This coarse graining captures all available usable information. 

The relative thermodynamic value of usable information is in between these two extremes for all other temperature ratios, $\tau$, and in this intermediate regime, optimal strategies do not coarse grain. To summarize our results, we focus on work medium geometries, for which the region in which observables contain no usable information (what we call the uncertain region) is less than half of the total size, because these have the most interesting behavior.

Optimal observers with two-state memories do not coarse grain the observable into two equally sized regions. Instead they assign data in the uncertain region into either memory state at random, incurring no encoding cost for data that contains no usable information. Therefore, if we compare the optimal observer strategy to totally naive coarse graining along the mid-line, it would appear that the cleverness of an optimal observer lies in being able to achieve the same thermodynamic gains at lower costs.

However, among those observers that are restricted to coarse grainings, the naive splitting into two equally sized regions is not the best. We find that it is less costly, and captures more usable information, to coarse grain asymmetrically: one region contains one of the certain regions, the other region contains the rest.

When we compare the resulting engine performance, we have to compare at the same value of $\tau$, and the soft partitions assign data in the certain regions probabilistically, with optimal probabilities changing with the relative value of usable information compared to memorized information, $\tau$, (at larger $\tau$, data in the certain regions are assigned with higher probability to the memory state that results in the correct action on the work medium). The probabilistic assignment further reduces memory making costs, but increases the inference error associated with each memory state. Using only two memory states a residual inference uncertainty is unavoidable. 

When the temperature ratio makes the use of three states worthwhile for an optimal observer, the strategy changes and the observer wants to be certain about when it has no knowledge: All data in the uncertain region get mapped to the same memory state with probability one. This memory state results in no action on the work medium. The option to do nothing only becomes available with the addition of the third memory state and mapping all data from the uncertain region into this new state ensures that the other two memory states allow the observer to infer the empty side of the work medium container with certainty. Cost-saving for three-state observers occurs by assigning data in certain regions to this same ``no-action" memory state with some probability, which decreases with increasing thermodynamic value of usable information, $\tau$. With remaining probability the data in the certain regions is mapped to that memory state which results in the correct work extraction (volume expansion into the empty side of the work medium), and never to the memory state that results in the wrong work extraction protocol (volume compression). 
As the thermodynamic value of usable information increases with increasing $\tau$, the probability with which data in the certain regions is mapped to the memory state that results in the correct action on the work medium, increases, until the three-state coarse graining is recovered at large values of $\tau$.

The novel soft partitioning strategies, while reasonable in hind sight, are not immediately obvious to guess. They were revealed by our analysis. We hope that this might inspire descriptions of other thermodynamic systems in similar situations, when not all pertinent information is available to the observer, or when data compression and low thermodynamic model costs are important for other reasons. 

\begin{acknowledgments} \noindent We thank Rob Shaw for extremely helpful discussions and comments. We are most grateful for funding from the Foundational Questions Institute, Grant Nos. FQXi-RFP-1820 (FQXi together with the Fetzer Franklin Fund) and FQXi-IAF19-02-S1. This publication was made possible through the support of the ID\# 62312 grant from the John Templeton Foundation, as part of the \href{https://www.templeton.org/grant/the-quantum-information-structure-of-spacetime-qiss-second-phase}{‘The Quantum Information Structure of Spacetime’ Project (QISS)}. The opinions expressed in this publication are those of the author(s) and do not necessarily reflect the views of the John Templeton Foundation.
\end{acknowledgments} 

\appendix
\section{Mapping binary decision problems to partially observable \LS\ engines}\label{App:BDP}
A binary decision problem with observable data $d \in \mathbb{R}^n$ can be fully specified by $p(u=1|d)$, since $p(u=-1|d) = 1-p(u=1|d)$ follows from normalization. To map the binary decision problem to a partially observable \LS\ engine the values of $p(u=1|d)$ are rearranged in increasing order and $d$ is mapped to $x \in \mathbb{R}$, such that the resulting function $f(x)$ is monotonically increasing in $x$. If the probability density $\rho(x)$ (or equivalently $\rho(d)$) is constant, the partially observable \LS\ engine corresponding to the decision problem results from trapping a single-particle gas in a cube-shaped container (without loss of generality we assumed the container to have unit length in every spatial direction) with the divider inside the container shaped as $f(x)$. Discontinuities in $f(x)$ are connected vertically if necessary. If the probability density over the data is not constant, the work medium container will no longer be rectangular, instead the cross-sectional area of the container has to reflect $\rho(x)$, being wider in regions where the density is larger and narrower in regions where the density is smaller, assuming that the containers depth is the same everywhere. Let the perimeter of the container in the $x$-$y$-plane be described by $y(x)$, then we require $y(x_1)/y(x_2) = \rho(x_1)/\rho(x_2)$. In this way any binary decision problem can be mapped unto a partially observable \LS\ engine.

Importantly this mapping procedure can be used for binary decision problems with intrinsic uncertainty, measurement errors or a combination thereof. Intrinsic uncertainty is present even if the measurements are perfect and there is no measurement error. Both types of uncertainties affect $p(u|x)$ and can be taken into account by using an appropriately shaped divider in the partially observable \LS\ engine. Observations in the information engine can then be assumed to be error free (since measurement errors are already accounted for by the divider shape). A more detailed discussion of the effects of measurement errors will be left to a follow-up article.

\section{Dependence of usable and total information in memory on the data representation strategy}\label{App:Info}
To show explicitly the dependencies of total information kept in memory, $I_m$, and usable information, $I_u$, on the data representation characterized by $p(m|x)$, we write out Eq. (\ref{Eq:Imem}) (visually highlighting ${\cob p(m|x)}$ in the following equations in blue)  
\begin{equation}
I_m \!=\! \int dx \rho(x) \sum_m {\cob p(m|x)} \ln\left[ {\cob p(m|x)} \over \int dx' \rho(x') {\cob p(m|x')} \right], \notag 
\end{equation}
and combine Eqs. (\ref{Eq:Irel-def})-(\ref{Eq:pm}):
\begin{eqnarray}
I_u &=& \!\!\int\!\! dx \rho(x) \!\sum_{m,u} {\cob p(m|x)} p(u|x) \ln\left[ \!\!\int\!\! dx'' \rho(x'') {\cob p(m|x'')} p(u|x'') \over \!\!\int\!\! dx' \rho(x') {\cob p(m|x')} \right] \nonumber \\&&+ \ln(2)~.\notag  
\end{eqnarray}   

\section{Maximizing work output per cycle regardless of the costs}\label{App:MaxWout}
\begin{figure}[ht!]
\centering
\includegraphics[width=0.95\linewidth]{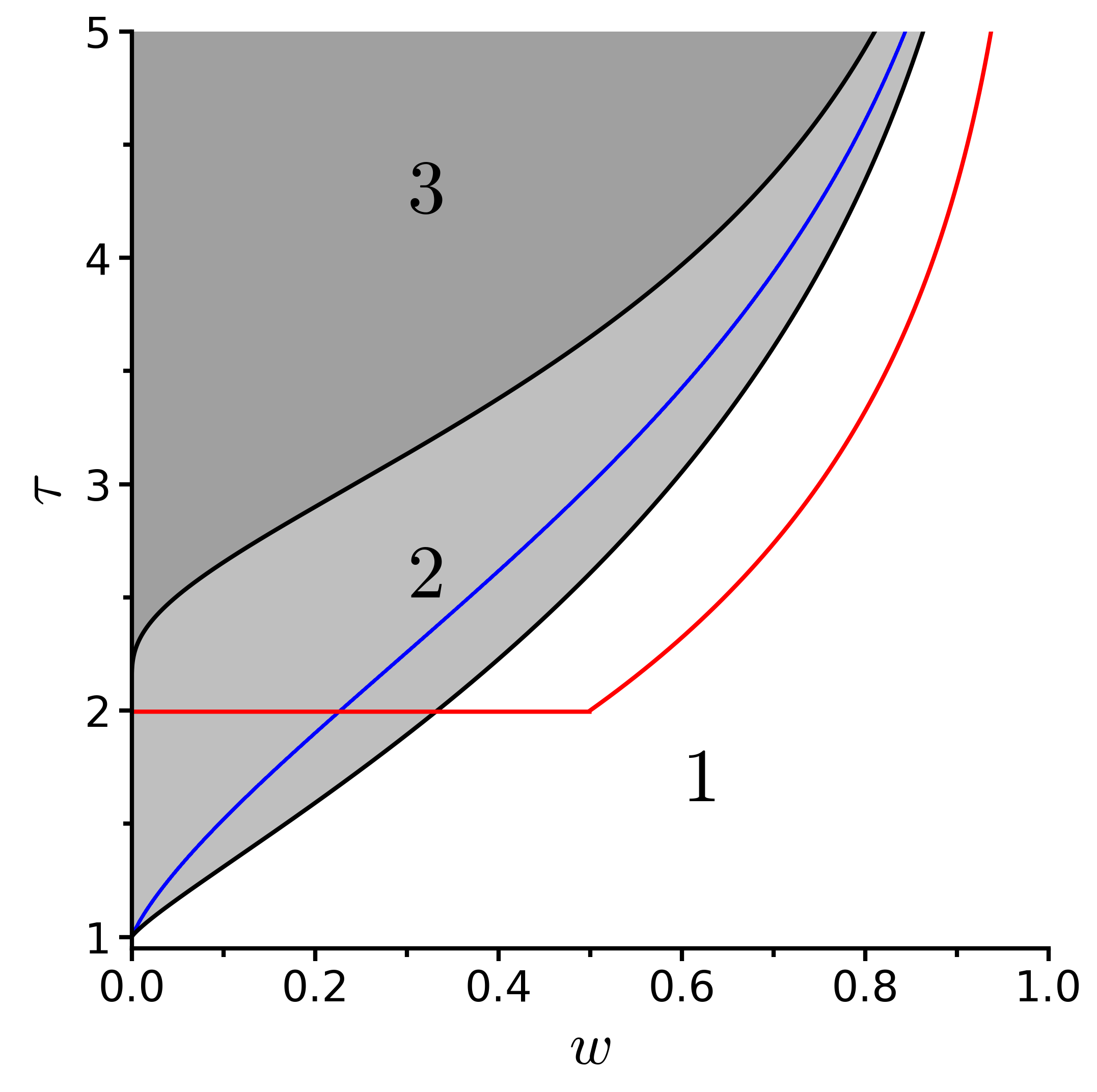}
\caption{Phase diagram for observers limited to coarse graining. 
Shaded regions correspond to the number of states: one (white), two (light grey), three (dark grey). Black lines mark critical $\tau$ values at transitions in the number of states (Eqs. (\ref{Eq:taudagger}) and (\ref{Eq:tau_det_2to3})). Red lines show when optimal observers start using three states and the blue line marks when maximizing the work output of the engine, regardless of the costs, produces positive net work output, Eq. (\ref{Eq:tau_zc}).}
\label{Fig:Det_PT_zc3}
\end{figure}

Observers that are only interested in maximizing the work output of the information engine per cycle, regardless of the operating costs, employ the three state coarse graining that captures all usable information (see Secs. \ref{Sec:Results}, \ref{Sec:CG} and App. \ref{App:DCG}). This data representation strategy categorizes the data based on two successive binary decisions: first the data is grouped into informative vs. uninformative and then informative data is further grouped into left vs. right side being empty ($u=\pm1$). The first decision provides no relevant information for work extraction, because the observer needs to know {\it which} side of the work medium container is empty. Since entropy is additive in this scenario and $I_{\rm m}^{(d3)} = H^{(d3)}[m]$, the total memorized information is the sum of two terms, $h(w)$ and $(1-w)\ln(2)$, one for each binary decision, see Eq. (\ref{Eq:3cg}). Only the second classification of the data is usable for work extraction and thus $I_{\rm u}^{(d3)} = (1-w) \ln(2)$ and $I_{\rm m}^{(d3)} - I_{\rm u}^{(d3)} = h(w)$, which is precisely the information gained by dividing the data in informative vs. uninformative samples. Since this information cannot be used for work extraction we referred to it as ``irrelevant" information in previous work \cite{CB, stilldaimer2022}, but here will denote it as ``useless" information: $I_{\rm ul} := I_{\rm m} - I_{\rm u}$. An observer can only produce positive net work output, if 
\begin{equation}
    \tau > \frac{I_{\rm m}}{I_{\rm u}} = 1 + \frac{I_{\rm ul}}{I_{\rm u}}
\end{equation}
and with this we immediately recover Eq. (\ref{Eq:tau_zc}) for observers that maximize the engine work output, regardless of the costs (blue line in Fig. \ref{Fig:Det_PT_zc3}).

Three coarse grained memory states are sufficient to produce the maximum possible work output for the class of information engines studied here, but how do observers using this data representation strategy compare to the thermodynamically optimal observers discussed in Secs. \ref{Sec:OptimalStrategies} and \ref{Sec:PSP}? The difference in {\it net} engine work output, measured in units of $kT$, between engines run by thermodynamically rational agents using three memory states and those that simply maximize the work output of the engine is
\begin{eqnarray}
    \frac{W_{\rm out}^{\rm (s3)} - W_{\rm out}^{(d3)}}{kT} &=& \tau(I_{\rm u}^{\rm (s3)} - I_{\rm u}^{\rm (d3)}) - (I_{\rm m}^{\rm (s3)} - I_{\rm m}^{\rm (d3)}) \notag \\ 
    &=& (\tau-1)(I_{\rm u}^{\rm (s3)} - I_{\rm u}^{\rm (d3)}) - (I_{\rm ul}^{\rm (s3)} - I_{\rm ul}^{\rm (d3)}) \notag \\
    &=& - q_3^*(1-w)\Delta \ln(2) - (I_{\rm ul}^{\rm (s3)} - I_{\rm ul}^{\rm (d3)}) \notag \\
    &=& \frac{ -w \Delta}{2^{\Delta}-1} \ln(2) \notag\\
    &&\!-\! w \left(\ln(2^{\Delta}\!-\!1) \!-\! \frac{2^{\Delta}}{2^{\Delta}\!-\!1} \ln(2^{\Delta})\right) \notag \\
    &=& -\frac{w}{2^{\Delta}-1}(\Delta \ln(2) - \ln(2^\Delta-1)) \notag \\
    &&-\frac{2^\Delta w}{2^\Delta-1}(\ln(2^\Delta-1)-\ln(2^\Delta)) \notag \\
    &=& \left(-\frac{w}{2^\Delta-1} + \frac{2^\Delta w}{2^\Delta-1}\right)\ln\left(\frac{2^\Delta}{2^\Delta-1}\right) \notag \\ 
    &=& w \ln\left(\frac{2^{\Delta}}{2^{\Delta} - 1} \right), \label{Eq:NetWorkDiff3}
\end{eqnarray}
where we used $\Delta \equiv \tau -1$ to shorten notation, $I_{\rm ul}^{\rm (d3)} = h(w)$, $q_3^* = \frac{w}{(1-w)(2^{\Delta}-1)}$ and
\begin{eqnarray}
    I_{\rm ul}^{(s3)} &=& h(w + (1-w)q_3^*) - (1-w)h(q_3^*) \\
    &=& w \left(\ln(2^{\Delta}\!-\!1) - \frac{2^{\Delta}}{2^{\Delta}-1} \ln(2^{\Delta})\right) \!+\! h(w).
\end{eqnarray}

The advantage of thermodynamically rational strategies, as quantified by Eq. (\ref{Eq:NetWorkDiff3}), is linear in the amount of uncertainty, $w$, and monotonically decreasing in the relative temperature difference $\Delta=(T'-T)/T$. For $w < 1/2$ the maximum difference between the two types of observers occurs at $\Delta_{2 \to 3} = 1$ (horizontal red line in Fig. \ref{Fig:PT_App}), where it is $w\ln(2)$, while for $w > 1/2$, the maximum difference is $-w\ln(w)$ at $\Delta_{1 \to 3} = -\log_2(1-w)$. For large uncertainties, $w>1/2$, the maximum difference between the two types of three-state observers is somewhat less interesting, as it is simply a lack of loss. Along $\Delta_{1 \to 3} = -\log_2(1-w)$ optimal three-state observers produce zero net work output, while the three-state coarse graining incurs a net average operating cost of $w \ln(w)$ per cycle. 
At $w=1/2$, $\tau=2$ ($\Delta=1$) the largest advantage can be achieved: it is $kT \ln(2)/2$, half of the maximally achievable gain of $kT \ln(2)$ at these parameter values.

\section{Coarse graining strategies}\label{App:DCG}
Here we provide detailed calculations for the inferences, $p(u \vert m)$, and the conditional entropies, $H[u \vert m]$, used to compute the usable information retained by the two coarse grainings in Sec. \ref{Sec:CG}. We also provide some additional details about those coarse grainings and compare them to a naive, symmetrical two-state coarse graining.

To compute the usable information retained in memory, Eq. (\ref{Eq:Irel-def}), $p(u \vert m)$ needs to be known. For any memory assignment, $p(m \vert x)$, and any engine geometry, $p(u \vert x)$, we have (see Eq. (\ref{Eq:pum_1})) 
\begin{equation} \label{Eq:pugmint}
    p(u \vert m) = {1\over p(m)} \int_{-1/2}^{1/2} dx \; p(u\vert x) p(m \vert x),
\end{equation}
where we used $\rho(x)=1$ for all $x$ inside the container.

The simplest two-state coarse graining is a symmetric partitioning of the observable with,
\begin{equation}\label{Eq:pmgx_LS}
    p^{\rm (d2)}(m \vert x) = \delta_{m \; {\rm sign}(x)}. 
\end{equation}
Equation (\ref{Eq:pmgx_LS}) describes the optimal memory assignment for a \LS\ engine for any value of the temperature ratio $\tau$. Such a symmetric coarse graining always captures $I_{\rm m}^{\rm (d2)} = \ln(2)$ nats or 1 bit of information, since $p^{\rm (d2)}(m) = 1/2$. 
To determine the usable part of the memorized information we first evaluate Eq. (\ref{Eq:pugmint}), using $p(u \vert x)$ defined in Eq. (\ref{Eq:pugx}) and $p^{\rm (d2)}(m \vert x)$ given in Eq. (\ref{Eq:pmgx_LS}):
\begin{eqnarray} \label{Eq:pugm_d2}
    p^{\rm (d2)}(u=1 \vert m=1) &=& 2 \!\left( \int_{-1/2}^{1/2} dx \; p(u=1 \vert x) \, p(m=1 \vert x) \right) \notag \\
    &=& 2 \!\left(\int_{-1/2}^{0}\!\!0 dx + \!\int_{0}^{w/2} \!{1 \over 2} dx + \!\int_{w/2}^{1/2}\!\! 1 dx \right) \notag \\
    &=& 1 - w/2.
\end{eqnarray}
Note that since $u$ is a binary random variable and since the two memory states are symmetric, it is sufficient to compute $p^{\rm (d2)}(u=1 \vert m=1)$, because we always have \mbox{$p(u=-1 \vert m=1) = 1- p(u=1 \vert m=1)$} and due to symmetry we also have \mbox{$p^{\rm (d2)}(u=1 \vert m=-1) = w/2 = p^{\rm (d2)}(u=-1 \vert m=1)$}.
Now we can evaluate the conditional entropy $H^{\rm (d2)}[u \vert m]$:
\begin{eqnarray} \label{Eq:condEnt_d2}
    H^{\rm (d2)}[u \vert m] &=& -\sum_{m,u} p^{\rm (d2)}(u\vert m) p^{\rm (d2)}(m) \ln[p^{\rm (d2)}(u \vert m)] \notag \\
    &=& -2p^{\rm (d2)}(m=1)\biggl[\sum_u p^{\rm (d2)}(u\vert m=1) \notag \\ 
    && \times \ln[p^{\rm (d2)}(u \vert m=1)]\biggr] \notag  \\
    &=& -p^{\rm (d2)}(u=1\vert m=1) \ln[p^{\rm (d2)}(u=1 \vert m=1)] \notag \\
    && -\!p^{\rm (d2)}(u\!=\!-1\vert m\!=\!1) \ln[p^{\rm (d2)}(u\!=\!-1 \vert m\!=\!1)] \notag \\
    &=& -(1-w/2) \ln(1-w/2) - w/2 \ln(w/2)\notag \\
    &=& h(w/2).
\end{eqnarray}
In the second line we used the fact that the two memory states are symmetric, $p^{\rm (d2)}(m=1) = 1/2 = p^{\rm (d2)}(m=-1)$, and for the final equality we expressed the result in terms of the entropy function $h$, introduced in Eq. (\ref{Eq:Entq}). Since we always have $H[u] = \ln(2)$, we get the usable information by inserting Eq. (\ref{Eq:condEnt_d2}) into Eq. (\ref{Eq:Irel-def}):
\begin{equation} \label{Eq:Irel_det2cg}
    I_{\rm u}^{\rm (d2)} = \ln(2) - h\left({w \over 2}\right).
\end{equation}

Interestingly, this is not the best two-state coarse graining for our engine class and it is outperformed by the asymmetrical coarse graining shown in the upper panel of Fig. \ref{Fig:CG} and described by Eq. (\ref{Eq:pmgx_da2}). 
This asymmetrical coarse graining maps one certain region and the uncertain region into one memory state with greater weight, while mapping the second certain region into the other, less probable memory state.

\begin{figure*}[t!]
\centering 
\begin{subfigure}{.45\linewidth}
    \centering
    \includegraphics[width=0.9\linewidth]{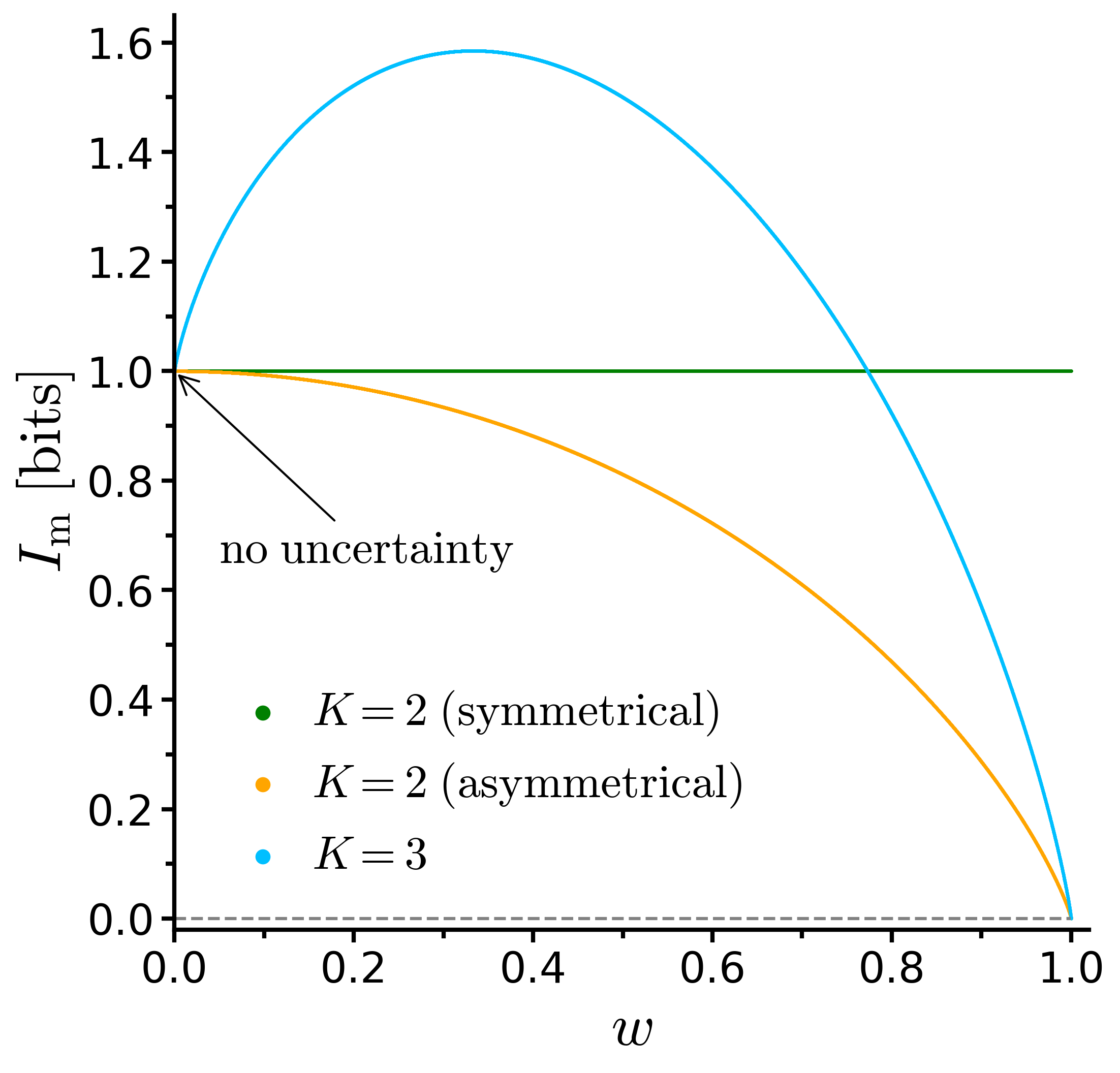}
    \phantomsubcaption
    \label{Fig:Imem}
\end{subfigure}
\hspace{1em}
\begin{subfigure}{.45\linewidth}
    \centering
    \includegraphics[width=0.9\linewidth]{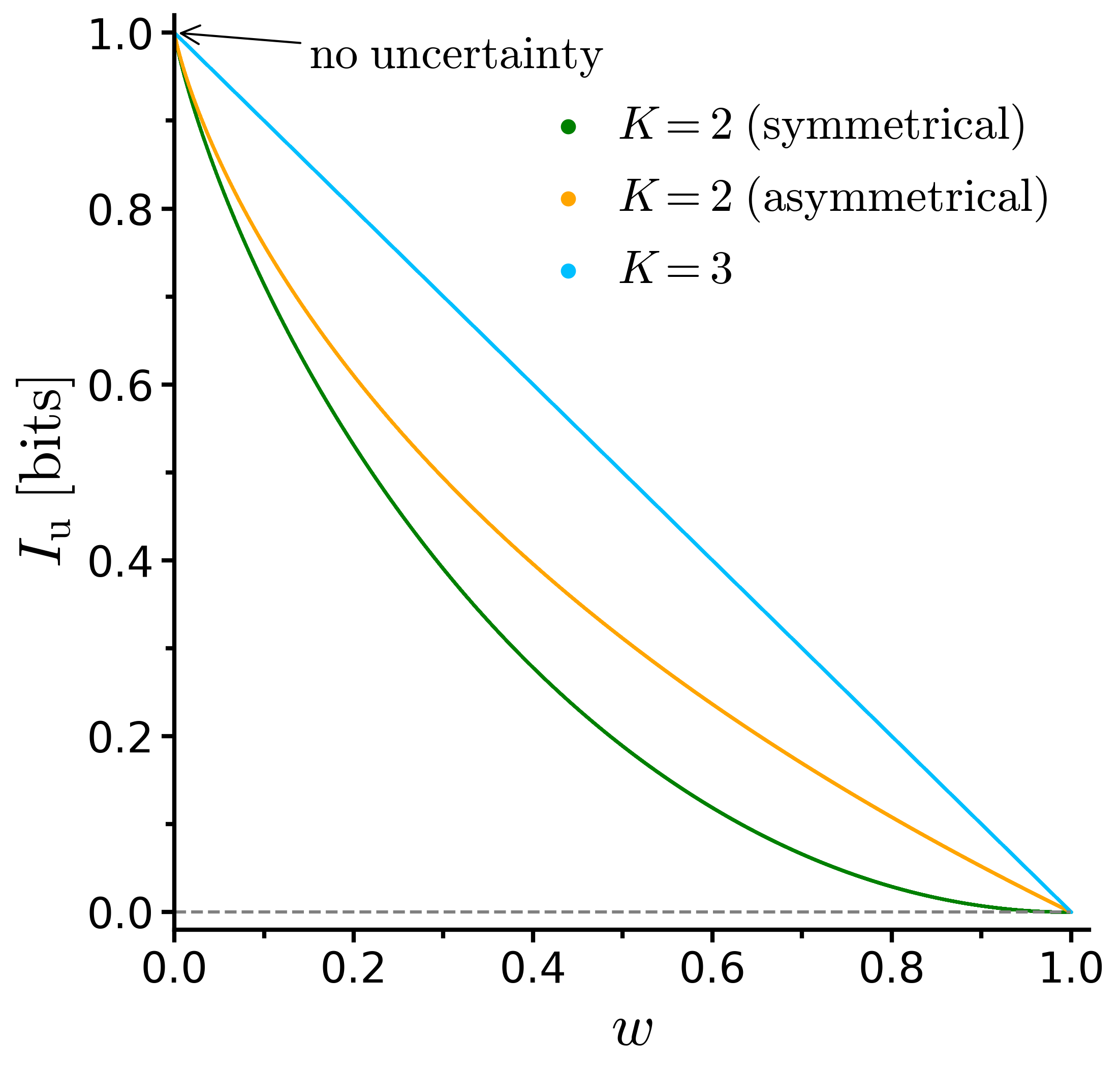}
    \phantomsubcaption
    \label{Fig:Irel}
\end{subfigure}
\caption{Total information (left) and usable information (right) in bits captured by the two coarse grainings discussed in Sec. \ref{Sec:CG} (orange and blue curves) as a function of the size of the uncertain region, $w$. Naive, symmetric two-state coarse graining for reference (green curve).}
\label{Fig:DetInfoCurves}
\end{figure*}

To compute the information quantities for the asymmetrical two-state coarse graining, recall that for a two-state memory we have $p(m=-1 \vert x) = 1-p(m=1 \vert x)$ and that $p(m) = \int_{-1/2}^{1/2}dx \, \rho(x) p(m \vert x)$. 
The probability of the more probable state, $m=1$, being realized is given by $p^{\rm (da2)}(m=1) = (1+w)/2$ and thus the total information stored in memory, Eq. (\ref{Eq:Imem_da2}), is:
\begin{eqnarray}
   I_{\rm m}^{\rm (da2)} &=& -\sum_m p^{\rm (da2)}(m) \ln[p^{\rm (da2)}(m)] \notag \\
   &=& -{1+w \over 2} \ln\left({1+w \over 2}\right)\!-\!{1-w \over 2} \ln\left({1-w \over 2}\right) \notag \\
   &=& h\left({1+w \over 2} \right).
\end{eqnarray}

By construction, the less probable memory state, in this case $m=-1$, carries no uncertainty about the relevant quantity $u$, 
\begin{equation}
    p^{\rm (da2)}(u \vert m=-1) = \delta_{-1\;u}.
\end{equation}
Removing the uncertainty from this state comes at the cost of increasing the inference error associated with the other memory state, making it larger than the error probability of a symmetric two-state coarse graining \footnote{Trivially, when $w=1$, the measurements do not carry any usable information, and the engine cannot be run.},
\begin{eqnarray}
    p^{\rm (da2)}(u=-1 \vert m=1) &=& {2 \over 1+w} \biggl(\int_{-1/2}^{-w/2} 0\;dx \notag \\
    &&+ \int_{-w/2}^{w/2} {1 \over 2} \; dx + \int_{w/2}^{1/2} 0\; dx \biggr) \notag \\
    &=& {2 \over 1+w}\int_{-w/2}^{w/2} {1 \over 2} \; dx \notag \\
    &=& {w \over 1+w} > {w \over 2} \quad \forall w < 1.
\end{eqnarray}
The usable information in memory is then given by Eq. (\ref{Eq:Irel_det2acg}) in the main text:
\begin{eqnarray}
    I_{\rm u}^{\rm (da2)} &=& \ln(2) + p^{\rm (da2)}(m=1) \notag \\
    &&\times\sum_u p^{\rm (da2)}(u \vert m=1) \ln[p^{\rm (da2)}(u \vert m=1)] \notag \\
    &=& \ln(2) + {1 + w \over 2} \notag \\
    && \times \left[{w \over 1+w} \ln\left({w \over 1+w}\right) +  {1 \over 1+w} \ln\left({1 \over 1+w}\right)\right] \notag \\
    &=& \ln(2) - {1 + w \over 2} h\left( {w \over 1+w} \right).
\end{eqnarray}
The asymmetrical coarse graining outperforms the symmetrical one, because combining the uncertain region with one of the certain regions not only reduces the costs, as seen in Fig. \ref{Fig:Imem}, but also results in one memory state without uncertainty about $u$, thus allowing for greater work extraction, see Fig. \ref{Fig:Irel}.

While two-state memories with even greater asymmetry between the two states cost even less to implement, the asymmetric memory discussed here captures the most usable information of any two-state coarse graining. Consequently it is the solution to Eqs. (\ref{Eq:IBalg}) for $\tau \to \infty$, if the number of memory states is constrained to two. 
We used parametric optimization to verify that for $\tau \leq \tau_{1 \to 2}^{\rm det}$ (see Eq. (\ref{Eq:taudagger})) deterministic observers cannot achieve positive net work output, thus their best option is to do nothing, i.e. use a one-state memory (see Fig. \ref{Fig:Det_PT}). 

The three-state coarse graining shown in blue in Fig. \ref{Fig:DetInfoCurves}, is the three-state coarse graining that captures all available usable information. This coarse graining is introduced in the beginning of Sec. \ref{Sec:Results}. The probability of realizing the individual memory states is given by the size of the three regions (left and right certain region and uncertain region) of the engine:
\begin{eqnarray}
    p^{\rm (d3)}(m=\pm 1) &=& {1-w \over 2}, \\
    p^{\rm (d3)}(m=0) &=& w.
\end{eqnarray}
Thus the total information stored in this memory is given by Eq. (\ref{Eq:3cg}). Since the $m=0$ state has maximum uncertainty about $u$, whereas the two other states have no uncertainty in their inference, the conditional entropy is:  
\begin{equation}
    H^{\rm (d3)}[u \vert m] = w\ln(2) = H[u \vert x].
\end{equation} 
Consequently this coarse graining retains all usable information available in the observable, 
Eq. (\ref{Eq:Irel_d3}).

\section{Parametric soft partitioning strategies} \label{App:PSP}
In this appendix we provide detailed calculations of the total information stored in memory as well as the usable part of it for the two soft partitionings studied in Sec. \ref{Sec:PSP} (see Fig. \ref{Fig:SP_param}), as well as for an asymmetrical soft partitioning with $K=2$. Additionally we compare thermodynamically optimal soft partitioning strategies to optimal solutions found algorithmically and to strategies limited to coarse graining. 

\subsection{Information quantities}
The calculation of the information quantities is done in analogy to the calculations for the deterministic assignments in Appendix \ref{App:DCG}. For soft partitionings $p(m \vert x)$ can in general no longer be expressed as a $\delta$-function and thus $H[m \vert x] \neq 0$, except in special cases.

The symmetric two-state partitioning shown in the upper panel of Fig. \ref{Fig:SP_param} still has \mbox{$p^{\rm (s2)}(m=1) = 1/2 = p^{\rm (s2)}(m=-1)$}, just like a symmetric two-state coarse graining. Consequently $H^{\rm (s2)}[m] = \ln(2)$. To determine the total amount of information stored in memory we compute the conditional entropy from the assignments given in Eq. (\ref{Eq:pmgx_2sp}),

\begin{eqnarray}
    H^{\rm (s2)}[m \vert x] &=&\!\! -\!\!\sum_m \int_{-1/2}^{1/2} dx \; p^{\rm (s2)}(m \vert x) \ln[p^{\rm (s2)}(m \vert x)] \notag \\
    &=& -2 \biggl({1-w \over 2} (1-q_2) \ln(1-q_2) \notag \\
    &&+ {w \over 2} \ln(1/2) + {1-w \over 2} q_2 \ln(q_2) \biggr) \notag \\
    &=& w \ln(2) + (1-w) h(q_2),
\end{eqnarray}
where we used the symmetry of the memory assignments to get rid of the sum over $m$ and the fact that $p^{\rm (s2)}(m \vert x)$ is constant within the three different regions (left certain region, uncertain region and right certain region). The total information stored in a symmetric two-state partitioning is thus given by Eq. (\ref{Eq:Imem_s2}).

Evaluating Eq. (\ref{Eq:pugmint}) for the assignments given in Eq. (\ref{Eq:pmgx_2sp}) we find,
\begin{eqnarray} \label{Eq:pugm_2sp}
p^{\rm (s2)}(u = 1 \vert m=-1) &=& \left(\int_{-w/2}^{w/2} {1 \over 2} \times {1 \over 2} \; dx  + \int_{w/2}^{1/2} q_2 \; dx \right)\notag \\
&=& w/2 + (1-w)q_2,
\end{eqnarray}
where we ignored the left certain region ($-1/2 \leq x < -w/2$), because in this region \mbox{$p(u=1 \vert x) = 0$}. Due to the symmetry of the memory, we have \mbox{$p^{\rm (s2)}(u=1 \vert m=-1) = p^{\rm (s2)}(u=-1 \vert m=1)$} and we arrive at Eq. (\ref{Eq:pugm_s2}).
Since $H[u] = \ln(2)$, we only need to calculate the conditional entropy $H^{\rm (s2)}[u \vert m]$ to quantify the usable information in memory:
\begin{eqnarray}
    H^{\rm (s2)}[u \vert m] &=& -2p^{\rm (s2)}(m=-1) \sum_u p^{\rm (s2)}(u \vert m=-1) \notag \\
    && \times \ln[p^{\rm (s2)}(u \vert m=-1)] \notag \\
    &=& -p^{\rm (s2)}(u\!=\!1 \vert m\!=\!-1) \ln[p^{\rm (s2)}(u\!=\!1 \vert m\!=\!-1)] \notag \\
    && -[1\!-\!p^{\rm (s2)}(u\!=\!1 \vert m\!=\!-1)]  \notag \\
    && \times \ln[1\!-\!p^{\rm (s2)}(u\!=\!1 \vert m\!=\!-1)] \notag \\
    &=& h(p^{\rm (s2)}(u \! \neq \! m \vert m)).
\end{eqnarray}
The amount of usable information in the symmetric two-state soft partitioning is then given by Eq. (\ref{Eq:Irel_s2}).

For the soft partitioning with three states we first compute $p(m) \!\!\!=\!\!\! \int_{-1/2}^{1/2} dx\,p(m \vert x)$ for each memory state. Using the memory assignments depicted in the lower panel of Fig. \ref{Fig:SP_param} and given by Eqs. (\ref{Eq:pmgx_3sp0}) - (\ref{Eq:pmgx_3sp}), we find $p^{\rm (s3)}(m=0) = w+(1-w)q_3$ and \mbox{$p^{\rm (s3)}(m=1) = (1-q_3)(1-w)/2 = p^{\rm (s3)}(m=-1)$}. Thus the entropy of the three-state soft partitioning is,
\begin{eqnarray} \label{Eq:Hm_s3}
    H^{\rm (s3)}[m] &=& -p^{\rm (s3)}(m=0) \ln[p^{\rm (s3)}(m=0)] - 2(1-q_3)\notag \\
    && \times (1-w)/2\biggl[\ln[(1-q_3)(1-w)] - \ln(2)\biggr] \notag  \\
    &=& -p^{\rm (s3)}(m=0) \ln[p^{\rm (s3)}(m=0)] \notag \\
    &&- [1\!-\!p^{\rm (s3)}(m=0)]\biggl[\ln[1\!-\!p^{\rm (s3)}(m=0)] \!-\! \ln(2)\biggr] \notag \\
    &=& [1-p^{\rm (s3)}(m=0)]\ln(2) \!+\! h(p^{\rm (s3)}(m=0)).
\end{eqnarray}
The conditional entropy is found to be,
\begin{eqnarray} \label{Eq:Hmgx_s3}
    H^{\rm (s3)}[m \vert x] &=& -\int_{-1/2}^{1/2} dx \; p^{\rm (s3)}(m=0 \vert x) \ln[p^{\rm (s3)}(m=0 \vert x)] \notag \\
    &&-2\int_{w/2}^{1/2} dx \; p^{\rm (s3)}(m=1 \vert x) \ln[p^{\rm (s3)}(m=1 \vert x)] \notag \\
    &=& -2\int_{w/2}^{1/2} dx \!\left[q_3 \ln(q_3) + (1-q_3) \ln(1-q_3)\right] \notag \\
    &=& (1-w) \left[-q_3\ln(q_3)-(1-q_3)\ln(1-q_3)\right] \notag \\ 
    &=& (1-w) h(q_3).
\end{eqnarray}
The total information retained by the three-state soft partitioning is given by the difference of Eq. (\ref{Eq:Hm_s3}) and Eq. (\ref{Eq:Hmgx_s3}), see Eq. (\ref{Eq:Imem_s3}) in the main text.

Since the three-state partitioning has $p^{\rm (s3)}(u=\pm 1 \vert m= \pm1) = 1$ and $p^{\rm (s3)}(u=1 \vert m=0) = 1/2$ finding the conditional entropy of the relevant quantity, $u$, given the memory state, $m$, is straightforward. Only the $m=0$ state contributes and since $p^{\rm (s3)}(u=1 \vert m=0) = 1/2 = p^{\rm (s3)}(u=-1 \vert m=0)$, it is sufficient to consider $u=1$ and multiply by 2:
\begin{eqnarray}
    H^{\rm (s3)}[u \vert m] &=& - 2 p^{\rm (s3)}(m=0)p^{\rm (s3)}(u=1 \vert m=0) \notag \\
    &&\times \ln[p^{\rm (s3)}(u=1 \vert m=0)] \notag \\
    &=& p^{\rm (s3)}(m=0) \ln(2).
\end{eqnarray}
Thus the usable information stored in the three-state partitioning is,
\begin{eqnarray}
    I_{\rm u}^{\rm (s3)} &=& \ln(2) \!-\! p^{\rm (s3)}(m=0) \ln(2) \!=\! (1\!-\!w\!-\!(1\!-\!w)q_3) \ln(2) \notag \\
    &=&(1-q_3)(1-w)\ln(2) = (1-q_3)I_{\rm u}^{\rm max}(w), 
\end{eqnarray}
which is the maximum relevant information, Eq. (\ref{Eq:TotalIrel}), reduced by $q_3I_{\rm u}^{\rm max}(w)$, since with probability $q_3$ the observer assigns an observation in one of the certain regions to the state $m=0$, corresponding to no action on the work medium.

\begin{figure}[h]
\centering 
\includegraphics[width=0.95\linewidth]{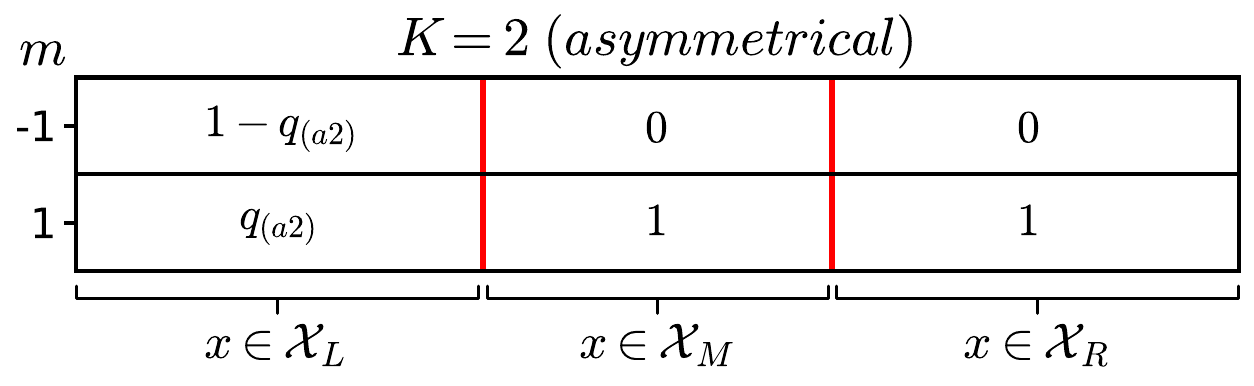}
\caption{Parameterization for asymmetric two-state soft partitioning strategy for an engine with $w=0.3$ (red lines at $\pm w/2$). Values are memory assignments, $p(m \vert x)$ for the corresponding regions of $x$.} 
\label{Fig:SP_asym}
\end{figure}

For completeness we also parameterize an asymmetrical two-state soft partitioning to compare it to the two other soft partitionings and the coarse grainings discussed in Sec. \ref{Sec:CG} and Appendix \ref{App:DCG}. The parameterization is shown in Fig. \ref{Fig:SP_asym}. Clearly there are two equivalent parameterization, since the memory state with greater weight can either be the state $m=1$ or $m=-1$. Here we calculate everything for $m=-1$ being the less probable state (as shown in Fig. \ref{Fig:SP_asym}).

The probabilities of being in either memory state are given by 
\begin{equation}
    p^{\rm (sa2)}(m= \pm 1) = (1 \pm \alpha)/2,
\end{equation} where we defined $\alpha \equiv (1-w)q_{(a2)}+w$ to simplify the notation.
The Shannon entropy of the two asymmetrical memory states is then
\begin{eqnarray}
    H^{\rm (sa2)}[m] &=& -{1 + \alpha \over 2} \ln\left({1 + \alpha \over 2}\right) - {1 - \alpha \over 2} \ln\left({1 - \alpha \over 2}\right) \notag \\
    &=& h\left({1 + \alpha \over 2}\right).
\end{eqnarray}
The conditional entropy is,
\begin{eqnarray}
    H^{\rm (sa2)}[m \vert x] &=& -{1 -w \over 2} \biggl[q_{(a2)}\ln(q_{(a2)}) \notag \\
    &&+ (1-q_{(a2)}) \ln(1-q_{(a2)})\biggr] \notag \\
    &=& {1 -w \over 2} h(q_{(a2)}),
\end{eqnarray}
and the information captured by the asymmetrical soft partitioning is,
\begin{eqnarray}
    I_{\rm m}^{\rm (sa2)} &=& h\left({1 + \alpha \over 2}\right) -  {1 -w \over 2} h(q_{(a2)})
\end{eqnarray}
If $m=-1$, the observer knows with certainty, that the right side of the container is empty, \mbox{$p^{\rm (sa2)}(u=1\vert m=-1) = 0$}. For the other memory state the inference error is given by \mbox{$p^{\rm (sa2)}(u=-1 \vert m=1)= \alpha/(1+\alpha)$}. Since only the more probable memory state has any uncertainty in the inference, the $m=-1$ state does not contribute to the conditional entropy of $u$ given $m$:
\begin{eqnarray}
    H^{\rm (sa2)}[u \vert m] &=& -{1 + \alpha \over 2}\biggl[{\alpha \over 1+\alpha} \ln\left({\alpha \over 1+\alpha}\right)  \notag \\
    &&+ {1 \over 1+\alpha} \ln\left({1 \over 1+\alpha}\right) \biggr]\notag \\
    &=& {1 + \alpha \over 2} \, h\left({\alpha \over 1+\alpha}\right).
\end{eqnarray}
Thus the usable information for the asymmetric two-state partitioning is given by
\begin{eqnarray}
    I_{\rm u}^{\rm (sa2)} = \ln(2) - {1 + \alpha \over 2} \, h\left({\alpha \over 1+\alpha}\right)
\end{eqnarray}

Note that we recover the corresponding deterministic quantities (see Appendix \ref{App:DCG}) for the asymmetric two-state and the three-state strategies by setting $q_K=0$. This is expected, since for $q_K=0$ the partitionings are no longer soft and we return to the coarse grained assignments (see Figs. \ref{Fig:SP_param} and \ref{Fig:SP_asym}). The symmetric two-state soft partitioning does not have a coarse grained counterpart, because, by design, observations in the uncertain region are assigned to either memory state completely at random.

\subsection{Comparison to numerical results for optimal observers and coarse graining}
\begin{figure}[h!]
\centering 
\includegraphics[width=0.95\linewidth]{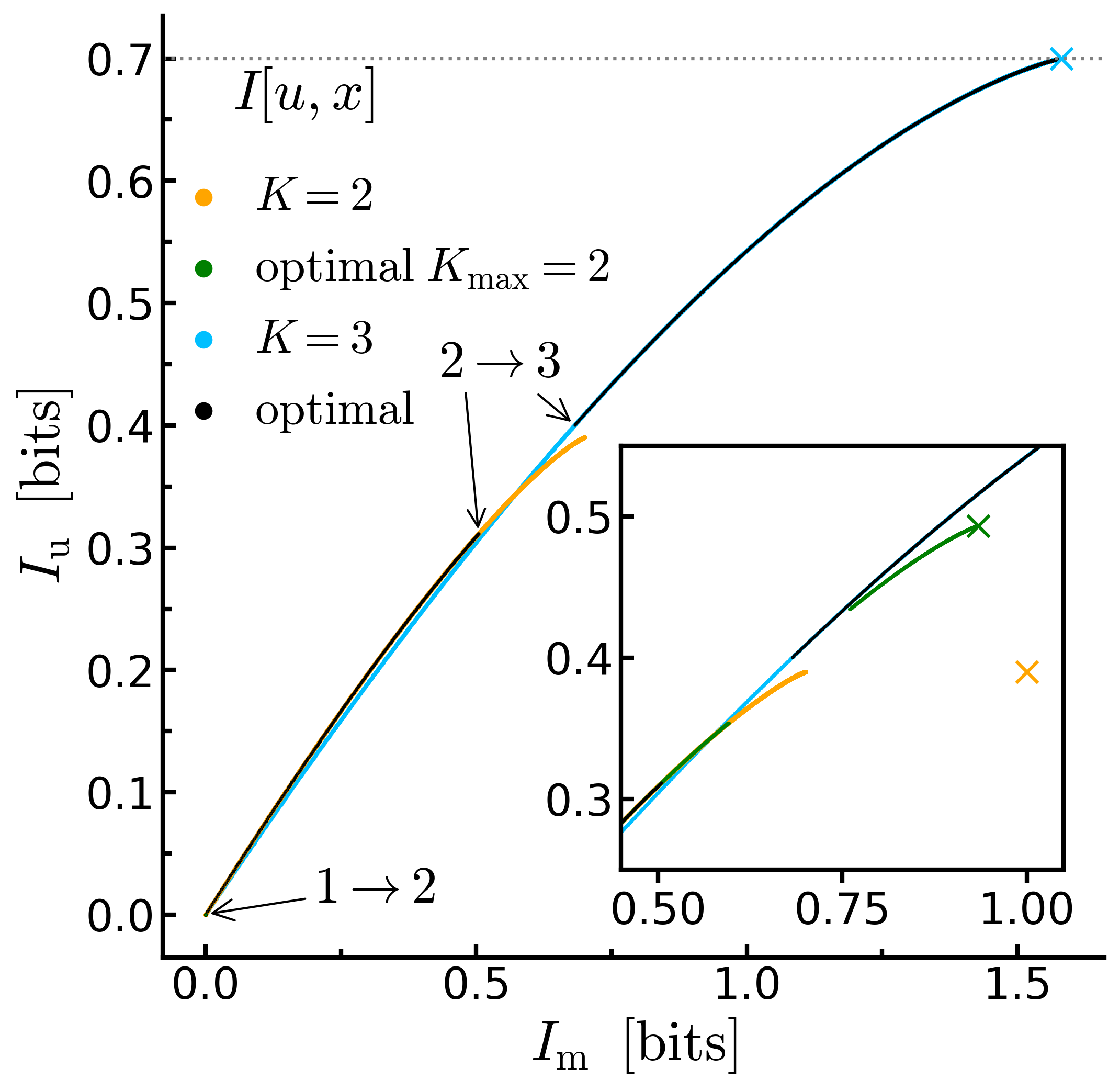}
\caption{Information plane representation of optimal solutions (black and green), compared to best parametric solutions (orange: $K=2$, blue: $K=3$), and deterministic coarse grainings (orange cross for naive, 2-state, symmetric coarse graining along the mid-line; green cross for best 2-state asymmetric coarse graining; blue cross for best 3-state coarse graining, see Sec. \ref{Sec:CG}). Work medium geometry: $w=0.3$. }
\label{Fig:Infoplane_c03}
\end{figure}

In Fig. \ref{Fig:Infoplane_c03} we compare the cost-benefit trade-off in the information plane between optimal memories, soft partitioning strategies and coarse grainings. It shows that optimal soft partitionings overlap with the optimal curve from the Information Bottleneck (IB) algorithm in the information plane.

When parametric soft partitionings are constrained to use at most two memory states ($K=3$ is prohibited), then the best solutions (orange curve in Fig. \ref{Fig:Infoplane_c03}) 
branch off from the optimal curve (black) when $\tau > 2$.  
The deterministic solutions are plotted with ``x" markers in Fig. \ref{Fig:Infoplane_c03}. Comparing the soft two-state partition with $q_2=0$ (end point of the orange curve), to the symmetric two-state coarse graining (orange ``x", further discussed in Appendix \ref{App:DCG}), we see that the latter captures no additional usable information, while incurring significant additional costs of about $1/4$ bit. 

Constraining the solutions found by the IB Algorithm to use at most two states (green curve in Fig. \ref{Fig:Infoplane_c03}), forces those solutions to be suboptimal at large enough $\tau$ ($\tau > 2$). For very large $\tau$, with this constraint in place, the asymmetric two-state coarse graining (green ``x" in Fig. \ref{Fig:Infoplane_c03}, see Sec. \ref{Sec:CG}) is found, which costs less bits to encode while simultaneously capturing more usable bits than the naive symmetric two-state coarse graining (this difference is further discussed in Appendix \ref{App:DCG}).

Symmetric and asymmetric two-state coarse grainings differ from optimal two-state soft partitions even when there is no residual probability in the assignments,$q_2=0$, because soft partitioning strategies assign observations in the uncertain region with equal probability to either of the two memory states. Therefore, by construction, symmetrical soft partitionings constrained to two states do not converge to a hard partitioning (a regular coarse graining). Since we also optimize over the number of memory states when finding thermodynamically optimal soft partitioning strategies, this is not an issue, as the best soft partitioning observers already use three memory states in the $\tau$ regime where the asymmetrical two-state coarse graining outperforms the symmetrical two-state soft partitioning.

Soft partitioning based memory making strategies save thermodynamic costs in the regime in which the thermodynamic value of the retained information is not large enough to warrant creating a summary of the data that is fully informative of the situation in the work medium.
These strategies are, of course, the same strategies as we observed in the algorithmically computed optimal observer memories, since the parameterization is inspired by those. The parameterization, however, eases interpretability of optimal observer strategies. 

The optimal parameterization at each $
\tau$ performs on-par with the optimal solutions, having less than $5\times10^{-5}$ \% of relative lost work output (relative to the maximum possible work output, $kT' I_{\rm u}^{\rm max}$) and the small existing difference is purely due to the numerical calculation of the parametric information quantities with a step size of $2\times10^{-6}$ in $q_K$. Note that for this comparison, we excluded all solutions that are within two annealing steps (annealing rate of $1.001$ was used) of a phase transition to a larger number of memory states. This was done, because phase transitions in the IB Algorithm contain a small degree of randomness and do not always happen at exactly the same value of $\tau$. Since analytic expressions for the information quantities of the soft partitioning solutions exist (see also Sec. \ref{Sec:PSP}), they do not suffer from this issue and so for some values of $w$, they might employ a solution with a larger number of memory states for one or two annealing steps before the solutions computed by the iterative algorithm also increase their number of states. These points would then dominate the difference in work output if they were included in the calculation.

More on phase transitions and computational issues of the numerical optimization for this model class can be found in Appendices \ref{App:CritTau} and \ref{App:AlgoIssues} respectively. 

Finding thermodynamically optimal soft partitioning strategies adds little computational hassle, compared to finding thermodynamically optimal coarse grainings, yet the soft partitioning strategies significantly outperform coarse graining in sizeable $\tau$ ranges (see. Fig. \ref{Fig:Det_diff}) and recover deterministic solutions, when they become optimal.

\section{Critical temperature ratios at transitions in the number of memory states}\label{App:CritTau}
\subsection{First transition} \label{App:FirstPT}
The value of the temperature ratio $\tau$, at which it becomes worthwhile for an optimal observer to memorize anything can be computed using the parametric soft partitions introduced in Sec. \ref{Sec:PSP}. 

\begin{figure}[h!]
\centering 
\includegraphics[width=0.95\linewidth]{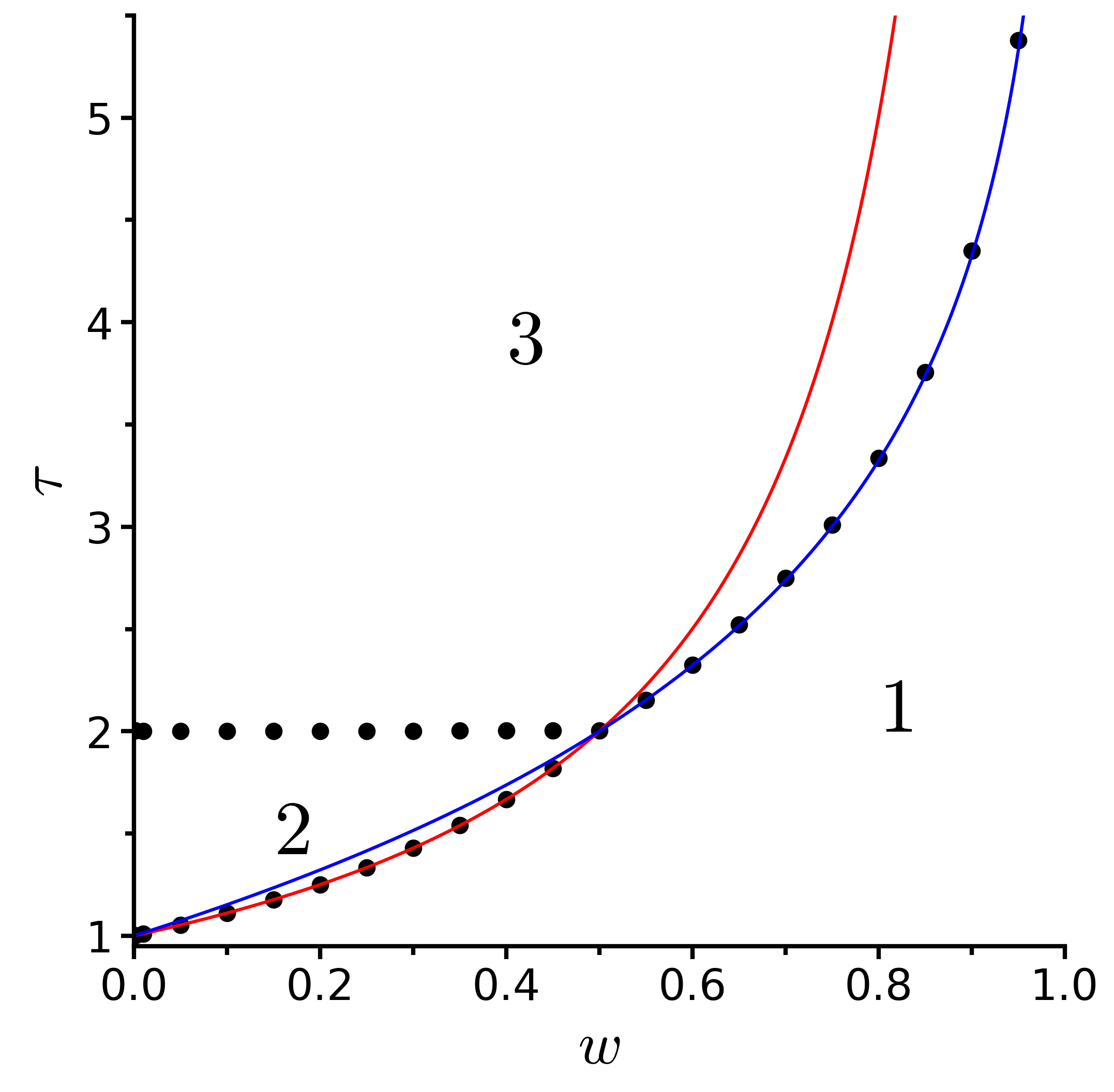}
\caption{Phase diagram for optimal observers. 
Numbers indicate number of memory states, black dots mark algorithmically found critical $\tau$ values (transitions to more memory states) for selected engine geometries. Colored curves show $\tau^*_{1 \to 2}(w)$ (red, Eq. (\ref{Eq:tau1-2_App})) and $\tau^*_{1 \to 3}(w)$ (blue, Eq. (\ref{Eq:tau1-3_App})).}
\label{Fig:PT_App}
\end{figure}

Before memorizing anything, an observer uses a data representation with $I_{\rm m}=I_{\rm u} = 0$, i.e. a one-state memory. To determine the critical $\tau$ value at which an observer can derive positive net work output from a two-state memory for the first time, we compare the net engine work output (in units of $kT'$) at the transition from one to two memory states:
\begin{equation}
    \lim_{q_2 \to {1 \over 2}} \left(I_{\rm u}^{\rm (s2)} - {1 \over \tau} I_{\rm m}^{\rm (s2)}\right) = 0,
\end{equation}
where the left hand side is the net engine work output of a parametric two-state soft partitioning right after the transition from a one-state memory and the right hand side is the net engine work output of an engine were nothing is memorized. The critical temperature ratio is thus given by,
\begin{eqnarray}
    \tau^*_{1 \to 2}(w) &=& \!\!\lim_{q_2 \to {1 \over 2}} {I_{\rm m}^{\rm (s2)} \over I_{\rm u}^{\rm (s2)}} \\
    &=& \!\!\lim_{q_2 \to {1 \over 2}} {(1-w)(\ln(2) - h(q_2)) \over \ln(2) \!-\! h(w/2\!+\!(1\!-\!w)q_2)} \!\to\! {0 \over 0}.
\end{eqnarray}
Since the limit approaches an indefinite expression of the form $0 / 0$, we can use L'H\^{o}pital's rule to evaluate the limit. The derivative of the binary entropy function with respect to its argument is 
\begin{equation}
    {d\over dx} h(ax) = a \ln\left({1-x \over x}\right).
\end{equation}
Applying L'H\^{o}pital's rule twice and using the above derivative we find 
\begin{eqnarray}
    \tau^*_{1 \to 2}(w) &=& \lim_{q_2 \to {1\over2}} {\ln\left({1-q_2 \over q_2}\right) \over \ln\left({1-w/2-(1-w)q_2 \over w/2+(1-w)q_2}\right)} \to {0 \over 0}\\
    &=& \lim_{q_2 \to {1\over2}} {{1 \over 1-q_2} + {1 \over q_2} \over {1-w \over 1-w/2-(1-w)q_2} + {1-w \over w/2+(1-w)q_2}} \\ 
    &=& {1 \over 1-w}. \label{Eq:tau1-2_App}
\end{eqnarray}
This expression describes the red curve in Fig. \ref{Fig:PT_App}.

Similarly we can compute the critical value of the temperature ratio, $\tau^*_{1 \to 3}$, for which a three-state memory outperforms doing nothing for the first time:
\begin{eqnarray}
    \tau^*_{1 \to 3}(w) &=& \lim_{q_3 \to 1} {I_{\rm m}^{\rm (s3)} \over I_{\rm u}^{\rm (s3)}} \\
    &=& \!\lim_{q_3 \to 1} \biggl({(1\!-\!q_3)(1\!-\!w)\ln(2) \over (1-q_3)(1-w)\ln(2)} \notag \\ &&- {(1\!-\!w)h(q_3) \!-\! h(w\!+\!(1\!-\!w)q_3) \over (1-q_3)(1-w)\ln(2)} \biggr) \!\to\! {0 \over 0}. \notag
\end{eqnarray}
As before, we use L'H\^{o}pital's rule to evaluate the limit,
\begin{eqnarray}
    \tau^*_{1 \to 3}(w) &=& \!\lim_{q_3 \to 1} \biggl({(1\!-\!w)\ln(2) + (1\!-\!w) \ln\left({1-q_3 \over q_3}\right) \over (1-w)\ln(2)} \notag \\
    && - {(1-w) \ln\left({(1-w)(1-q_3) \over w + (1-w)q_3}\right) \over (1-w) \ln(2)}\biggr) \notag \\
    &=& \!\lim_{q_3 \to 1}  \biggl(\!1\!+\!{\ln(1\!-\!q_3) \!-\! \ln(q_3) \!-\! \ln((1\!-\!w)(1\!-\!q_3)) \over \ln(2)} \notag \\
    && + {\ln(w\!+\!(1\!-\!w)q_3) \over \ln(2)} \biggr) \notag \\
    &=& 1 \!+\! \lim_{q_3 \to 1} \!\left({\ln(1\!-\!q_3) \over \ln(2)} - {\ln(1\!-\!q_3) \!+\! \ln(1\!-\!w) \over \ln(2)} \right) \notag \\
    &=& 1 - {\ln(1-w) \over \ln(2)}. \label{Eq:tau1-3_App}
\end{eqnarray}
This expression describes the blue curve in Fig. \ref{Fig:PT_App}. 

Note that at $w=1/2$ we have \mbox{$\tau^*_{1 \to 2}(1/2) = 2 = \tau^*_{1 \to 3}(1/2)$} and for $w > 1/2$, transitioning to a three-state memory becomes worthwhile at lower temperature ratios than transitioning to a two-state memory, $\tau^*_{1 \to 2}(w>1/2) > \tau^*_{1 \to 3}(w>1/2)$. Thus there are no optimal two-state memories for $w > 1/2$. Conversely  for $w < 1/2$, optimal observer first transition to using two memory states ($\tau^*_{1 \to 2}(w<1/2) < \tau^*_{1 \to 3}(w<1/2)$) and at $\tau=2$ they switch to three memory states, regardless of the value of $w$, see Fig. \ref{Fig:PT_App}. This universal second phase transition is analyzed in detail in Appendix \ref{App:SecondPT}.

\subsection{Second transition - two to three states} \label{App:SecondPT}
For engine geometries, where the uncertain region spans less than half of the total volume, $w < 1/2$, the parametric soft partitions can be used to derive analytic expressions for the net engine work output of optimal two- and three-state observers at the critical temperature ratio $\tau^*_{2 \to 3}=2$. For the reader's convenience detailed calculations of all quantities can be found in Appendix \ref{App:FullCalc}. 

At $\tau=2$, the net engine work output is given by
\begin{equation}
    W_{\rm out}^{(sK)}(q_K, w) \!=\! k T'\! \left(\!I_{\rm u}^{(sK)}(q_K, w) - \frac{1}{2} I_{\rm m}^{(sK)}(q_K, w) \!\right),
\end{equation}
with $K\in \{2,3\}$ (see Eqs. (\ref{Eq:W_out_2_general}) and (\ref{Eq:W_out_3_general}) for the full expression for each $K$). To find the value of $q_K$ that maximizes the work output, we take the derivative of $W_{\rm out}^{(sK)}(q_K, w)$ with respect to $q_K$ and set it to zero. For two states we have
\begin{eqnarray}
    \frac{d W_{\rm out}^{(s2)}(q_2, w)}{d q_2} &\overset{!}{=}& 0 \label{Eq:s2_diff}\\
    \implies \,\, q_2 = \frac{1}{2}  \,\, \lor \,\, q_2 = \frac{1}{2} &\pm& \frac{\sqrt{1+5w^2-4w-2w^3}}{2(1-w)^2}.\notag
\end{eqnarray}
Using the second derivative, we confirmed that in the region of interest, $0 < w < 1/2$, $q_2=1/2$ is a minimum of the net engine work output (detailed calculations in Appendix \ref{App:FullCalc}). This is expected, since it corresponds to a one-state memory, allowing for no net work output at all. For $w>1/2$, $q_2=1/2$ is the only solution to Eq. (\ref{Eq:s2_diff}) and in this region it corresponds to a maximum of the net engine work output, because at $\tau=2$ it is best to do nothing for two-state observers, if the uncertain region spans more than half of the work medium geometry (see also Fig. \ref{Fig:PT_App}). The other two solutions exist only for $w < 1/2$ and they correspond to just one probability, because by definition $0 \leq q_2 \leq 1/2$. (A memory with $q_2 > 1/2$ is identical to a memory with the labels of the memory states switched and $q_2 \leq 1/2$, see Sec. \ref{Sec:PSP}). 

Thus, at $\tau=2$, two-state observers achieve maximum net engine work output with 
\begin{equation} \label{Eq:q2*}
    q_2^*(w, \tau=2) = \frac{1}{2} - \frac{\sqrt{1+5w^2-4w-2w^3}}{2(1-w)^2}.
\end{equation} 

Similarly we can derive the probability $q_3^*(w, \tau=2)$, that maximizes net engine work output of three-state observers at $\tau=2$:
\begin{equation} \label{Eq:q3*}
    \frac{d W_{\rm out}^{(s3)}(q_3, w)}{d q_3} \overset{!}{=} 0 \, \implies \, q_3^*(w, \tau=2) = \frac{w}{1-w}.
\end{equation}
Again this result is only valid for $w < 1/2$. We confirmed that $q_3^*=w/(1-w)$ maximizes the net engine work output of three-state observers at $\tau=2$ using the second derivative (see Appendix \ref{App:FullCalc}).

\begin{figure}[h!]
\centering 
\includegraphics[width=0.95\linewidth]{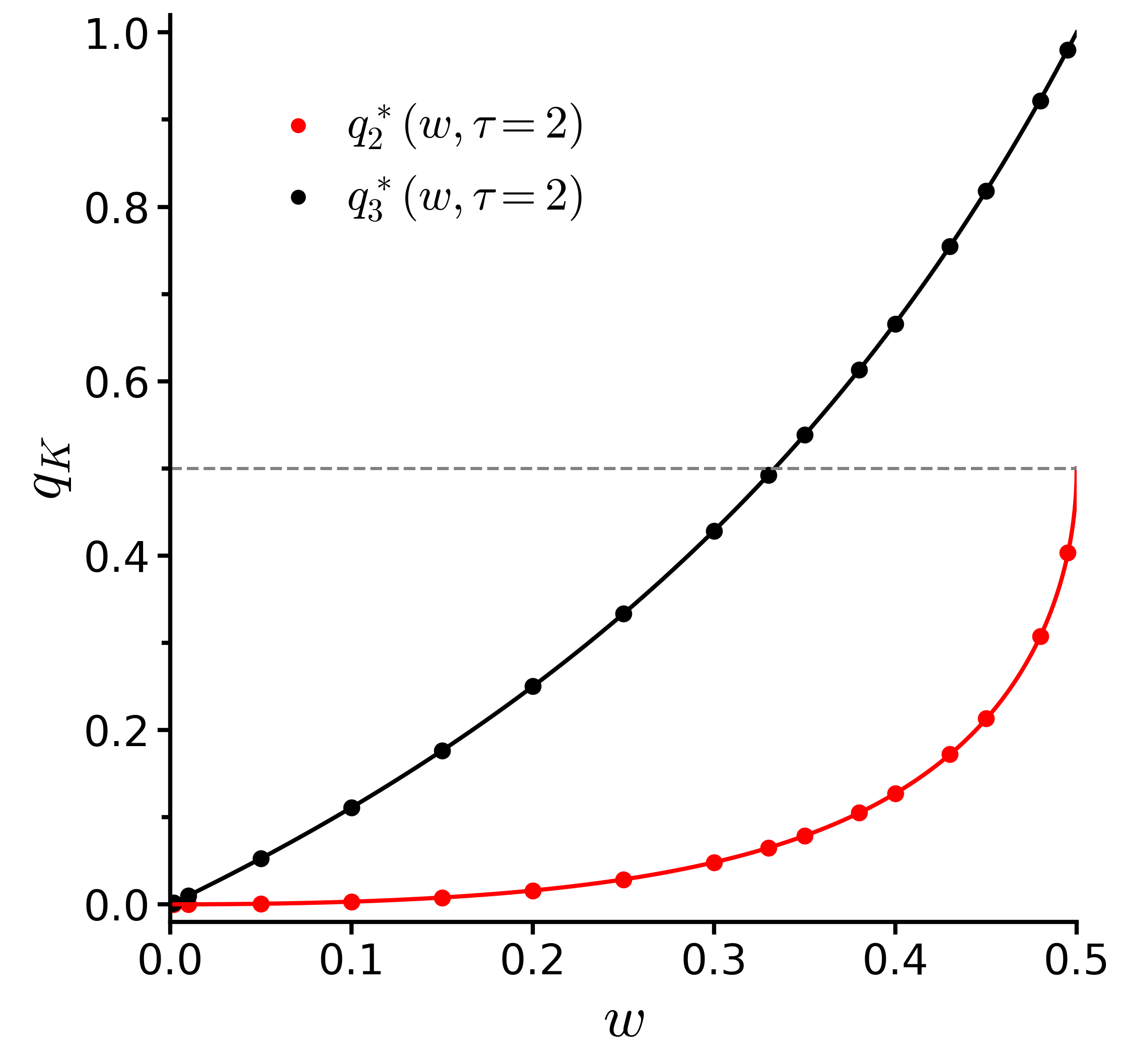}
\caption{Residual probabilities $q_2^*$ and $q_3^*$ that optimize the net engine work output of two- (red) and three-state (black) observers at $\tau=2$. Colored dots are numerical values for selected engine geometries, colored curves are analytical results (Eqs. (\ref{Eq:q2*}) and (\ref{Eq:q3*})). Dashed grey line at $q_K = 1/2$.}
\label{Fig:qK}
\end{figure}

The optimal residual probabilities for two- and three-state observers, $q_K^*(w, \tau=2)$, are depicted in Fig. \ref{Fig:qK}.

For engines with uncertain regions that span less than half of the work medium container, $w < 1/2$, the maximum net engine work output for optimal two- and three-state observers (Eqs. (\ref{Eq:W_out_2_general}) and (\ref{Eq:W_out_3_general}) evaluated at $q_2^*$ and $q_3^*$ respectively) is identical at $\tau=2$:
\begin{equation}
    W_{\rm out}^{(s2)}(q_2^*, w) = \frac{kT'}{2}(\ln(2) - h(w)) = W_{\rm out}^{(s3)}(q_3^*, w).
\end{equation} 

\begin{figure}[h]
\centering 
\includegraphics[width=0.95\linewidth]{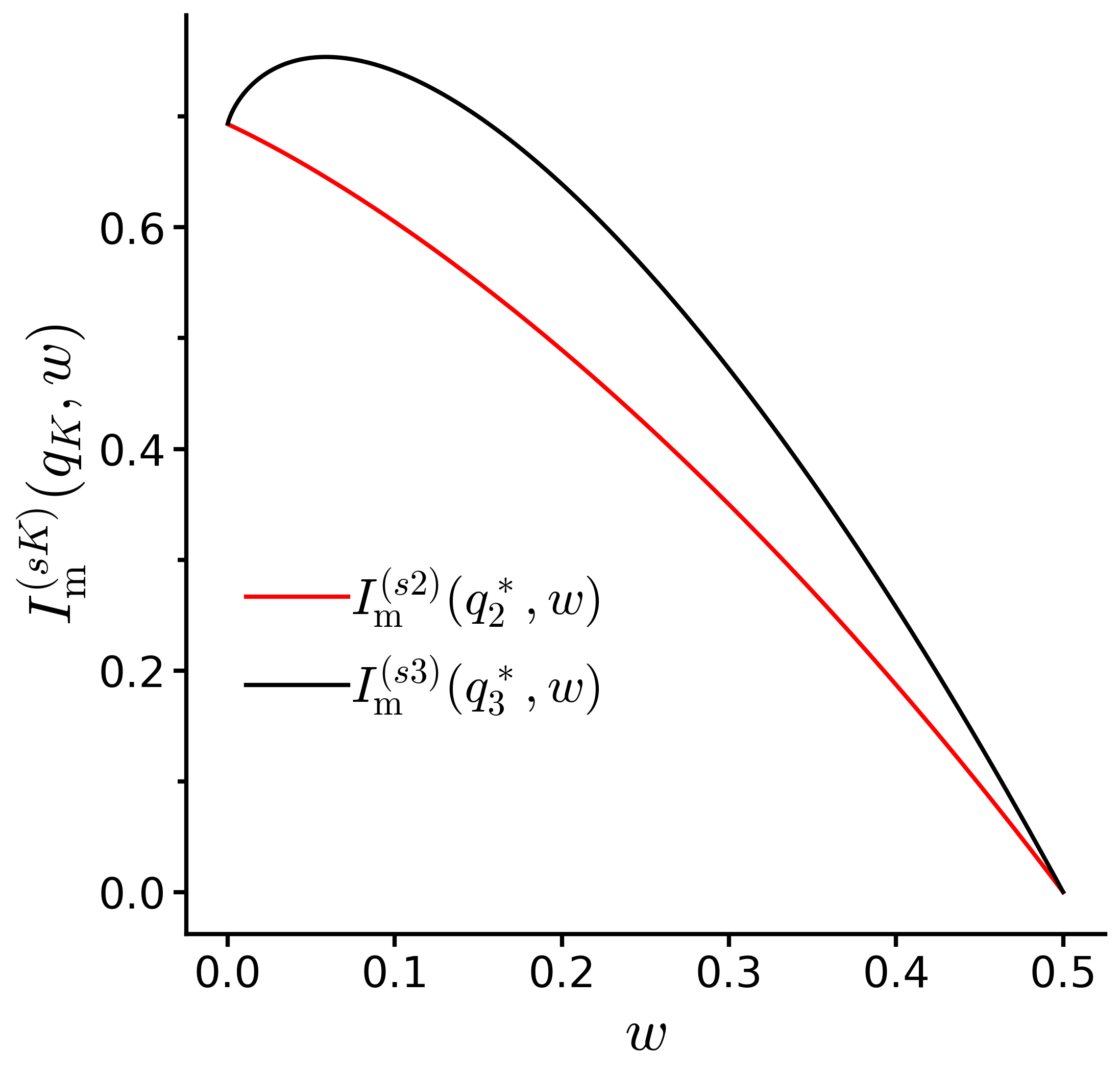}
\caption{Memorized information $I_{\rm m}^{(sK)}(q_K^*, w)$ of optimal observers as a function of $w$ at $\tau=2$. The black line marks optimal three-state observers, while the red line denotes optimal two-state observers.}
\label{Fig:Slopes}
\end{figure}

To see that for $\tau >2$ optimal three-state observers outperform their two-state counterparts, we will consider the slope of the net engine work output in units of $kT'$ as a function of $\tau$:
\begin{eqnarray}
    {W_{\rm out} \over kT'} &=& I_{\rm u} - \frac{1}{\tau} I_{\rm m} \\
    \implies \frac{\partial}{\partial \tau}\frac{W_{\rm out}}{kT'} &=& \frac{1}{\tau^2} I_{\rm m}.
\end{eqnarray}
The slope is proportional to $I_{\rm m}$, so if $I_{\rm m}^{\rm (s3)}(q_3^*, w) > I_{\rm m}^{\rm (s2)}(q_2^*, w)$, then optimal three-state observers will outperform optimal two-state observers for $\tau > 2$, while they are outperformed by optimal two-state observers for $\tau < 2$. Analytically it is hard to compare $I_{\rm m}^{\rm (s2)}(q_2^*, w)$ and $I_{\rm m}^{\rm (s3)}(q_3^*, w)$ (Eqs. (\ref{Eq:Imem_s2}) and (\ref{Eq:Imem_s3}) evaluated at $q_2^*$ and $q_3^*$ respectively), but graphically, in Fig. \ref{Fig:Slopes}, one can easily see that optimal three-state observers indeed memorize more information at $\tau=2$.

\subsubsection{Limit of vanishing uncertainty}
The critical value $\tau_{2\to 3}^{\rm det}(w)$, for which deterministic three-state observers begin to outperform minimally dissipative, deterministic two-state observers, can be found using Eq. (\ref{Eq:taustar}). If there is no uncertain region, $w=0$, we recover the \LS\ engine and since three-state observers are never optimal for it $\tau_{2\to 3}^{\rm det}(w)$ diverges at $w=0$. 

But what happens for $w \rightarrow 0$? Let us consider the limit of Eq. (\ref{Eq:taustar}) as $w$ approaches zero,
\begin{eqnarray}
\underset{w \to 0}{\lim} \tau_{2\to 3}^{\rm det}(w) \!=\! \underset{w \to 0}{\lim} {{1-w \over 2} \ln\left({2\over1-w}\right) \!-\! w\ln(w) \!-\! {1+w\over2} \ln\left({2\over1+w}\right) \over {1+w \over 2} h\!\left({w \over 1+w}\right) -w \ln(2)} \notag \\ 
\!\!= \underset{w \to 0}{\lim} {{1-w \over 2} \ln\left({2\over1-w}\right) \!-\! w\ln(w) \!-\! {1+w\over2} \ln\left({2\over1+w}\right) \over -w\ln(2) -w/2\ln(w) + {1+w \over 2} \ln(1+w)} \rightarrow {0 \over 0}. \notag
\end{eqnarray}
We can employ L'H\^{o}pital's rule when evaluating the limit, since the limit tends towards an undefined expression of the form $0/0$. Differentiating numerator and denominator separately, with respect to $w$, we find,
\begin{eqnarray}
    \underset{w \to 0}{\lim}\tau_{2\to 3}^{\rm det}(w) &=& \underset{w \to 0}{\lim} {\ln(2) + \ln(w) - \ln(1-w^2)/2 \over \ln(2) +\ln(w)/2 - \ln(1+w)/2} \notag \\&=& \underset{w \to 0}{\lim} {\ln(w) \over \ln(w)/2}  = 2,
\end{eqnarray}
where we used the fact that for $w \to 0$ all terms except the two $\ln(w)$ terms either vanish or become negligible since $\lim_{w \to 0}\ln(w) \to -\infty$.

As soon as there is any uncertainty in the information engine, no matter how small, deterministic three-state observers will not be able to outperform the best (least dissipative) deterministic two-state observers in terms of net engine work output for trade-off parameters $\tau < 2$. Since for $w \to 0$ the difference between probabilistic and deterministic observers vanishes, this bound also holds for optimal, probabilistic observers as can be seen in Figs. \ref{Fig:Obj}, \ref{Fig:PT} and \ref{Fig:PT_App}.

\section{Algorithmic details} \label{App:AlgoIssues}
To iteratively solve Eqs. (\ref{Eq:IBalg}) we used the same algorithm as in \cite{stilldaimer2022}. Pseudocode and a detailed description of the algorithm can be found in the appendix of \cite{stilldaimer2022}. Here we instead focus on algorithmic issues that occurred for our model class, due to the discontinuous shape of the divider in the work medium.

In a small region around $\tau=2$ there is a large degeneracy in the space of optimal solutions to Eqs. (\ref{Eq:IBalg}). We used the soft partitions discussed in Sec. \ref{Sec:PSP} and Appendix \ref{App:PSP} to verify that around $\tau=2$, all three types of soft partitioning strategies (symmetric and asymmetric two-state as well as three-state) yield (almost) identical net engine work output, Eq. (\ref{Eq:Woutengine}).
Consequently it is possible for the Information Bottleneck algorithm to arrive at an almost degenerate but suboptimal solution in that $\tau$-region. This effect is especially pronounced for uncertain regions spanning half the work medium container, $w_c = 1/2$, as the optimal strategies qualitatively change their behaviour for uncertain regions of that size (transitioning from one state immediately to three states, skipping two-state solutions). 

While the iterative algorithm is theoretically guaranteed to find the optimal solution at each value of $\tau$ \cite{rose1998deterministic}, the actual implementation relies on four important input parameters: 
\begin{itemize}
\item \emph{annealing rate}, used during the deterministic annealing procedure (default value $1.001$)
\item \emph{convergence threshold}, maximum difference below which solutions to two consecutive iteration steps are deemed converged (default value $10^{-32}$)
\item \emph{perturbation}, maximum perturbation applied when increasing the number of memory states (default value $5 \times 10^{-4}$)
\item \emph{merge tolerance}, minimum difference in the inference $p(u \vert m)$ above which two memory states are considered different from each other (default value $5\times 10^{-2}$).
\end{itemize}

To better understand the impact of these parameters, the splitting and merging of memory states during the annealing procedure has to be understood. States are split and merged in terms of their inference, $p(u \vert m) \in [0,1]$. 
For each value of $\tau$ (starting with $\tau=1$) the number of possible memory states is doubled, $K \to 2K$, and a small, random perturbation (between 0 and ${\emph perturbation} / \tau$, where ${\emph perturbation}$ is an input parameter) is applied to each new state. Then Eqs. (\ref{Eq:IBalg}) are iteratively solved for the $2K$ memory states until the solutions are deemed to be converged, which implies ${\rm DKL}[p(u \vert m) \vert \vert p_{\rm old}(u \vert m)] \leq \textit{convergence threshold}/\tau$. Here ${\rm DKL}$ denotes the Kulback-Leibler divergence, $p_{\rm old}(u \vert m)$ is the inference from the previous iteration and \emph{convergence threshold} is an input parameter. 

After convergence, all states $m$ and $m'\neq m$ for which the element-wise difference $\vert p(u \vert m) - p(u \vert m') \vert$ never exceeds the \emph{merge tolerance} are combined into a single average state, reducing the number of states from $2K$ to $K' \in [K, 2K]$. Equations (\ref{Eq:IBalg}) are then iteratively solved for the new number of states, $K'$, and the trade-off parameter, $\tau$, is increased to $\textit{annealing rate} \times \tau$. 

Depending on the initial choices for the four different parameters, the iterative algorithm will transition to a larger number of memory states at slightly different values of the trade-off parameter, $\tau$. While this effect is negligible for a broad range of input parameters and continuous $p(u \vert x)$, it becomes an issues for the discontinuous divider shapes considered here. 

For $0.4 < w < 0.6$, the iterative algorithm sometimes returns asymmetrical two-state memories as optimal solutions for one or two annealing steps around $\tau=2$.
Careful investigation and numerical comparison to the soft partitioning strategies shows that these solutions are not in fact optimal, but their net engine work output is (almost) identical to the optimal solutions that have three memory states. As $\tau$ increases the algorithm quickly arrives at the optimal three-state solution for $\tau \geq 2+\epsilon$. 

Since for $w > w_c$ there are no optimal two-state memories, the splitting procedure has to be adjusted. Instead of always doubling the number of states, i.e. going from $K=1$ to $K=2$ for the first transition, effectively forcing the algorithm to find a suboptimal two-state solution for $w > w_c$, we allowed the algorithm to use $K_{\rm max} > 2$ states in each annealing step. With this adjustment optimal solutions for $w \geq 0.6$ transition from one to three states, skipping the suboptimal two-state solutions, while optimal solutions for $w \leq 0.4$ correctly transition from one to two states and only use three states for $\tau = 2+\epsilon$. In the region $0.4 < w < 0.6$, where the degeneracy between optimal two- and three-state memories is the most pronounced, algorithmic solutions will still occasionally get stuck in suboptimal, asymmetric two-state solutions for a small number of annealing steps ($<5$). 

\newpage
\onecolumngrid
\section{Full calculation of maximum net engine work output at second phase transition} \label{App:FullCalc}
\setlength{\parindent}{0pt}
The net engine work output is given by Eq. (\ref{Eq:Woutengine}) in the main text:
\begin{equation}
    W_{\rm out} = kT'I_u - kTI_m~,
\end{equation}
or, in units of $kT'$, $W_{\rm out}/kT' = I_u - I_m/\tau$. For better readability, all work quantities in this section are given in units of $kT'$: $W_{\rm out}^{(sK)} := W_{\rm out}/kT'$.

The net engine work output for parametric two-state soft partitionings is:
\begin{eqnarray} 
     W_{\rm out}^{(s2)}(q_2, w, \tau) &=& I_{\rm u}^{(s2)}(q_2, w) - {1 \over \tau} I_{\rm m}^{(s2)}(q_2, w) \\
     &=& \ln(2) - h\left(\frac{w}{2} + (1-w)q_2\right) - \frac{1}{\tau}(1-w)(\ln(2)-h(q_2)) \label{Eq:W_out_2_general} ~,
\end{eqnarray}
where $I_{\rm m}^{(s2)}$ and $I_{\rm u}^{(s2)}$ are given in Eqs. (\ref{Eq:Imem_s2}) and (\ref{Eq:Irel_s2}) respectively. 
Take the derivative with respect to $q_2$ to find $q_2^*$, which maximizes the net engine work output at each value of $\tau$:
\begin{eqnarray}
    \frac{dW_{\rm out}^{(s2)}}{dq_2} &=& -(1-w) \ln\left(\frac{1-w/2-(1-w)q_2}{w/2+(1-w)q_2}\right) + \frac{1-w}{\tau} \ln\left(\frac{1-q_2}{q_2}\right) \overset{!}{=}0 \\
    \implies && \tau \ln\left(\frac{1-w/2-(1-w)q_2}{w/2+(1-w)q_2}\right) = \ln\left(\frac{1-q_2}{q_2}\right) \\
    &&\left(\frac{1-w/2-(1-w)q_2}{w/2+(1-w)q_2}\right)^{\tau} = \frac{1-q_2}{q_2}
\end{eqnarray}
To continue we set $\tau=2$, because we are interested in the behaviour at this critical value (also note that for general, non-integer $\tau$ no analytic solution exists):
\begin{eqnarray}
    \frac{(1-w/2-(1-w)q_2)^2}{(w/2+(1-w)q_2)^2} &=& \frac{1-q_2}{q_2} \\
    q_2(1+(w/2+(1-w)q_2)^2-2(w/2+(1-w)q_2)) &=& (1-q_2)(w^2/4+(1-w)^2q_2^2+(1-w)wq_2) \\
q_2 \!+\! \frac{w^2}{4}q_2\!+\!(1\!-\!w)^2q_2^3\!+\!(1\!-\!w)wq_2^2\!-\!wq_2\!-\!2(1\!-\!w)q_2^2 &=& \frac{w^2}{4} \!+\! (1\!-\!w)^2q_2^2 \!+\! (1\!-\!w)wq_2\!-\!\frac{w^2}{4}q_2\!-\!(1\!-\!w)^2q_2^3\!-\!(1\!-\!w)w^2q_2 \notag\\
    2(1-w)^2q_2^3 -3(1-w)^2q_2^2+\left(1-2w+\frac{3w^2}{2}\right)q_2 - \frac{w^2}{4} &=& 0 \\
    q_2 \left(2(1-w)^2q_2^2 -3(1-w)^2q_2+\left(1-2w+\frac{3w^2}{2}\right) \right) - \frac{w^2}{4} &=& 0.
\end{eqnarray}
The three roots of the cubic equation are:
\begin{equation}
    q_2 = \frac{1}{2}  \,\, \lor \,\, q_2 = \frac{1}{2} \pm \frac{\sqrt{1+5w^2-4w-2w^3}}{2(1-w)^2}.
\end{equation}
The term inside the square root is only positive for $0 \leq w < 1/2$. At $w=1/2$, it is zero, so the only root is $q_2=1/2$. 

The second derivative of the net engine work output with respect to $q_2$ is
\begin{eqnarray}
    \frac{d^2 W_{\rm out}^{(s2)}(q_2, w, \tau)}{d q_2^2} &=& \frac{(1-w)^2}{1-w/2/(1-w)q_2} + \frac{(1-w)^2}{w/2+(1-w)q_2}-\frac{1-w}{\tau} \frac{q_2+1-q_2}{q_2(1-q_2)}\\
    &=& \frac{(1-w)^2}{w/2-w^2/4+(1-w)^2(q_2-q_2^2)}-\frac{1-w}{\tau} \frac{1}{q_2(1-q_2)}.
\end{eqnarray}
For $\tau=2$ and $q_2 =1/2$ we have
\begin{equation}
    \frac{d^2 W_{\rm out}^{(s2)}(q_2=1/2, w, \tau=2)}{dq_2^2} = 4(1-w)^2 - 2(1-w) = \begin{cases}
        > 0 \,\, \text{for} \, w < 1/2 \\
        < 0 \,\, \text{for} \, w > 1/2,
    \end{cases}
\end{equation}
so $q_2 = 1/2$ is a minimum of the net engine work output at $\tau=2$ if $w < 1/2$, but a maximum if $w>1/2$. This matches Figs. \ref{Fig:PT} and \ref{Fig:PT_App}, where we see that for $w > 1/2$ doing nothing is the optimal strategy at $\tau =2$, while for $w < 1/2$, optimal observers can derive positive net engine work output from a two-state memory. 

To verify that the other root of the first derivative maximizes the net engine work output for $0 \leq w < 1/2$, let us fist rewrite the second derivative at $\tau=2$:
\begin{eqnarray}
    \frac{d^2 W_{\rm out}^{(s2)}(q_2, w, \tau\!=\!2)}{dq_2^2} \!&=&\! \frac{(1-w)^2}{w/2-w^2/4+(1-w)^2(q_2-q_2^2)}-\frac{1-w}{2} \frac{1}{q_2(1-q_2)} \\
    \!&=&\! \frac{2(1-w)^2(q_2-q_2^2) - (1-w)(w/2-w^2/4+(1-w)^2(q_2-q_2^2))}{2(q_2-q_2^2)(w/2-w^2/4+(1-w)^2(q_2-q_2^2))}\\
    \!&=&\! \frac{(1-w)\left((1+w)(1-w)(q_2-q_2^2)-w/2\left(1-w/2\right)\right)}{2(q_2-q_2^2)(w/2-w^2/4+(1-w)^2(q_2-q_2^2))}\\
    \!&=&\! \frac{(1-w)}{2(q_2-q_2^2)(w/2-w^2/4+(1-w)^2(q_2-q_2^2))} \left(\!(q_2-q_2^2) \!+\! w^2\!\left(\frac{1}{4} \!-\! (q_2\!-\!q_2^2)\right)\!-\! \frac{w}{2} \right)\!. \label{Eq:W2SecondDeriv}
\end{eqnarray}
Since the fraction in Eq. (\ref{Eq:W2SecondDeriv}) is always greater than zero, it is sufficient to consider the second factor. Moreover since $0\leq q_2 \leq 1/2$ by definition, we need only consider 
\begin{equation}\label{Eq:q2_star}
    q_2^* = \frac{1}{2} - \frac{\sqrt{1+5w^2-4w-2w^3}}{2(1-w)^2}.
\end{equation}
This gives us a maximum of the net engine work output if 
\begin{equation}
    (q_2^*-q_2^{*2}) + w^2\left(\frac{1}{4} - (q_2^*-q_2^{*2})\right) - \frac{w}{2} < 0 \qquad \forall w \in \left(0, 1/2\right).
\end{equation}
Inserting the expression for $q_2^*$, Eq. (\ref{Eq:q2_star}), and using
\begin{equation}
    (q_2^*-q_2^{*2}) = \frac{1}{4} - \left(\frac{\sqrt{1+5w^2-4w-2w^3}}{2(1-w)^2}\right)^2~,
\end{equation}
we get
\begin{eqnarray}
    (q_2^*\!-\!q_2^{*2}) \!+\! w^2\left(\frac{1}{4} \!-\! (q_2^*\!-\!q_2^{*2})\right) \!-\! \frac{w}{2}  &=& \frac{1-2w}{4} - (1-w^2) \left(\frac{\sqrt{1+5w^2-4w-2w^3}}{2(1-w)^2}\right)^2\\
    &=& \frac{(1-2w)(1-w)^4 - (1-w^2)(1+5w^2-4w-2w^3)}{4(1-w)^4}\\
    &=& \frac{w^4\!-\!4w^3\!+\!6w^2\!-\!4w\!-\!2w^5\!+\!8w^4\!-\!12w^3\!+\!8w^2\!-\!2w\!-\!5w^2\!+\!4w\!+\!2w^3\!+\!w^2\!+\!5w^4\!-\!4w^3\!-\!2w^5}{4(1-w)^4} \notag \\
    &=& \frac{-4w^5 + 14w^4-18w^3+10w^2-2w}{4(1-w)^4} \\
    &=& \frac{-2w(w-1)^3(2w-1)}{4(1-w)^4} \, < 0 \qquad \forall w \in(0, 1/2). \label{Eq:W2SecondDeriv2}
\end{eqnarray}
The denominator in Eq. (\ref{Eq:W2SecondDeriv2}) is always positive. The numerator is zero at $w=0$ and $w=1/2$ (and $w=1$, but this is not in the region of interest). For $0< w < 1/2$, every term in the numerator is less than zero and since the three terms are multiplied the whole numerator is less than zero. Thus we have shown that $q_2^*$ maximizes the net engine work output for $0 < w < 1/2$.

For three-state observers the net engine work output is 
\begin{eqnarray}
    W_{\rm out}^{(s3)}(q_3, w, \tau) &=& I_{\rm u}^{(s3)}(q_3, w) - {1 \over \tau} I_{\rm m}^{(s3)}(q_3, w) \\
    &=& (1-q_3)(1-w)\ln(2) - \frac{1}{\tau}\left((1-w)(1-q_3)\ln(2) - (1-w) h(q_3) + h(w+(1-w)q_3)\right), \label{Eq:W_out_3_general}
\end{eqnarray}
where $I_{\rm m}^{(s3)}$ and $I_{\rm u}^{(s3)}$ are given in Eqs. (\ref{Eq:Imem_s3}) and (\ref{Eq:Irel_s3}) respectively. 
Taking the first derivative with respect to $q_3$ and setting it to zero to find the local extrema, we have
\begin{equation}
        \frac{dW_{\rm out}^{(s3)}}{dq_3} = -(1-w) \ln(2) - \frac{1}{\tau}\left(-(1-w)\ln(2)-(1-w)\ln\left(\frac{1-q_3}{q_3}\right) + (1-w)\ln\left(\frac{1-w-(1-w)q_3}{w+(1-w)q_3}\right)\right) \overset{!}{=}0
\end{equation}
    \begin{eqnarray}
    \implies \ln(2)\left(1-\frac{1}{\tau}\right) &=& \ln\left(\frac{1-q_3}{q_3}\right) - \ln\left(\frac{(1-w)(1-q_3)}{w+(1-w)q_3}\right)\\
    (\tau-1)\ln(2) &=& \ln\left(\frac{(1-q_3)(w+(1-w)q_3}{q_3(1-w)(1-q_3)}\right)\\
    2^{\tau-1} &=& \frac{(1-q_3)w+q_3(1-w)(1-q_3)}{q_3(1-w)(1-q_3)}\\
    2^{\tau-1} -1 &=& \frac{w}{q_3(1-w)} \\
    q_3^* &=& \frac{w}{(1-w)(2^{\tau-1} -1)}  \label{Eq:General_q3*}
\end{eqnarray}

To verify that $q_3^*$ maximizes the net engine work output for a three-state memory at any value of $\tau$, we consider the second derivative:
\begin{equation}
    \frac{d^2W_{\rm out}^{(s3)}(q_3, w, \tau)}{dq_3^2} = -\frac{1}{\tau}\left(-(1-w)\left(-\frac{1}{1-q_3} - \frac{1}{q_3}\right) + (1-w) \left(\frac{-(1-w)}{(1-w)(1-q_3)}-\frac{(1-w)}{w+(1-w)q_3}\right)\right).
\end{equation}
Inserting Eq. (\ref{Eq:General_q3*}) we find:
\begin{eqnarray}
    \frac{d^2W_{\rm out}^{(s3)}(q_3^*, w, \tau)}{dq_3^2} \!&=&\! -\frac{1}{\tau}\left(\frac{(1-w)^2(2^{\tau-1}-1)}{2^{\tau-1}(1-w)-1} \!+\! \frac{(1-w)^2(2^{\tau-1}-1)}{w} \!-\! \frac{(1-w)^2(2^{\tau-1}-1)}{2^{\tau-1}(1-w)-1}\!-\!\frac{(1-w)^2}{w + \frac{w}{2^{\tau-1}-1}}\right) \\
    \!&=&\! -\frac{(1-w)^2(2^{\tau-1}-1)}{\tau} \left( \frac{1}{w} - \frac{1}{(2^{\tau-1}-1)w + w}\right) \\
    \!&=&\! -\frac{(1-w)^2(2^{\tau-1}-1)^2}{2^{\tau-1}\tau w} < 0 \qquad \forall w < 1 \wedge \tau > 1.
\end{eqnarray}
Thus $q_3^*$ maximizes the net engine work output for any value of $\tau > 1$ and $w < 1$. At $\tau = 2$ we have
\begin{equation}
    q_3^*(w, \tau=2) = \frac{w}{1-w}~,
\end{equation}
which is the ratio of completely uninformative observables, $w$, to completely informative ones, $1-w$.

Let us compute the maximum net engine work output (in units of $kT'$) of engines run by two- and three-state observers at a temperature ratio of $\tau=2$. For three-state observers we have:
\begin{eqnarray}
    W_{\rm out}^{(s3)}\left(q_3^*\!=\!\frac{w}{1-w}, w, \tau\!=\!2\right) \!&=&\! (1\!-\!q_3^*)(1\!-\!w)\ln(2) \!-\! \frac{1}{2}\left((1\!-\!w)(1\!-\!q_3^*)\ln(2) \!-\! (1\!-\!w) h(q_3^*) \!+\! h(w\!+\!(1\!-\!w)q_3^*)\right) \notag \\
    \!&=&\! (1-2w) \ln(2) - \frac{1}{2}\left((1-2w)\ln(2)-(1-w)h\left(\frac{w}{1-w}\right) + h(2w) \right) \notag \\
    \!&=&\! \frac{1\!-\!2w}{2}\ln(2) \!+\! \frac{1\!-\!w}{2}\left(-\frac{w}{1\!-\!w}\ln\left(\frac{w}{1\!-\!w}\right)\!-\!\frac{1\!-\!2w}{1\!-\!w}\ln\left(\frac{1\!-\!2w}{1\!-\!w}\right)\right) \notag \\
    &&- \frac{1}{2}\left(-2w\ln(2w)\!-\!(1\!-\!2w)\ln(1\!-\!2w)\right) \notag \\
    \!&=&\! \frac{1\!-\!2w}{2}\ln(2) \!-\! \frac{w}{2}\ln\left(\!\frac{w}{1\!-\!w}\!\right) \!-\! \frac{1\!-\!2w}{2}\ln\left(\!\frac{1\!-\!2w}{1\!-\!w}\!\right) \!+\! w\ln(2w) \!+\! \frac{1\!-\!2w}{2}\ln(1\!-\!2w) \notag \\
    \!&=&\! \frac{1}{2}\biggl((1-2w)\ln(2) -w\ln(w) +(1-w)\ln(1-w) + 2w\ln(2w)\biggr) \notag \\
    \!&=&\! \frac{1}{2} \biggl(\ln(2) + w\ln(w) + (1-w)\ln(1-w)\biggr) \notag \\
    \!&=&\! \frac{1}{2}\biggl(\ln(2) - h(w)\biggr).
\end{eqnarray}

For two-state observers the maximum net engine work output at $\tau=2$ is achieved with residual probability
\begin{equation}
    q_2^* = \frac{1}{2} - \frac{\sqrt{1+5w^2-4w-2w^3}}{2(1-w)^2}.
\end{equation}
To simplify the notation we use $a \equiv \sqrt{1+5w^2-4w-2w^3}$. The maximum net work output (in units of $kT'$) achievable by two state observers at $\tau=2$ is,
\begin{eqnarray}
    W_{\rm out}^{(s2)}(q_2^*, w, \tau\!=\!2) &=& \ln(2)-h(w/2+(1-w)q_2^*) - \frac{1}{2}(1-w)\biggl(\ln(2) - h(q_2^*)\biggr) \\
    &=& \frac{1+w}{2}\ln(2) - h\biggl(\frac{w}{2} + \frac{(1-w)^2-a}{2(1-w)}\biggr) + \frac{1-w}{2}h\biggl(\frac{1}{2}-\frac{a}{2(1-w)^2}\biggr)\\
    &=& \frac{1+w}{2}\ln(2) - h\biggl(\frac{1}{2}-\frac{a}{2(1-w)}\biggr) + \frac{1-w}{2}h\biggl(\frac{1}{2}-\frac{a}{2(1-w)^2}\biggr)\\
    &=& \frac{1+w}{2}\ln(2)+\biggl(\frac{1}{2}-\frac{a}{2(1-w)}\biggr)\ln\biggl(\frac{1}{2}-\frac{a}{2(1-w)}\biggr) + \biggl(\frac{1}{2}+\frac{a}{2(1-w)}\biggr)\ln\biggl(\frac{1}{2}+\frac{a}{2(1-w)}\biggr) \notag \\
    &&-\frac{1\!-\!w}{2}\left(\biggl(\frac{1}{2}\!-\!\frac{a}{2(1\!-\!w)^2}\biggr)\ln\biggl(\frac{1}{2}\!-\!\frac{a}{2(1\!-\!w)^2}\biggr) \!+\! \biggl(\frac{1}{2}\!+\!\frac{a}{2(1\!-\!w)^2}\biggr)\ln\biggl(\frac{1}{2}\!+\!\frac{a}{2(1\!-\!w)^2}\biggr)\right) \\
    &=& \frac{1+w}{2}\ln(2)-\ln(2(1-w))+\frac{1-w}{2}\ln(2(1-w)^2) +\biggl(\frac{1}{2}-\frac{a}{2(1-w)}\biggr)\ln(1-w-a) \notag \\ &&+\biggl(\frac{1}{2}+\frac{a}{2(1-w)}\biggr)\ln(1-w+a) - \biggl(\frac{1-w}{4} - \frac{a}{4(1-w)} \biggr) \ln((1-w)^2-a)\notag \\ &&-\biggl(\frac{1-w}{4} + \frac{a}{4(1-w)} \biggr) \ln((1-w)^2+a) \\
    &=& -w\ln(1-w) + \frac{1}{2}\ln((1-w)^2-a^2) - \frac{1-w}{4}\ln((1-w)^4-a^2) \notag \\
    &&+ \frac{a}{2(1-w)}\ln\biggl(\frac{1-w+a}{1-w-a}\biggr) - \frac{a}{4(1-w)}\ln\biggl(\frac{(1-w)^2+a}{(1-w)^2-a}\biggr). \label{Eq:Wout2_App}
\end{eqnarray}
The second and third term in Eq. (\ref{Eq:Wout2_App}) can be simplified by using:
\begin{equation}
    (1-w)^2-a^2 = 2w(1-w)^2 \qquad (1-w)^4-a^2 = w^2(1-w)^2.
\end{equation}
Using these identities, we can work out a concise expression:
\begin{eqnarray}
    W_{\rm out}^{(s2)}(q_2^*, w, \tau\!=\!2) &=& -w\ln(1-w) + \frac{1}{2}\ln(2w(1-w)^2) - \frac{1-w}{4}\ln(w^2(1-w)^2) \notag \\
    &&+ \frac{a}{4(1-w)}\biggl(2\ln\biggl(\frac{1-w+a}{1-w-a}\biggr)-\ln\biggl(\frac{(1-w)^2+a}{(1-w)^2-a}\biggr)\biggr)\\
    &=& \frac{1}{2}\biggl(\ln(2) \!+\! w \ln(w) \!+\!(1-w)\ln(1-w)\biggr) \!+\! \frac{a}{4(1-w)}\ln\biggl(\frac{(1-w+a)^2((1-w)^2-a)}{(1-w-a)^2((1-w)^2+a)}\biggr)\\
    &=& \frac{1}{2}\biggl(\ln(2) - h(w)\biggr) + \frac{a}{4(1-w)}\ln\biggl(\frac{(1-w+a)^2((1-w)^2-a)}{(1-w-a)^2((1-w)^2+a)}\biggr).\label{Eq:Wout2Final}
\end{eqnarray}
Let us consider the numerator and denominator of the argument of the natural log in Eq. (\ref{Eq:Wout2Final}) separately:
\begin{eqnarray}
    (1-w+a)^2((1-w)^2-a) &=& ((1-w)^2+a^2+2(1-w)a)((1-w)^2-a)\\
    &=& (1-w)^4 + a^2(1-w)^2 + 2(1-w)^3a - (1-w)^2a - a^3 - 2(1-w)a^2\\
    &=& (1-w)^4 + a^2((1-w)^2-2(1-w)) + a(2(1-w)^3-(1-w)^2) - a^3\\
    &=& (1-w)^4 + a^2 (w^2-1),
\end{eqnarray}
where we used $2(1-w)^3-(1-w)^2 = a^2$ to cancel the $a^3$ term.
\begin{eqnarray}
    (1-w-a)^2((1-w)^2+a) &=& ((1-w)^2+a^2-2(1-w)a)((1-w)^2+a)\\
    &=& (1-w)^4 + a^2(1-w)^2 - 2(1-w)^3a + (1-w)^2a + a^3 - 2(1-w)a^2\\
    &=& (1-w)^4 + a^2((1-w)^2-2(1-w)) - a(2(1-w)^3-(1-w)^2) + a^3\\
    &=& (1-w)^4 + a^2 (w^2-1).
\end{eqnarray}
Since the numerator and denominator of the argument of the natural log are equal, the last term in Eq. (\ref{Eq:Wout2Final}) is zero ($\ln(1)=0$) and the maximum net engine work output at $\tau=2$, in units of $kT'$, is
\begin{equation}
    W_{\rm out}^{(s2)}(q_2^*, w, \tau=2) = \frac{1}{2}\biggl(\ln(2) - h(w)\biggr) =  W_{\rm out}^{(s3)}(q_3^*, w, \tau=2).
\end{equation}

\twocolumngrid
\bibliographystyle{unsrt}
\bibliography{CB_Box}
\end{document}